\documentclass[prx,english,superscriptaddress,floatfix,longbibliography,nofootinbib]{revtex4-2}

\usepackage[utf8]{inputenc} 
\usepackage[T1]{fontenc} 
 
\usepackage{amsmath}
\usepackage{amsfonts}
\usepackage{amssymb}
\usepackage{graphicx}
\usepackage{bm} 
\usepackage{dcolumn}

\usepackage{xspace}
\usepackage{placeins}

\usepackage{hyperref}
\usepackage[all]{hypcap} 
\hypersetup{
 pdftitle={ stochastic accleration },%
 pdfauthor={R. Zamansky},%
 colorlinks={true}, 
 linkcolor={black}, 
 citecolor={black},
 pagecolor={black},
 urlcolor={black},
 unicode={true},
}

\DeclareMathOperator\erf{erf} 

\begin{document}
	
	\title{Acceleration scaling and stochastic dynamics of a fluid particle in turbulence}

	\author{Rémi Zamansky}%
	 \email{remi.zamansky@imft.fr}
	\affiliation{%
	 Institut de Mécanique des Fluides de Toulouse (IMFT), Université de Toulouse, CNRS-INPT-UPS, 31400 Toulouse France
	}%

	\date{\today}
	
	\begin{abstract}
		
		It is well known that the fluid-particle acceleration is intimately related to the dissipation rate of turbulence, in line with the Kolmogorov assumptions. On the other hand, various experimental and numerical works have reported as well its dependence on the kinetic energy, which is generally attributed to intermittency and non-independence of the small-scale dynamics from large-scale ones. The analyses given in this paper focus on statistics of the fluid-particle acceleration conditioned on both the local dissipation rate and the kinetic energy. It is shown that this quantity presents an exponential dependence on the kinetic energy with a growth rate independent of the Reynolds number, in addition to the expected power law behavior with the dissipation rate. The exponential growth, which clearly departs from the previous propositions, reflects additional kinematic effects of the flow structures on the acceleration. Regarding intermittency, to account for the persistence of the effect of the large-scales on the dissipation rate, it is further proposed scaling laws for the Reynolds number dependence of the conditional and unconditional acceleration variance using Barenblatt's incomplete similarity framework. It is then shown that both these intermittency and kinematic effects can be combined in a multiplicative cascade process for the acceleration depending on the kinetic energy and the dissipation rate. On the basis of these observations, we introduce a vectorial stochastic model for the dynamics of a tracer in turbulent flows. This model incorporates a fractional log-normal process for the dissipation rate recently proposed, as well as an additional hypothesis regarding non-diagonal terms in the diffusion tensor which naturally leads to the decomposition between tangential and centripetal acceleration. This model is shown to be in good agreement with direct numerical simulations and presents the essential characteristics of the Lagrangian turbulence highlighted in recent years, namely (i) non-Gaussian acceleration, (ii) scale separation between the norm of the acceleration and its components, (iii) anomalous scaling law for the Lagrangian velocity spectra, and (iv) negative skewness of the increments of the mechanical power, reflecting the temporal irreversibility of the dynamics. 

	\end{abstract}
	
	\maketitle

	\section{Introduction}

	With the advances of experimental techniques and the increase in computing power of the last decades, remarkable features of the dynamics of fluid particles in turbulent flows have been discovered.	
	Among other things, the measurement of the probability distribution of the acceleration of these tracers has been shown to be very clearly non-Gaussian with a high frequency of observing very intense events \cite{La-Porta:2001,Voth:2002,Mordant:2004,Mordant:2004b}. 
	Even for moderate Reynolds numbers, it is relatively common to observe accelerations more than 100 times greater than its standard deviation.
	In addition, the components of acceleration and its norm present very different correlation times, the ratio of these characteristic times increasing with the Reynolds number \cite{Pope:1990b,Mordant:2002,Mordant:2004c} showing that the dynamics of the tracers is influenced by the full spectrum of turbulence scales.
	On one hand, the short-time correlation of the acceleration component is connected to the centripetal forces in intense vorticity filaments \cite{Biferale:2005,Mordant:2004c}.
	On the other hand, the acceleration norm has been shown to be directly correlated with the local dissipation rate of turbulence \cite{Reynolds:2005,Yeung:2006,Homann:2011,Biferale:2006}, in accordance with Kolmogorov's hypotheses. 
	Nevertheless, various experimental and numerical works have also reported its dependence on local kinetic energy \cite{Mordant:2004b,Sawford:2003,Biferale:2004,Crawford:2005,Borgas:1998,Reynolds:2005,Aringazin:2004},	which is generally attributed to non-independence of the small-scale dynamics from large-scale ones.
	 In this view, the Lagrangian acceleration is essentially given by the local gradients of the velocity, but the latter present correlation with the kinetic energy caused by direct energy transfers between large- and small-scales in the energy cascade \cite{Tsinober:2009}.
	The absence of proper scale separation explains that the Lagrangian correlation functions present power laws with anomalous exponents which can be described by the multifractal formalism \cite{Chevillard:2003,Arneodo:2008,Biferale:2008,Sawford:2013,Lanotte:2013} as the signature of intermittency and persistence of viscous effects.
	To end this list, we mention the asymmetry of the fluctuations of the mechanical power received or given up by a fluid particle reflecting the temporal irreversibility of its dynamics \cite{Falkovich:2012,Xu:2014,Pumir:2014,Dubrulle:2019,Drivas:2019}.

	Such complex phenomenology must be attributed to the collective and dissipative effects.
	Indeed according to the Navier-Stokes equation, the acceleration of a fluid particle is essentially given by the pressure field which is determined by the motion of all the other particles \cite{Douady:1991,Tsinober:2001}.
	Moreover, although the Laplacian term in the Navier-Stokes equation is of order $Re^{-1}$ smaller than the pressure gradient term, the viscosity cannot be neglected. 
	Indeed, as a small force integrated over a long period could be significant, the viscosity insidiously affects the fluid tracer velocity.
	Which in turn influences the particle acceleration through modification of the pressure gradient and local interactions are intrinsically inseparable from the nonlocal ones.
	This is manifested by the persistence of the Reynolds number effect on the acceleration statistics, even for very large Reynolds numbers.
	Such a scenario is supported by \cite{Constantin:2008,Peskin:1985,Chorin:1994} who showed that adding noise to an inviscid Lagrangian flow leads to irreversibility of the dynamics.
	
	Following the Kolmogorov first hypothesis \cite{Kolmogorov:1941b,Kolmogorov:1962} stating that locally homogenous turbulent flows
	are universal, it should be possible, in principle, to propose a stochastic model that reproduces the dynamics of a single fluid particle by effectively accounting for the interactions with all the other fluid particles.
	Let us note that the Kolmogorov first hypothesis received some support from recent studies \cite{Tang:2020,Lawson:2019,Bos:2019}.
	In order to propose such a stochastic model, our main assumption in this paper is to write the increments of the acceleration vector of a fluid particle as $da_i = M_i dt + D_{ij} dW_j$.
	Both $M$ and $D$ depend on the particle acceleration $a$ and velocity $u$. The latter is simply given by the kinematic relation of a fluid particle $ u_i = \int a_i dt$.
	It is indeed a necessary condition that $a$ depends on $u$ to present a restoring effect that can counteract the diffusion in velocity space and have statistically stationary dynamics of the fluid particle.
	We will propose closed expressions for $M$ and $D$ from basic consideration using as a starting point the acceleration statistics conditioned on both the local values of the dissipation rate and the kinetic energy observed from direct numerical simulations (DNS) and presented as well in this paper.
	It will be shown that introducing a "maximal winding hypothesis" associated to a non-diagonal diffusion tensor, this simple stochastic model reproduces all the statistical feature of the Lagrangian dynamics presented above without any adjustable parameter.
	
	Let us first review some previous works on the stochastic modeling for the Lagrangian dynamics (see also \cite{Aringazin:2004}). 
	Among the pioneering works, Sawford \cite{Sawford:1991} proposed a scalar Gaussian model for the acceleration presenting a feedback term proportional to the velocity.
	Pope and Chen \cite{Pope:1990} devised a Langevin like equation for the velocity coupled with a log-normal model for the dissipation through the introduction of conditional statistics.
	Similarly \cite{Reynolds:2003,Reynolds:2003b,Beck:2003} proposed an extension of the Sawford model leading to a non-Gaussian scalar model for the acceleration.
	This work was further refined by \cite{Lamorgese:2007} who also advanced a non-Gaussian scalar model for the dynamics by prescribing an ad hoc shape of the conditional acceleration statistics with the dissipation along with a linear dependence on the velocity.	
	The model introduced in \cite{Reynolds:2004b} describes increments of the derivative of acceleration in a so-called third-order model to better account for the Reynolds number dependence on the acceleration statistics. 
	Recently \cite{Viggiano:2020} proposed generalization to an infinite number of layers leading to smooth 1D trajectory along with a multifractal correction to account for intermittency, as introduced in \cite{Bacry:2001,Mandelbrot:1974,Kahane:1976}.
	 An acceleration vector model has been proposed in \cite{Reynolds:2004} by imposing an empirical correlation between velocity and acceleration, with additive noise leading to Gaussian statistics for the acceleration.
	Likewise, \cite{Pope:2002b} presented a 3D Gaussian model, with linear dependence on the velocity as well as an extension to non-homogenous flows.
	In order to account for intermittency effect, in \cite{Gorokhovski:2018,Sabelnikov:2011,Zhang:2019b,Sabelnikov:2019,Barge:2020,Gorokhovski:2022} the 3D acceleration vector is given by the product of two independent stochastic processes, one for the acceleration norm the other for its orientation. 
	In these models the velocity feedback on the dynamics is realized by a coupling with a large eddy simulation framework.
	To summarize, to our knowledge, a 3D vectorial model for the tracer dynamics that is autonomous and reproducing the essential features of Lagrangian turbulence (irreversibility, non-Gausianty, multifractality) has not yet been proposed.

	The essential building block of previously cited models is the conditional acceleration statistics.
	Previous studies have focused on conditional statistics with either the velocity or the dissipation rate separately.
	From the extensive analysis of \cite{Yeung:2006}, one can conclude that the acceleration variance conditioned on the dissipation rate $\varepsilon$ presents a power law behavior for large values of $\varepsilon$ with a Reynolds number dependent exponent reflecting that the small-scale dynamics are not independent of large-scales.
	
	Regarding the links between the fluid particle acceleration and their velocity, Biferale et al. \cite{Biferale:2004} argue that according to the Heisenberg-Yaglom scaling for the acceleration $\langle a^2 \rangle \sim a_\eta^2 = \langle \varepsilon \rangle^{3/2} \nu ^{-1/2} = \langle K \rangle ^{9/4} L^{-3/2} \nu ^{-1/2}$, with $\nu$ the kinematic viscosity, $ K = 1/2 u_iu_i $ the kinetic energy and $ L $ the characteristic size of large structures, one should expect that the variance of the velocity-conditioned acceleration behaves like: 	$\langle a^2 | K \rangle \propto K^{9/4}$. 
	Then on the basis of the multifractal formalism, they proposed a very close scaling law, $ \langle a^2 | K \rangle \sim K^{2.3} $.
	The proposed relation was observed to be in agreement with DNS for large velocity, typically $ |u|> 3 \sigma_u $ with $ \sigma_u = \sqrt {2/3 \langle K \rangle} $. 
	These events remain very rare since the PDF of the fluid velocity is Gaussian so the range of validity of the power law is, at best, very limited.
	 On the other hand, Sawford et al. \cite{Sawford:2003} propose that $ \langle a_x ^ 2 | u_x \rangle \sim u_x^6 $ based on a mechanism involving vorticity tubes.
	This scaling law which seems compatible with the first measurements of the acceleration conditioned on velocity in \cite{Mordant:2004b}, is confirmed neither by the DNS of \cite{Biferale:2004} nor in a second experimental paper by Crawford et al. \cite{Crawford:2005} which gave more credit to the $ K ^ {9/4} $ law.
	As mention above, it is been proposed that the dependence of the acceleration on the velocity arises through the dependence of the dissipation rate on the kinetic energy  due to intermittency effect \cite{Biferale:2004}. 
	Additionally, \cite{Tsinober:2009} propose that the dependence on velocity is a consequence of direct and bi-directional coupling of large- and small-scales caused by kinematic relations related to non-local interactions.

	In this paper we study the acceleration statistics conditional on both the kinetic energy and the dissipation rate.
	To our knowledge such doubly-conditional statistics of the acceleration have never been presented. 
	It will be shown that the variance can be expressed as $\langle a^2 | \varepsilon, K \rangle \sim \exp( \alpha K/\langle K \rangle + \gamma \ln \varepsilon/ \langle \varepsilon \rangle )$.
	This result is clearly in contrast with the previously proposed power law dependence on velocity.
	 It shows that the influence of the large-scales through the intermittent distribution of the dissipation rate, which manifests through the Reynolds number dependence of the coefficient $\gamma$, is supplemented by an explicit dependence on the local kinetic energy. 
	This direct dependence on the large-scale characteristics is of a kinematic nature as it appears independent of the Reynolds number.
	The behavior of the doubly-conditional acceleration can be interpreted as a consequence of scaling symmetry for the fluid-particle acceleration incorporating both the intermittency and the kinematic effects of the flow structure.
	We also propose to apply the incomplete similarity framework introduced by Barenblatt to explain the dependence of the statistics of the acceleration conditional to the dissipation rate on the Reynolds number and to account for the intermittency effect.
	That enables to provide as well new scaling relations for the unconditional variance in good agreement with the DNS.
	Eventually the doubly-conditional statistic of the acceleration which gives a relation between the force, the energy and the power will serve as a corner stone to build the stochastic model for the dynamics of a fluid particle mentioned above.
	Although such a model could be of interest for practical applications, its construction is relevant to study the specificities of the Lagrangian description of turbulence by linking the cascade picture to the fluid particles dynamics on the basis of the behavior of the conditional statistics obtained by the DNS of the Navier-Stokes equations.

	In section \ref{sec:stat_acc} we present the statistic of the acceleration conditioned on the local values of the dissipation rate and kinetic energy obtained from DNS.
	Then we show that the Reynolds number dependence on the acceleration conditioned on the dissipation rate can be described using the Barenblatt incomplete similarity. We deduce a new relation for the unconditional acceleration variances. To end this section, we show that these new results can be interpreted as a multiplicative cascade for the acceleration with scale dependent kinematic effects. 
	Then in section \ref{sec:stoch_model} we give the derivation of the stochastic model for the single fluid particle dynamics taking as an initial step the doubly-conditional acceleration variance, and present the outcome of the model for the Reynolds number up to $Re_\lambda=9000$ along with comparison with DNS results when available.


	\section{Scaling laws of the acceleration \label{sec:stat_acc}}
	
	\subsection{Methodology}

	We present in this section results concerning the statistics of the acceleration of a fluid particle.
	These results have been obtained from 5 direct numerical simulations (DNS) of isotropic turbulence in a periodic box with Taylor-scale Reynolds numbers of $Re_\lambda = 50$, 90, 150, 230 and 380.
	We used pseudo-spectral code as detailed in \cite{Zamansky:2016,Zhang:2019b,Le-Roy-De-Bonneville:2021}.
	The DNS was carried with resolutions of $128^3$, $256^3$, $512^3$, $1024^3$ and $2048^3$ with the large scale forcing proposed by \cite{Kumar:2014}. For each simulation we have $\eta/\Delta x = 1$ with $\eta=\langle \varepsilon \rangle^{-1/4}\nu^{3/4}$ the Kolmogorov length scale and $\Delta x$ the grid size.
	The statistics are computed from 40 3D fields sampled at roughly each large-eddy-turnover time.
	
	We will show statistics of the acceleration conditioned by the dissipation rate and the kinetic energy.
	Note that in this paper we consider the pseudo-dissipation $\tilde{\varepsilon} = \nu (\partial_j u_i)^2 $, which is the second invariant of the velocity gradient tensor multiply by the viscosity rather than the dissipation $\varepsilon = \dfrac{1}{2}\nu (\partial_j u_i + \partial_i u_j)^2$. 
	We prefer to show here the statistics of the pseudo-dissipation to be consistent with the next section of the paper, in which we will use the log-normal distribution hypothesis for the dissipation. 
	Indeed, this property is very well verified for the pseudo-dissipation whereas it is only approximate for the dissipation \cite{Yeung:2006}. 
	Nevertheless, the statistics presented below have also been computed considering the dissipation, $\varepsilon$, and no significant differences were observed. 
	To lighten the paper, in the sequel, we will drop the tilde in the notation of the pseudo-dissipation, as well, in the text, we will write dissipation instead of pseudo-dissipation.
		
	\subsection{Conditional statistics given the dissipation and the kinetic energy }	

\begin{figure*}[h!]
	\centering
	\includegraphics[height=0.30\textheight,keepaspectratio, clip]{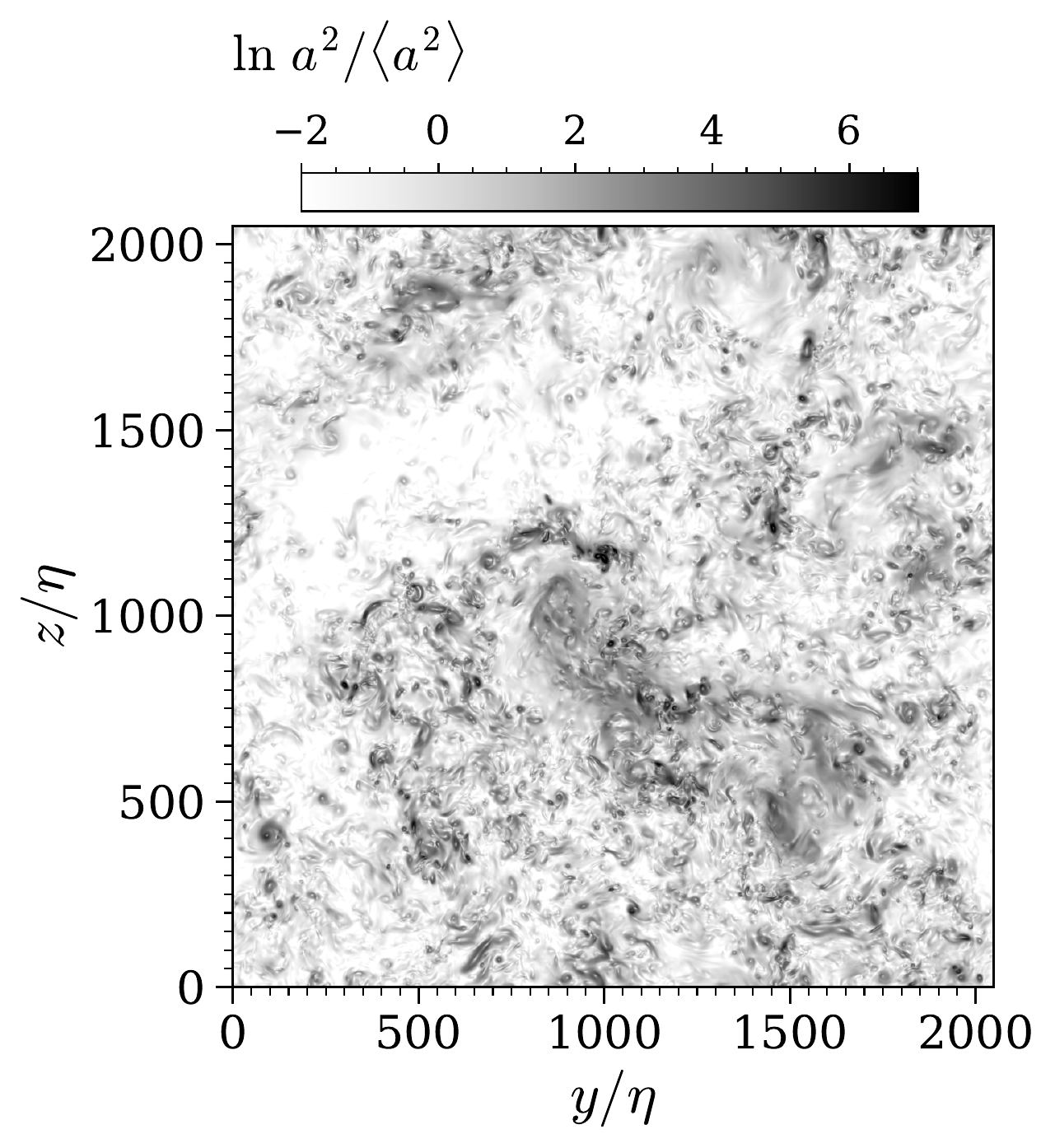}
	\includegraphics[height=0.30\textheight,keepaspectratio, clip]{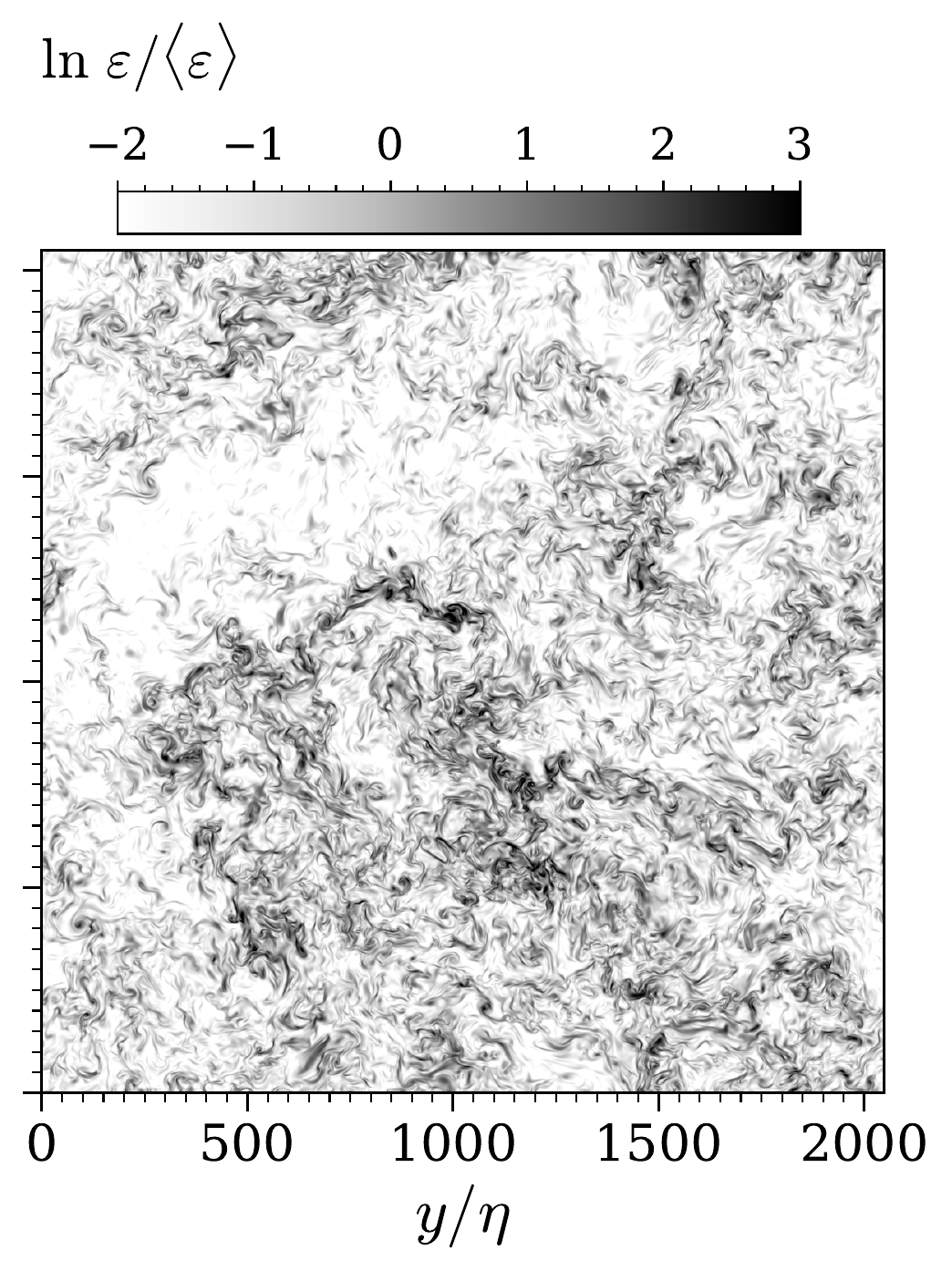}
	\includegraphics[height=0.30\textheight,keepaspectratio, clip]{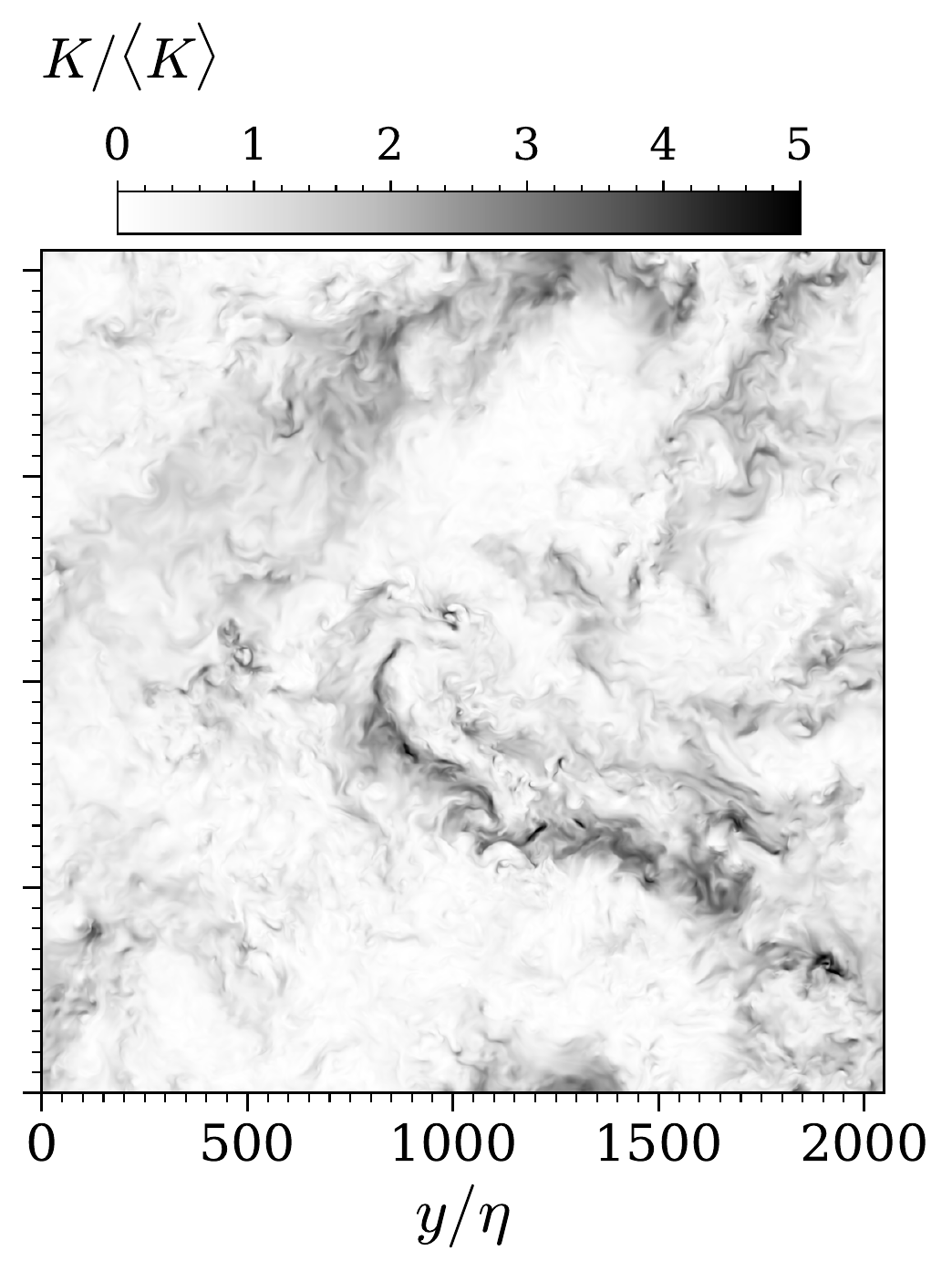}

	\caption{Visualization of the instantaneous fields of the square of the acceleration, of the dissipation and of the kinetic energy in a cut $ y-z $ of the flow by DNS at 
	$Re_\lambda=380$.
	(Left): $\ln(a^2/<a^2>)$; (Middle): $\ln(\varepsilon/<\varepsilon>)$ and (Right): $K/<K>$. } 
	\label{fig:visu}
	\end{figure*}

	In order to illustrate the relationships between acceleration, energy dissipation and kinetic energy, we show in Fig. \ref{fig:visu} visualizations of these quantities at the same instant obtained from our DNS.
We notice that $ \ln a^2 / \langle a^2 \rangle $ and $ \ln \varepsilon / \langle \varepsilon \rangle $ show a fairly marked correlation although the acceleration appears more diffuse than dissipation.
We also notice that to some extent the kinetic energy and the dissipation rate appears correlated.
In addition, it seems that some areas of the flow where the kinetic energy is high also correspond to regions of high acceleration magnitude.
	
\begin{figure*}[h]
	\centering	
	\includegraphics[width=0.49\textwidth,height=0.4\textheight,keepaspectratio, clip]{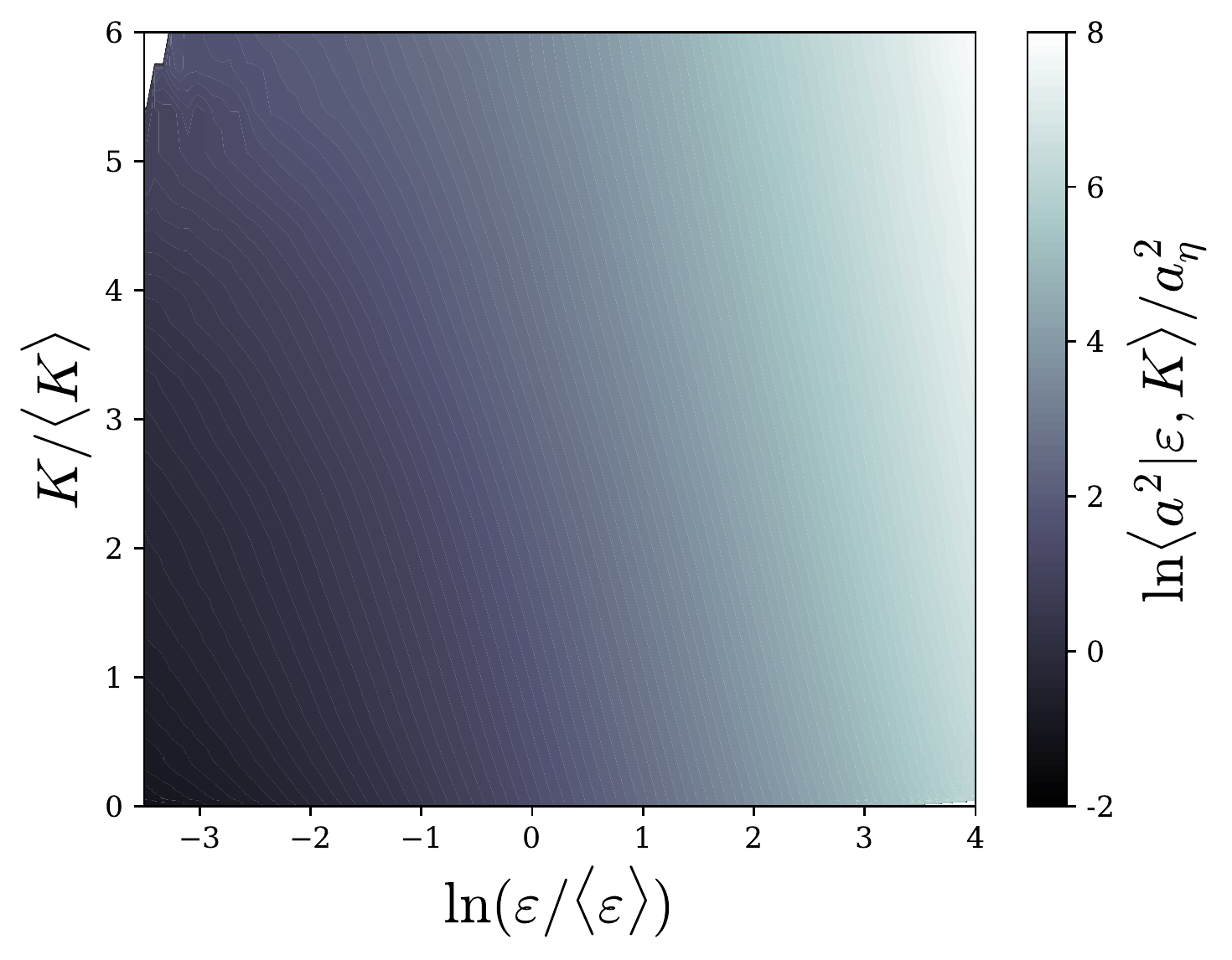}\\
	\includegraphics[width=0.49\textwidth,height=0.4\textheight,keepaspectratio, clip]{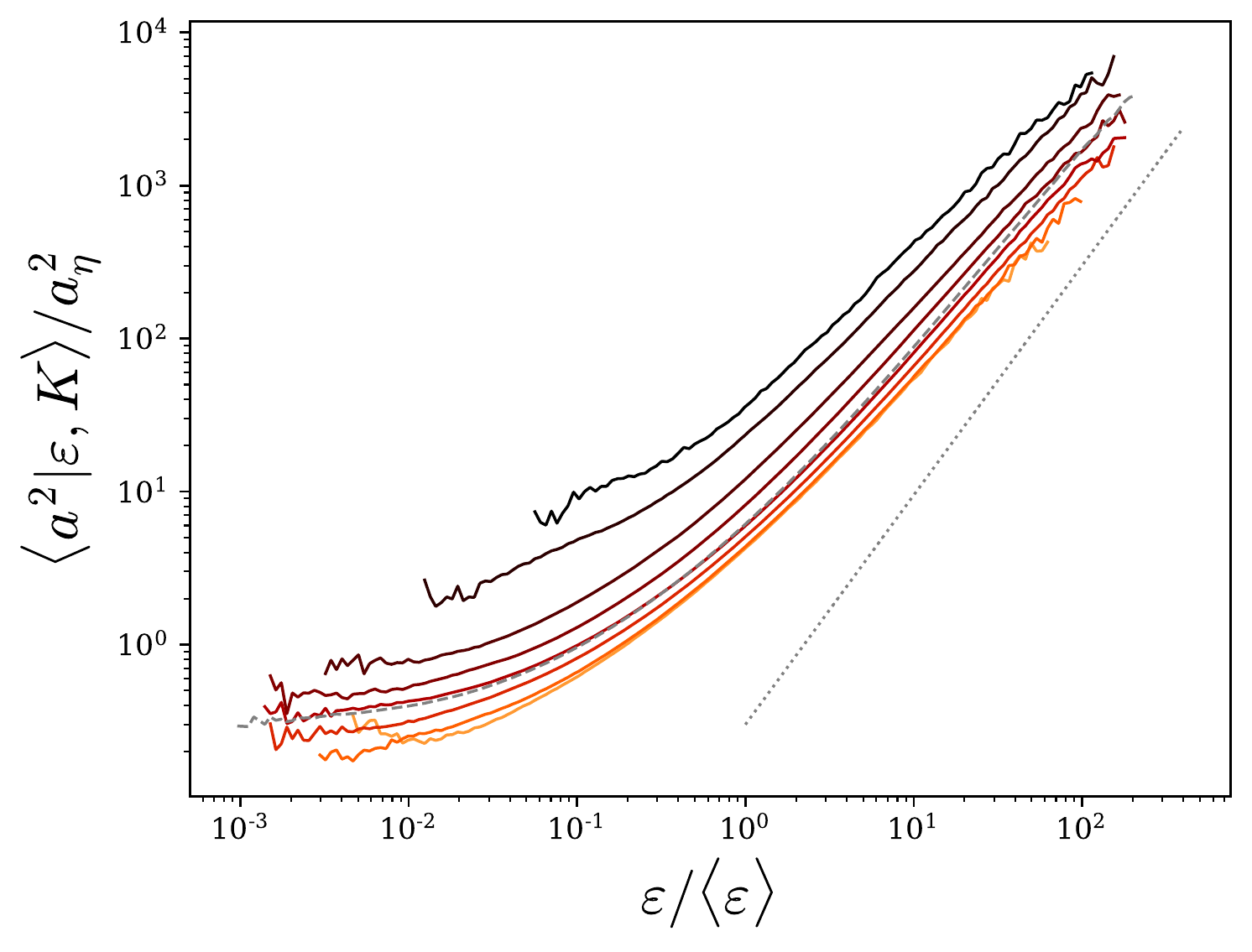}
	\includegraphics[width=0.49\textwidth,height=0.4\textheight,keepaspectratio, clip]{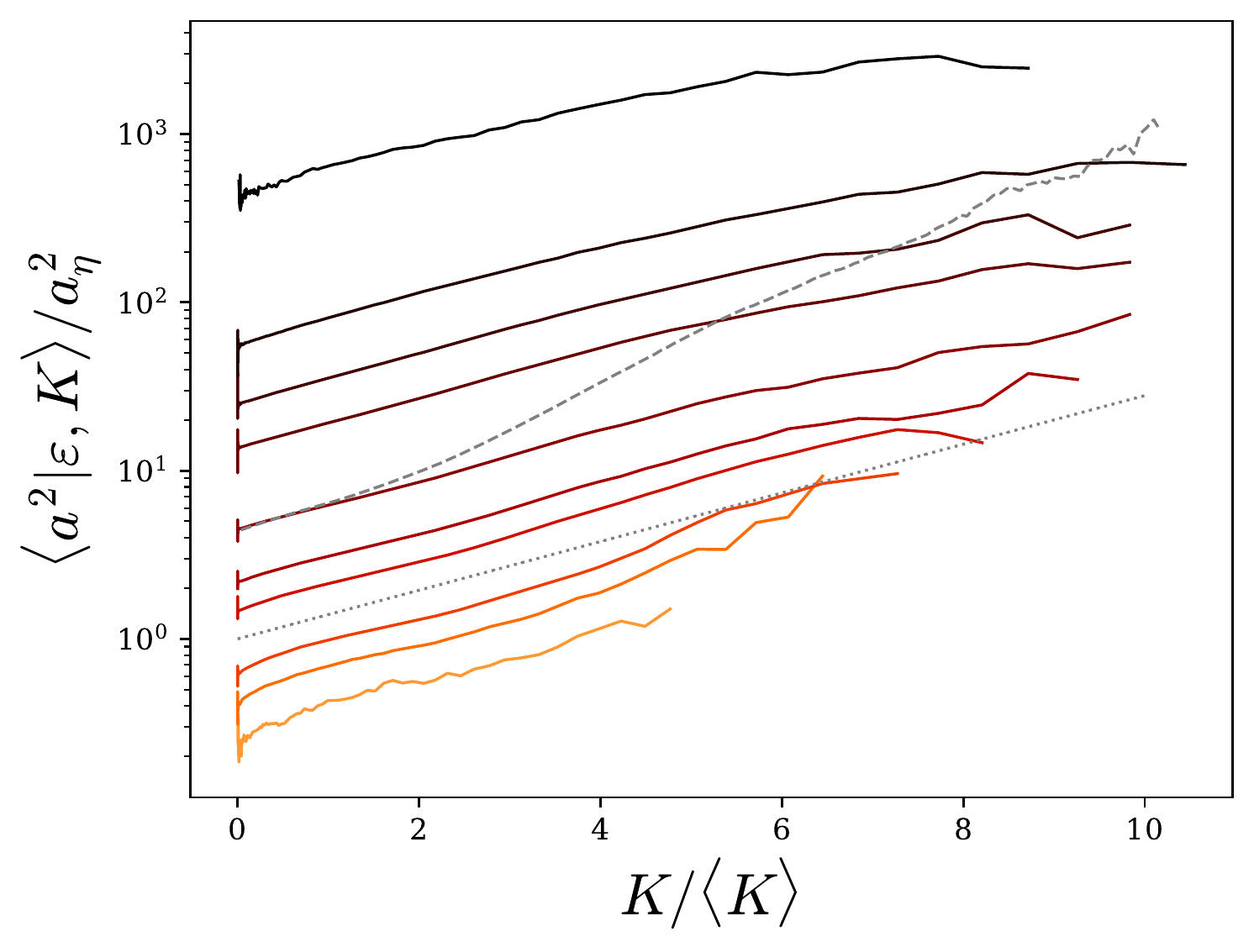}
	\caption{ Variance of acceleration conditioned on the local dissipation rate and kinetic energy obtained from DNS at 
	$ Re_\lambda = 380$.
	(Top) Map of $\ln \langle a^2 | \varepsilon, K \rangle / a_\eta^2  $ versus $\ln\varepsilon/\langle \varepsilon\rangle $ and $K /\langle K\rangle$.
	(Left) Plot, in logarithmic scales, of $\langle a^2 | \varepsilon, K \rangle / a_\eta^2 $ against $\varepsilon/\langle \varepsilon\rangle$ for $K/ \langle K\rangle = 0.025, 0.1, 0.5, 1 , 2, 3,5, 6.5 \pm30\% $ from orange to black. 
	Comparison with $\langle a^2 | \varepsilon \rangle / a_\eta^2$ in gray dashed line and with the power law $(\varepsilon/\langle \varepsilon\rangle)^{3/2}$ in gray dotted line.
	(Right) Plot, in semi-logarithmic scales, of $\langle a^2 | \varepsilon, K \rangle / a_\eta^2 $ against $K/ \langle K\rangle $ for $\varepsilon/\langle \varepsilon\rangle = 0.01, 0.05, 0.1, 0.3, 0.5, 1, 3, 5, 10, 50 \pm30\% $ from orange to black. 
	Comparison with $\langle a^2 | K \rangle / a_\eta^2$ in gray dashed line and with $\exp(\alpha K/\langle K \rangle )$ with $\alpha=1/3$ in gray dotted line.
	}
	\label{fig:cond_acc_dissip_tke} 
	\end{figure*}

	Figure \ref{fig:cond_acc_dissip_tke} presents the variance of the acceleration of a fluid particle conditioned to the local values of the kinetic energy and the dissipation rate: 
	$\langle a^2 | \varepsilon, K \rangle $.
In Fig. \ref{fig:cond_acc_dissip_tke} (Top), the levels of the logarithm of the conditional variance are shown as a function of $ K $ and of $ \varepsilon $.
We see that the conditional variance of the acceleration depends on these two quantities and that the dependence on $ K $ seems somewhat similar to that of $ \ln \varepsilon $.
In a more quantitative way, we show in Fig. \ref{fig:cond_acc_dissip_tke} (Left) the variance of the acceleration as a function of $ \varepsilon $ for different values of $ K $. 
We can see that the shape of the curves remains globally unchanged when $ K $ varies and also presents the same shape as the variance conditioned by $ \varepsilon $ only as also presented in this figure.
Essentially, it is observed that the conditional variance presents power law behavior for $\varepsilon \gg \langle \varepsilon \rangle $ with an exponent close to $3/2$ and a prefactor depending on $K$. 
	As discussed in more details below, we observe a slight deviation of the scaling law compared to the acceleration conditioned only by the dissipation. 
	
	Figure \ref{fig:cond_acc_dissip_tke} (Right) shows the variance of the acceleration as a function of $ K $ for different values of $ \varepsilon $.
	As expected, we find that the variance of the acceleration increases with $ K $. 
	We clearly notice an exponential growth of the variance over the whole range of $ K $ with a growth rate $ \alpha $ which appears independent of $ \varepsilon $:
	\begin{equation}
		\langle a^2|\varepsilon, K\rangle = c_\varepsilon a_\eta^2 \exp(\alpha K/\langle K \rangle)
		\label{eq:c_eps}
	\end{equation} 
	with $a_\eta^2= \langle \varepsilon\rangle^{3/2} \nu^{-1/2} = \langle \varepsilon\rangle /\tau_\eta $ the so-called Kolmogorov acceleration and the prefactor $ c_\varepsilon $ depending on $ \varepsilon $.
	From our DNS it appears that $ \alpha \approx 1/3 $ for all the Reynolds numbers considered here.
	We also find the same value of $ \alpha $ from the database of \cite{Bec:2010,Lanotte:2011} obtained for $ Re_\lambda = 400 $
suggesting that the value of $ \alpha $ is independent of the Reynolds number. 
	
	This exponential behavior contrasts with the references mentioned in the introduction in which power laws behavior for the variance conditioned on $ K $ solely had been proposed.
	Nevertheless, we can notice that exponential growth does not seem to disagree with the data presented in these references.
	Interestingly, this relationship only depends on a characteristic velocity, (not a time and a length scale separately).
	The absence of characteristic time is attributed to the scale separation between large structures and small ones (the large structures of the flows appear as quasi stationary and infinite to the smallest ones such that only their relative velocity matters).
	The independence of the coefficient $\alpha$ on the Reynolds number tends to confirm that the velocity scale used for the nondimensionalization of the argument of the exponential is appropriate.

	\begin{figure*}[h]
		\centering
		\includegraphics[width=0.49\textwidth,height=0.4\textheight,keepaspectratio, clip]{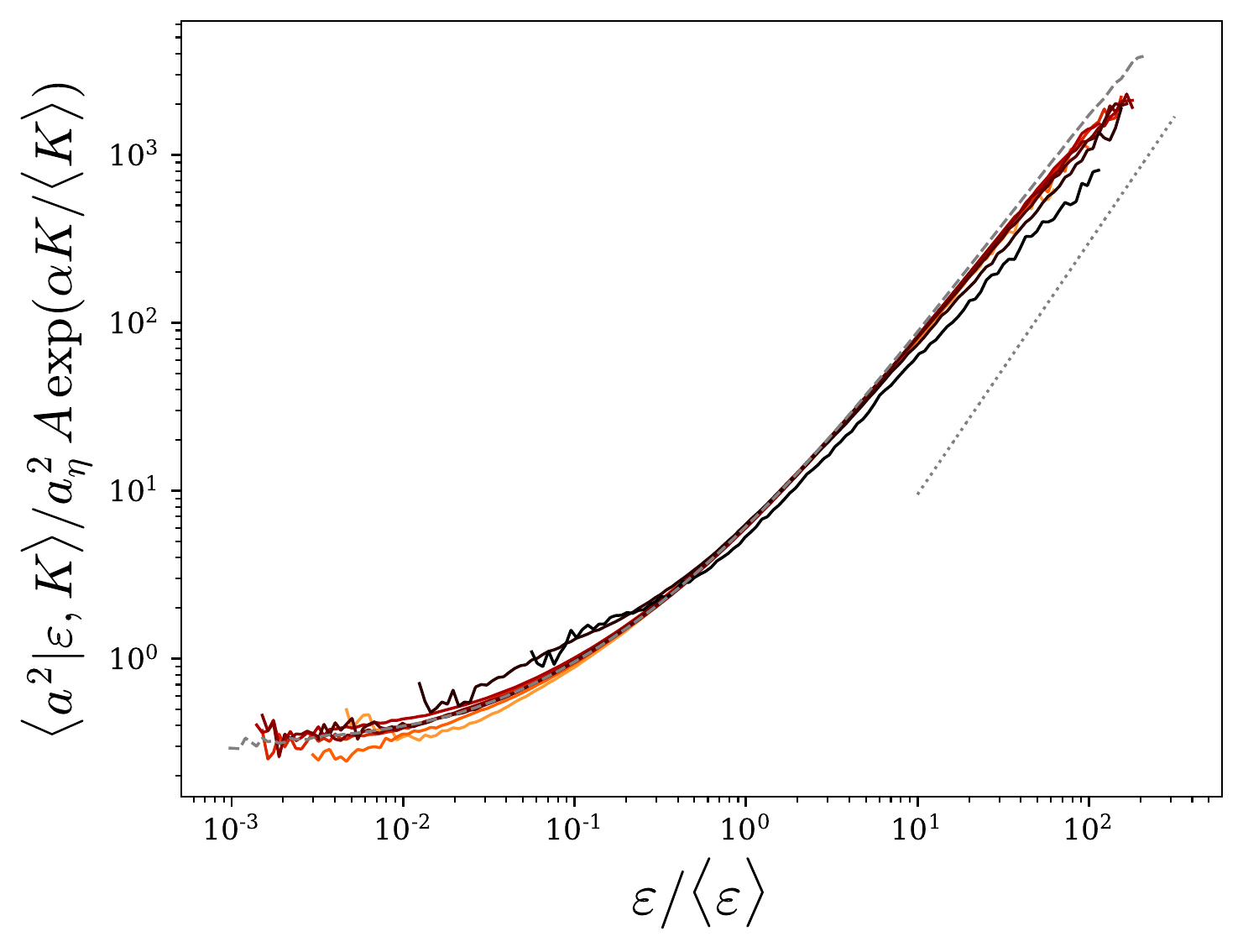}
		\includegraphics[width=0.49\textwidth,height=0.4\textheight,keepaspectratio, clip]{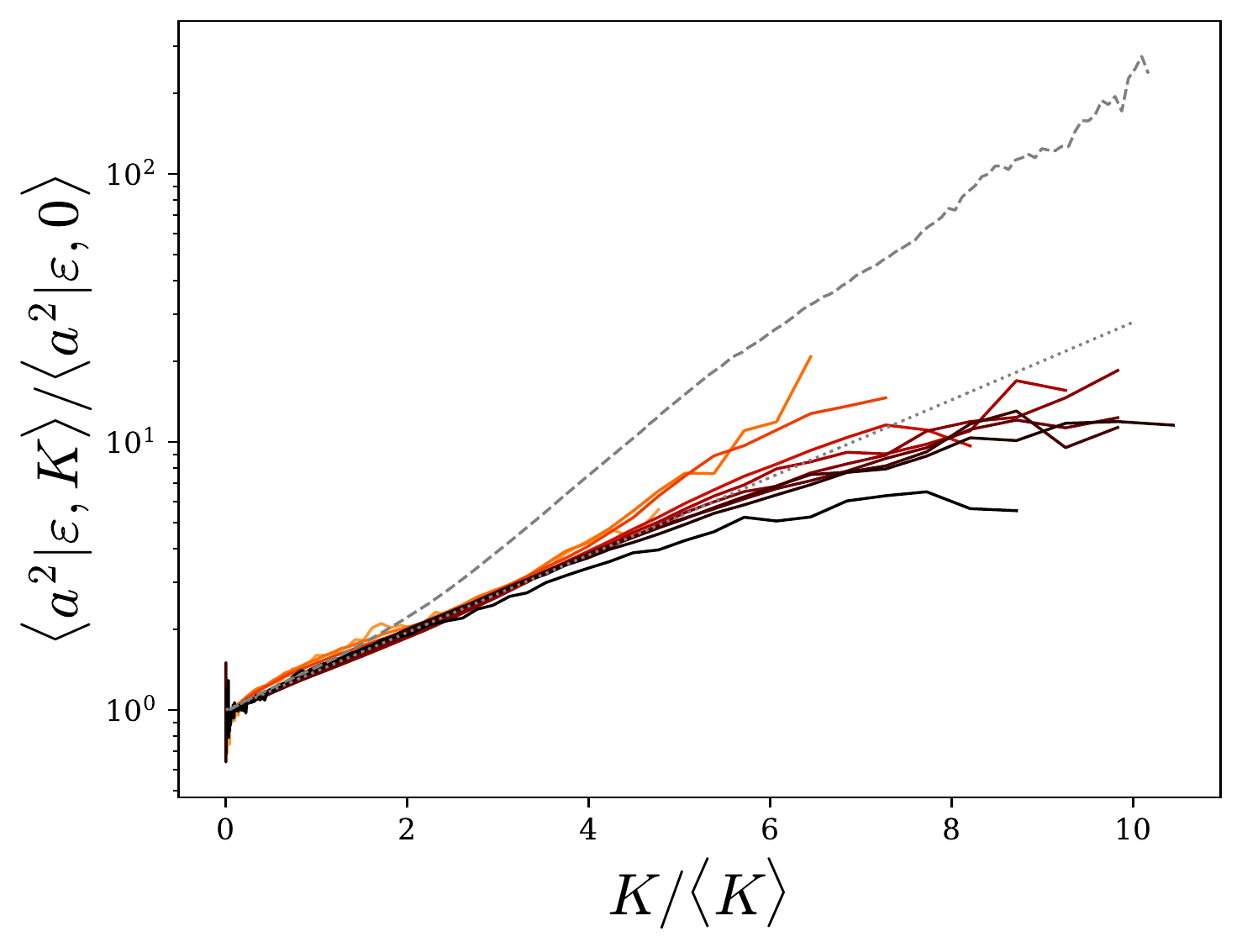}
		
		\caption{ Normalized variance of acceleration conditioned on the local dissipation rate and kinetic energy obtained from DNS at
		 $ Re_\lambda = 380$.
		(Left) Plot of $\langle a^2 | \varepsilon, K \rangle / A a_\eta^2 \exp(\alpha K / \langle K \rangle ) $ against $\varepsilon/\langle \varepsilon\rangle$ for various values of $K$. 
		Comparison with $\langle a^2 | \varepsilon \rangle / a_\eta^2$ in gray dashed line and with the power law $(\varepsilon/\langle \varepsilon\rangle)^{3/2}$ in gray dotted line.
		(Right) Plot of $\langle a^2 | \varepsilon, K \rangle / \langle a^2|\varepsilon, K = 0\rangle $ against $K/ \langle K\rangle $ for various values of $\varepsilon/\langle \varepsilon\rangle$. 
		Comparison with $\langle a^2 | K \rangle / a_\eta^2$ in gray dashed line and with $\exp(\alpha K/\langle K \rangle )$ with $\alpha=1/3$ in gray dotted line
		The ranges for the fixed values of $K$ and $\varepsilon$ for both plots are the same as in Fig. \ref{fig:cond_acc_dissip_tke}.
		}
		\label{fig:cond_acc_dissip_tke_norm} 
	\end{figure*}

	In appendix \ref{sec:c_eps} we propose to estimate  the factor $c_\varepsilon$ as:
 	\begin{equation}
 		c_\varepsilon \approx A \, \langle a^2|\varepsilon \rangle / a_\eta^2
 	\end{equation}
 	 where $A= \left(1-\dfrac{2 }{3} \alpha \right)^{3/2}$, which is equal to $A=7\sqrt{7}/27\approx 0.686$, for $\alpha = 1/3$, neglecting a small logarithmic dependence on $ \varepsilon / \langle \varepsilon \rangle $.
		 
	 Consequently, for large Reynolds numbers, the doubly-conditioned variance of the fluid-particle acceleration is expressed as 
	 \begin{equation}
	 	\langle a^2|\varepsilon, K\rangle = A\, \langle a^2|\varepsilon\rangle \exp(\alpha K/\langle K \rangle) \, .
		\label{eq:cond_acc_B}
	 \end{equation}
	 This relation is confirmed in Fig. \ref{fig:cond_acc_dissip_tke_norm} which presents the conditional variance of the acceleration normalized by 
	 $A a_\eta^2 \exp(\alpha K / \langle K \rangle )$ as a function of $ \varepsilon $ for different values of $ K $
	 as well as 
	 normalized by $A \langle a^2 | \varepsilon \rangle= \langle a^2|\varepsilon, K = 0\rangle$ as a function of $ K $ for different values of $ \varepsilon $.
	 It can be seen that a fairly good overlap of the various curves is obtained, confirming the self-preserving character of the acceleration conditioned on both the kinetic energy and the dissipation rate.
	 We see in this relation an explicit effect of the local kinetic energy on the acceleration. 
	 Since the argument of the exponential depends on $K/\langle K \rangle$ not on a local Reynolds number, it suggests  pure kinematic effects for the acceleration which is likely associated to the divergence free constrain and the non-locality of the pressure gradient.
	  There is also indirect effect through the dependence of the dissipation rate on the large-scale structures. 
	  The later is manifested as Reynolds number dependence of the conditional acceleration on the dissipation rate solely.
	  This intermittency effect is analyzed further in the next section.
	 We postpone to section \ref{sec:cascade} further comments on the behavior of the doubly-conditioned variance.

	\subsection{Similarity of the conditional statistics given the dissipation }	\label{sec:cond_dissip}

\begin{figure*}[b]
		\centering
		
		\includegraphics[width=0.49\textwidth,height=0.4\textheight,keepaspectratio, clip]{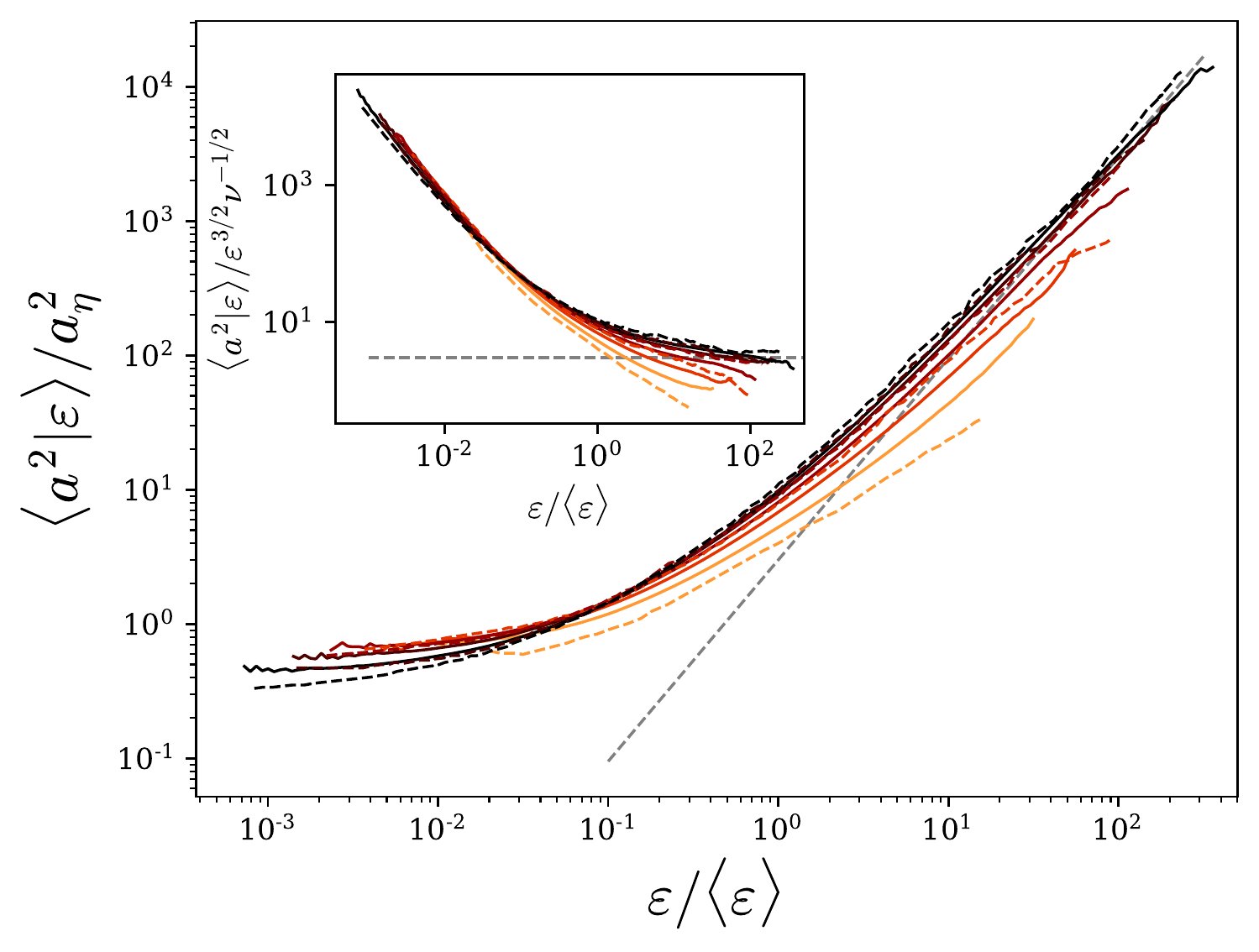}
		\includegraphics[width=0.49\textwidth,height=0.4\textheight,keepaspectratio, clip]{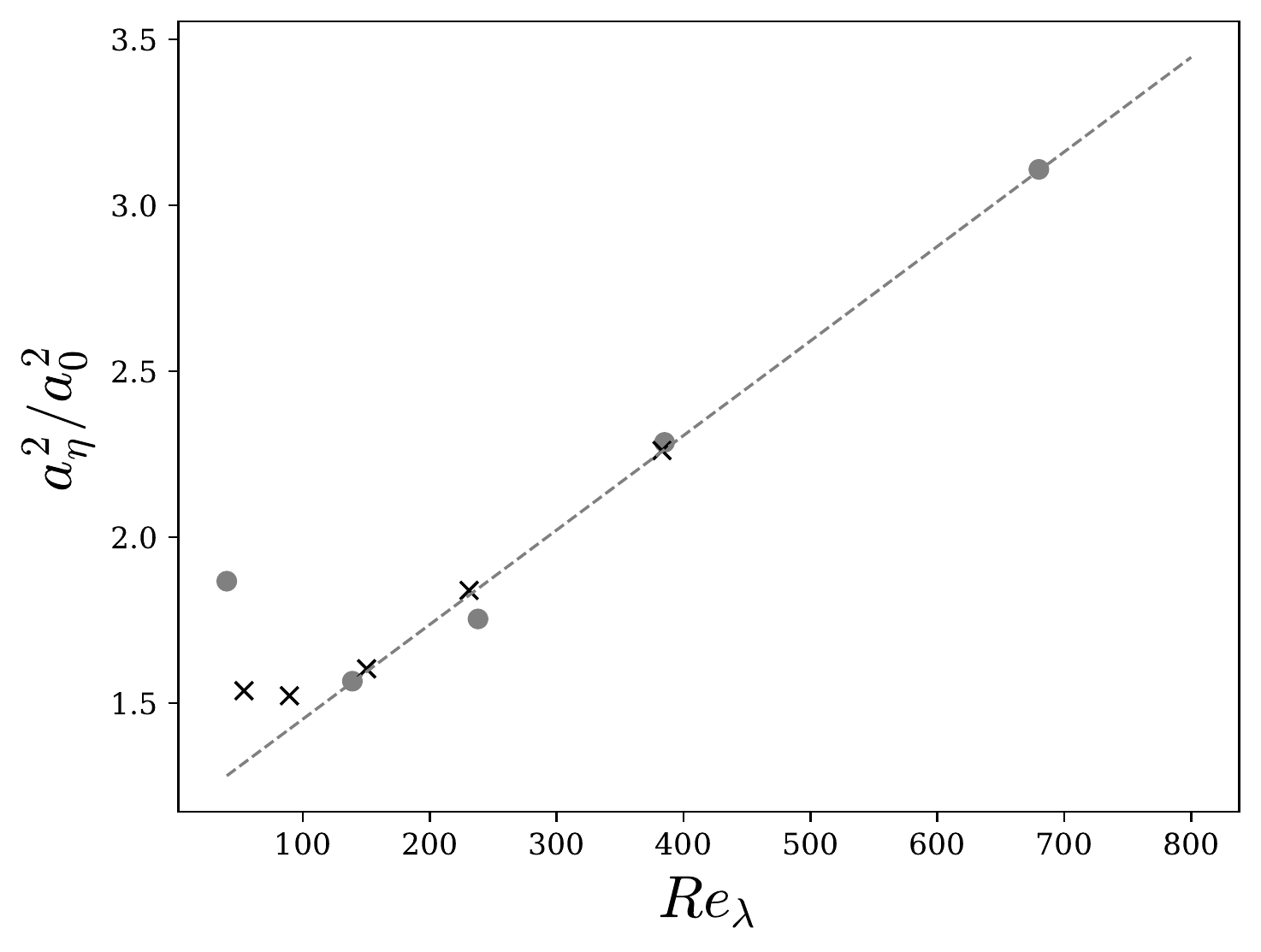}
		
		\caption{
		 (Left) Acceleration variance conditioned on the local dissipation rate normalized by the Kolmogorov acceleration $\langle a^2 | \varepsilon \rangle / a_\eta^2$. 
		 The continuous line are for our DNS for $Re_{\lambda}=50$, 90, 150, 230 and 380 from orange to black; 
		 the dashed lines correspond to the DNS of Yeung et al. \cite{Yeung:2006} for $Re_{\lambda}=40$, 139, 238, 385, 680, from orange to black.
		 Comparison with the power law $(\varepsilon/\langle \varepsilon \rangle )^{3/2}$.
		 Inset conditional acceleration normalized by $\varepsilon^{3/2}\nu^{-1/2}$.
		 (Right) $ a^2_{\eta}/a^2_0 $ as a function of $Re_\lambda$ with $ a^2_0 = \lim_{\varepsilon \rightarrow 0}\langle a_i^2 | \varepsilon \rangle$ the acceleration variance in low dissipative regions. Gray dots for the  DNS of Yeung et al., black crosses for our DNS.
		 Comparison with the line 
		 $0.0028Re_\lambda+1.16$
		  in dashed lines.
		 }
		\label{cond_acc_1} 
	\end{figure*}

	We propose now to focus with more details on the scaling law of the acceleration variance conditioned on the dissipation rate only, $\langle a^2 | \varepsilon \rangle$.
	For that we consider the DNS data from Yeung et al. \cite{Yeung:2006}, along with our DNS data.
	Figure \ref{cond_acc_1}(Left) presents the conditional acceleration variance for Reynolds numbers in the range $ Re_{\lambda} = 40 $ to 680.
	We first notice that for weak values of the dissipation rate ($ \varepsilon \ll \langle \varepsilon \rangle $) the value of the conditional acceleration variance tends towards an asymptotic value, which depends on the Reynolds number. 
	The saturation of the conditional acceleration shows that the local acceleration is not only determined by the microstructure of the flow, and that it presents somehow effects of the large structures of the flow which dominates in low dissipative regions.
	We denote by $ a_0^2 $ the asymptotic value of the conditional variance: 
	\begin{equation}
			a^2_0 = \lim_{\varepsilon \rightarrow 0}\langle a^2 | \varepsilon \rangle \, .
	\end{equation}
	Assuming that the acceleration of fluid particles in low dissipative regions is mainly influenced by large scales, we can estimate $ a_0^2 $ as $ a_0^ 2 \sim \langle K \rangle / \tau_L ^ 2 $ with $ \tau_L $ the integral time scale of the flow.	
	This leads to the following estimate: 
	\begin{equation}
		a_0^2/a_\eta^2 \sim \tau_\eta / \tau_L \sim Re^{-1}_{\lambda}\, .
		\label{eq:a0_scaling}
	\end{equation}
	We test this scaling law for $ a_0 $ in Fig. \ref{cond_acc_1}(Right) by presenting $ a_\eta^2/ a_0^2 $ as a function of $ Re_\lambda $ from the different DNS datasets. 
	We observe a linear growth rate of $ a_0^2 / a_{\eta}^2 $ with $ 1 / Re_\lambda $.
	
	For large values of $ \varepsilon $, we notice in Fig. \ref{cond_acc_1}(Left), as already reported in \cite{Yeung:2006}, that the conditional variance presents a power law behavior with $ \varepsilon$.
	The exponent of this scaling law is seen to evolve continuously with the Reynolds number, and seems to tend asymptotically towards $ \varepsilon^{3/2}$.
	From dimensional analysis we define $f$ as: 
	\begin{equation}
		\dfrac{\langle a^2 | \varepsilon \rangle}{\varepsilon^{3/2} \nu^{-1/2}} = f(\varepsilon/ \langle \varepsilon \rangle, Re_\lambda) \, .
		\label{eq:a2_cond_adim}
	\end{equation}	 
	In the inset of Fig. \ref{cond_acc_1}(Left), it is seen that $f$ seems to admit an asymptotic constant value for $\varepsilon \gg \langle \varepsilon \rangle$ only in the limit of very large Reynolds number.
	For intermediate Reynolds numbers,  $f$ presents power-law behavior with $\varepsilon$ for $\varepsilon \gg \langle \varepsilon \rangle$ but with a Reynolds number dependent exponent. 
	This implies an absence of similarity of the flow when the Reynolds number is changed and the persistence of the Reynolds number effect, even for large Reynolds numbers, which highlights an absence of proper scale separation suggesting direct coupling between large- and small-scales.
This is reminiscent of the incomplete similarity framework proposed by Barenblatt \cite{Barenblatt:1998,Barenblatt:2002,Barenblatt:2004}. 
Following Barenblatt, we assume that $f$ presents an incomplete similarity in $\varepsilon/ \langle \varepsilon \rangle$ and absence of similarity in $Re_\lambda$. 
	Accordingly we write 
	\begin{equation}
	 f(\varepsilon/ \langle \varepsilon \rangle, Re_\lambda) = B \left( \varepsilon/ \langle \varepsilon \rangle \right)^{\beta}
	 \label{eq:f_anomalous}
	\end{equation}
	where the anomalous exponent $\beta$, and the prefactor $B$ are both functions of $Re_\lambda$.
	Arguing for a vanishing viscosity principle, it can be assumed that the critical exponent becomes independent of the Reynolds number in the limit of asymptotically large Reynolds number.
	Finally arguing that the dependence of $B$ and $\beta$ on the Reynolds number is small, Barenblatt further proposed that they presents inverse logarithmic dependence on $Re_\lambda$,
	which is also in agreement with the log-similarity proposed by \cite{Castaing:1993,Gagne:1991}.
	Expending $\beta$ and $B$ in power of $1/\ln( Re_\lambda)$ yields, keeping only the leading-order term in $Re_\lambda$:
	\begin{eqnarray}
		\beta &=& \beta_0 + \beta_1/\ln Re_\lambda \label{eq:crit}
\\
		 B &=& B_0 + B_1/\ln Re_\lambda 		\label{eq:crit_B}
	\end{eqnarray}
	
\begin{figure*}[h]
 	\centering	
 	\includegraphics[width=0.49\textwidth,height=0.4\textheight,keepaspectratio, clip]{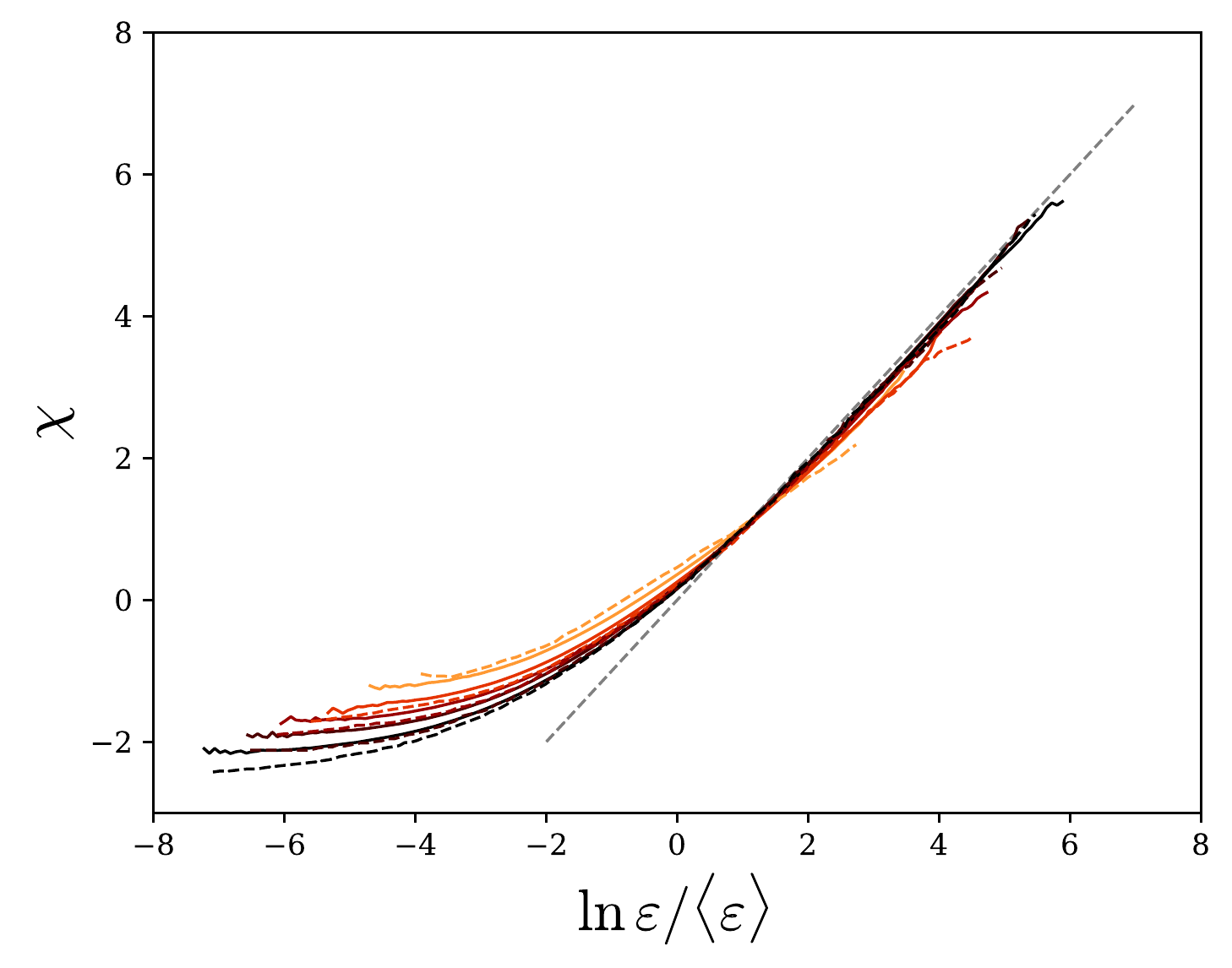}
 	\includegraphics[width=0.49\textwidth,height=0.4\textheight,keepaspectratio, clip]{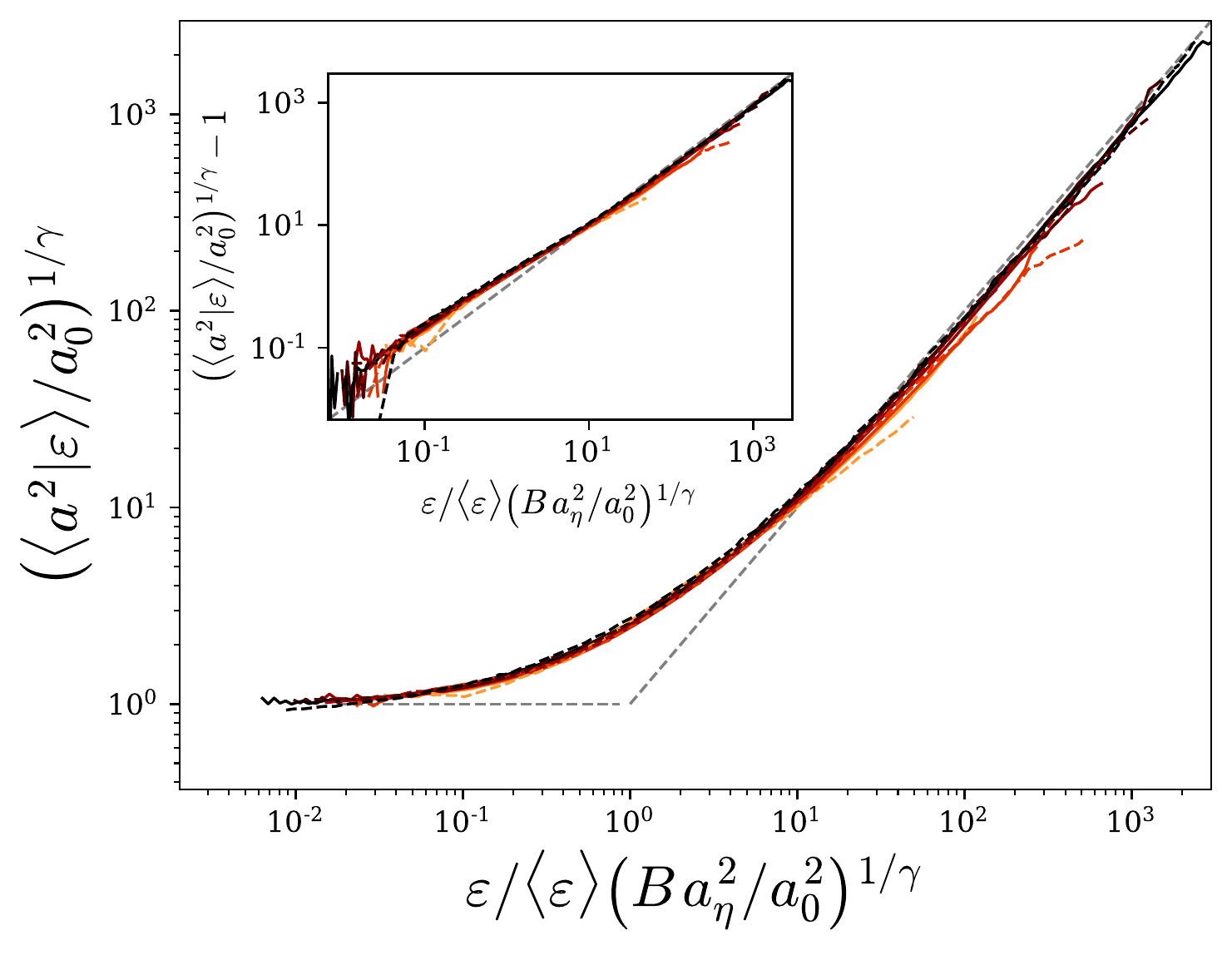}
	
 	\caption{ (Left) Evolution of $ \chi = \dfrac{1}{\gamma} \ln(1/B \, \langle a^2 | \varepsilon \rangle / a_\eta^2 )$, with $\gamma=3/2 + \beta$ versus $\ln\varepsilon/ \langle \varepsilon \rangle$ with $B=B_0+B_1/\ln(Re_\lambda)$ and $\beta=\beta_1/\ln(Re_\lambda)$ and 
	$B_0=17.1$, $B_1=-54.7$ and $\beta_1=-1$,
	for various $Re_\lambda$: 
  continuous line for our DNS at $Re_{\lambda}=50$, 90, 150, 230 and 380 from orange to black; and dashed lines correspond to the DNS of Yeung et al. \cite{Yeung:2006} for $Re_{\lambda}=40$, 139, 238, 385, 680, from orange to black.
	Comparison with the line $\chi=\ln\varepsilon/ \langle \varepsilon \rangle$ in gray dashed line.
 	(Right) Evolution of $\left( \langle a^2 | \varepsilon \rangle / a_0^2 \right)^{1/\gamma}=\exp(\chi-\chi_0)$ against $\zeta=\varepsilon / \langle \varepsilon \rangle \left( B a_\eta^2 / a_0^2 \right)^{1/\gamma} $ for the various Reynolds numbers. 
	Inset: plot of $\left( \langle a^2 | \varepsilon \rangle / a_0^2 \right)^{1/\gamma} - 1 $ against $\zeta=\varepsilon / \langle \varepsilon \rangle \left( B a_\eta^2 / a_0^2 \right)^{1/\gamma} $.
 }
 	\label{cond_acc_2} 
 	\end{figure*}

	To have a finite limit, consistently with the vanishing viscosity principle, we need $\beta_0=0$. 
	The remaining constants $B_0$, $B_1$ and $\beta_1$ are then determined by comparison with the DNS data. 
	From the inset of Fig. \ref{cond_acc_1}(Left) we see that both $\beta$ and $B$ are increasing functions of $Re_\lambda$ implying that both $B_1$ and $\beta_1$ are negative.
	In Fig. \ref{cond_acc_2}(Left) we assess the relations \eqref{eq:f_anomalous}-\eqref{eq:crit_B} by plotting 
	\begin{equation}
		\chi = \dfrac{1}{\gamma} \ln(1/B \, \langle a^2 | \varepsilon \rangle / a_\eta^2 )	\, ,
		\label{eq:def_chi}
	\end{equation}
	with 
	\begin{equation}
		\gamma = 3/2 + \beta	\, ,
		\label{eq:def_gamma}
	\end{equation}
	against $\ln(\varepsilon/ \langle \varepsilon \rangle) $ for various Reynolds numbers. 
	It is observed that with 
	$B_0=17.1$, $B_1=-54.7$ and $\beta_1=-1$,
	all the DNS data collapse on the line $\chi=\ln(\varepsilon/ \langle \varepsilon \rangle) $ (the bisectrix of the graph) for $\varepsilon \gg \langle \varepsilon \rangle$, validating the scaling relation.

	We can go a step further by using the low dissipative limit of the conditional acceleration.
	For that we introduce $\chi_0=\lim_{\varepsilon \rightarrow 0}\chi = \dfrac{1}{\gamma} \ln(1/B \, a_0^2 / a_\eta^2 )$.
	With this definition, $\chi-\chi_0 = \ln \left[ \left( \langle a^2 | \varepsilon \rangle / a_0^2 \right)^{1/\gamma} \right]$ 
	tends to $0$ in the low dissipative regions ($\varepsilon \ll \langle \varepsilon \rangle$ ).
	On the other hand, for $\varepsilon \gg \langle \varepsilon \rangle$, $\chi-\chi_0$ should be equal to $\chi-\chi_0 = \ln( \varepsilon / \langle \varepsilon \rangle) -\chi_0 = 	\ln \left[ \varepsilon / \langle \varepsilon \rangle \left( B a_\eta^2 / a_0^2 \right)^{1/\gamma} \right] $.
	This is seen in Fig. \ref{cond_acc_2}(Right) that presents the evolution of $ \left( \langle a^2 | \varepsilon \rangle / a_0^2 \right)^{1/\gamma} $ against 
	\begin{equation}
		\zeta = \varepsilon / \langle \varepsilon \rangle \left( B a_\eta^2 / a_0^2 \right)^{1/\gamma} 
	\end{equation}
	 for the various Reynolds numbers considered here.

	It is interesting to note that the curves are all overlapping even for intermediate values of $\varepsilon$, suggesting that the conditional acceleration variance can be cast in a self-similar form: 
	\begin{equation}
		 \langle a^2 | \varepsilon \rangle = a_0^2 \left(\phi( \zeta ) \right)^{\gamma} 
	\end{equation} 
	 with $\phi$ a universal function of only one argument $\phi=\phi(\zeta)$ with the asymptotics $\phi(\zeta)=1$ for $\zeta\ll 1$ and $\phi(\zeta)=\zeta$ for $\zeta \gg 1$.
	 Making a Taylor expansion of $\phi$ around $\zeta=0$ and using a matching asymptotic argument, simply yields to 
	 \begin{equation}
	 	\phi( \zeta ) = 1+\zeta \, .
	 \end{equation}
	 It is seen in the inset of the Fig. \ref{cond_acc_2}(Right) that the proposed expression for $\phi$ gives a good approximation of the data over the whole range of $\varepsilon$ and $Re_\lambda$. 
	 We can indeed observe more than 5 decades of quasi-linear growth of $ \left( \langle a^2 | \varepsilon \rangle /  a_0^2\right)^{1/\gamma}-1=\exp(\chi-\chi_0)-1$ with $\zeta$.
	 
	 The non-dimensional function $f$ introduced in \eqref{eq:a2_cond_adim} can, in consequence, be expressed as:
	 \begin{equation}
	 	f(\varepsilon/ \langle \varepsilon \rangle, Re_\lambda) = B \left( \varepsilon/ \langle \varepsilon \rangle \right)^{\beta} \left( 1+ \dfrac{1}{\zeta}\right)^{3/2+\beta} \, ,
	 \end{equation}
	 where the term within the brackets is interpreted as a correction factor for small dissipative regions.
	 Accordingly, we obtain the following expression for the conditional acceleration variance: 
	\begin{equation}
		\langle a^2 | \varepsilon \rangle = B a_\eta^2 \left( \left( \dfrac{1}{B}\dfrac{a_0^2}{a_\eta^2}\right )^{1/\gamma} + \dfrac{\varepsilon}{\langle \varepsilon \rangle} \right )^{\gamma}  \, .
		\label{eq:avar_cond_dissip_final}
	\end{equation}
	As  $\gamma=3/2+\beta$ and $B$ evolve slowly with $Re_\lambda$, their expressions remain speculative and would require a much larger range of Reynolds numbers to be validated.

	\subsection{Reynolds number dependence of the unconditional acceleration variance} 
	\label{sec:acc_var}
	
	Assuming the distribution of  the dissipation, we can integrate relation \eqref{eq:avar_cond_dissip_final} to obtain the unconditional variance 
	\begin{equation}
	\langle a^2 \rangle = \int \langle a^2| \varepsilon \rangle P(\varepsilon) d\varepsilon \ , 
	\label{eq:avar_integral}	
	\end{equation}
	   and thus propose an alternative formula to the empirical relations proposed in \cite{Yeung:2006, Sawford:2003,Hill:2002}.
	We consider that $ \varepsilon/\langle \varepsilon \rangle $ presents a log-normal distribution with parameter $ \sigma^2 \approx 3/8 \ln Re_\lambda/R_c $ with $R_c=10$ as shown by \cite{Yeung:2006} from DNS data, consistently with the proposition of Kolmogorov and Oboukhov \cite{Kolmogorov:1962,Oboukhov:1962}. 
	Notice nevertheless that other expressions for $\sigma^2$ have been proposed in the literature reflecting the vanishing viscosity limits \cite{Castaing:1990}.
	Taking for $\langle a^2| \varepsilon \rangle$ the expression \eqref{eq:avar_cond_dissip_final}, we perform the integration numerically with the expression \eqref{eq:crit} and \eqref{eq:crit_B} for $\beta$ and $B$ with the values of $B_0$, $B_1$ and $\beta_1$ and the expression of $a_0^2/a_\eta^2$ proposed above.
	The resulting evolution of the acceleration variance with the Reynolds number is presented in Fig. \ref{fig:avar_vs_Re}.
	It is seen that the predicted acceleration variance is in very good agreement with the DNS of \cite{Yeung:2006} and our DNS and is also very close to the relation proposed by \cite{Sawford:2003} for $Re_\lambda<1000$.
	
	The first term within the brackets in \eqref{eq:avar_cond_dissip_final} is the footprint of the large-scale structures whose effects is vanishing if the local dissipation rate is larger than $ \varepsilon/\langle \varepsilon \rangle  \gg (B \, a_\eta^2/a_0^2)^{-1/\gamma}$ and therefore can be neglected when the Reynolds number is large since $a_\eta^2 / a_0^2 \sim Re_\lambda$.
	Hence, for large Reynolds numbers, equation \eqref{eq:avar_cond_dissip_final} reduces to:
	\begin{equation}
	 	\langle a^2|\varepsilon\rangle / a_\eta^2= B \left(\dfrac{\varepsilon}{\langle \varepsilon \rangle} \right)^{\gamma} \, .
		\label{eq:a2_cond_eps_asymptote}
	\end{equation}
	With this expression, the acceleration variance is simply estimated from the moments of the log-normal distribution as:
	\begin{equation}
		\dfrac{\langle a^2 \rangle }{a_\eta^2} = B \left( Re_\lambda/R_c \right)^{9/64+3\beta(1+\beta/2)/8 } \, .
		\label{eq:a_var_RZ}
	\end{equation}	
	This expression, also presented in Fig. \ref{fig:avar_vs_Re}, is shown to converge to the previous estimate as the Reynolds number increases.
	
	In appendix \ref{sec:acc_var_serie_expension}, we show that the integral \eqref{eq:avar_integral} can be expressed from the generalized binomial series expansion.
	We further obtain the following estimation for the acceleration variance keeping only the first two terms:
	\begin{equation}
		\dfrac{\langle a^2 \rangle }{a_\eta^2} = B \left( \dfrac{Re_\lambda}{R_c} \right)^{3/16 \gamma (\gamma-1)} \left( 1+ \gamma \left( \dfrac{1}{B}\dfrac{a_0^2}{a_\eta^2}\right )^{1/\gamma} \left( \dfrac{Re_\lambda}{R_c} \right)^{-3/8 (\gamma-1)}  \right)  \, .
		\label{eq:avar_1storder}
	\end{equation}  
	In Fig. \ref{fig:avar_vs_Re}, it is seen that this relation almost overlaps with the direct numerical calculation of the variance though \eqref{eq:avar_integral}.
	The term within the brackets enables to measure the contribution from small Reynolds number effects.
	At $Re_\lambda\approx 100$, the two estimates \eqref{eq:a_var_RZ} and \eqref{eq:avar_1storder} for the variance differ by about 20\%, while there is about 8\% in difference at $Re_\lambda\approx 500$.
	That confirms that the term containing $a_0^2$  is indeed vanishing at large Reynolds numbers.
	
	The previous estimations of the acceleration variance tend asymptotically to the following power law:
	\begin{equation}
		\langle a^2 \rangle /a_\eta^2 = 7.62 Re_\lambda^{9/64} \, ,
		\label{eq:a_var_RZ_asymptotic}
	\end{equation}
	where we have used \eqref{eq:crit} to obtain the value of the prefactor,  
	$R_c^{-9/64}  B_0 \exp(3 \beta_1/8 ) \approx 7.62$.
	This expression is presented as well in Fig. \ref{fig:avar_vs_Re}, confirming that the convergence toward the power law is very slow, and that \eqref{eq:a_var_RZ} should be considered as an intermediate asymptotic expression for the acceleration variance.

		\subsection{Multiplicative cascade for the acceleration} \label{sec:cascade}

Substituting \eqref{eq:a2_cond_eps_asymptote} into \eqref{eq:cond_acc_B}, we can eventually estimate the doubly-conditional acceleration variance for large Reynolds numbers as:
	\begin{equation}
	 	\langle a^2|\varepsilon, K\rangle / a_\eta^2= C \exp(\alpha K/\langle K \rangle) \left(\dfrac{\varepsilon}{\langle \varepsilon \rangle} \right)^{\gamma} \, ,
		\label{eq:a2_cond_eps_k_asymptote}
	\end{equation}
	where $C=A\, B$,  with expressions of $A$ and $B$ determined above.

As mentioned above and apparent in the previous formula, the acceleration depends on the local value of the kinetic energy, along with the local dissipation rate.
The acceleration, being mainly due to the pressure gradient, it present a non-local behavior.
The fact that the acceleration depends on the local kinetic energy but not on a local Reynolds number reflects that its non-locality is a purely kinematic effect.
Further, the exponential dependence on the kinetic energy suggests that the acceleration can respond to the structures of all sizes. 
	
	To illustrate this point, we discuss a multiplicative cascade model for the acceleration that incorporates effect of the full spectrum of the flow strucutre.
	Fluctuations of the locally-space-averaged dissipation rate can be modeled by multiplicative cascades \cite{Yaglom:1966,Kolmogorov:1941,Monin:1981b,Mandelbrot:1974,Benzi:1984,Frisch:1978}.
	Such model proposes to express the local dissipation over a volume of size $\ell= L \lambda^n$, with $\lambda<1$ and $L$ being the large-scale of the flow, as the product of $n$ random numbers $ \xi_i$:
	\begin{equation}
		\varepsilon_\ell = \langle \varepsilon \rangle \prod_{i=1}^{n} \xi_i \, .
		\label{eq:cascade_dissip}
	\end{equation}
	Typically, for $n$ large, this yields log-normal distribution of $\varepsilon_\ell$ assuming the $\xi_i$ are independent and identically distributed (and have as well finite variance).
	
	We propose likewise to write the squared acceleration, coarse-grained at scale $\ell$, as :
	\begin{equation}
		a_\ell^2 = a_0^2 \prod_{i=1}^{n} \theta_i\, .
	\end{equation} 
	The scale-to-scale factor $\theta_i$ is given by:
	\begin{equation}
		\theta_i 
		=  \exp\left(\dfrac{\alpha}{\langle K \rangle} \dfrac{1}{2}v_i^2 \right) (\xi_i)^{\gamma} 
		=   \exp\left(\dfrac{\alpha}{\langle K \rangle} \dfrac{1}{2}v_i^2 + \gamma \ln \xi_i \right) \, , 
	\end{equation}
	where $v_i$ is here the velocity of eddies of size $\ell_i=L\lambda^i$, which is also a fluctuating quantity.
	The exponential modulation is then interpreted as an entrainment acceleration due to these structures.

	With this expression we obtain:
	\begin{equation}
		a_\ell^2 
		= a_0^2 \exp\left(\dfrac{\alpha}{\langle K \rangle} \sum_{i=1}^n \dfrac{1}{2}v_i^2   +\gamma \sum_{i=1}^n \ln \xi_i \right) \, .
		\label{eq:cascade_a2}
	\end{equation}	
	Setting $n=\ln(\eta/L)/\ln(\lambda)\sim \ln{Re_\lambda}$, $\eta$ being the Kolmogorov length scale, we have $K =\sum_{i=1}^n \dfrac{1}{2}v_i^2 $ due to the additive nature of the kinetic energy. Thereby 
	 using \eqref{eq:cascade_dissip}, we obtain back \eqref{eq:a2_cond_eps_k_asymptote} by taking the conditional average of \eqref{eq:cascade_a2}.
	The order of magnitude of the eddy velocities can be estimated from the Kolmogorov relation, $(\varepsilon_\ell \ell)^{1/3}$, showing that the sum is  a priori dominated by the large-scales but, on the other hand, because of the intermittent behavior of $\varepsilon_\ell$, it may well happen that the 
	the inertial-scale structures can be dynamically important.
	
	Note that in this multiplicative model, we have transposed the statistical relation \eqref{eq:a2_cond_eps_k_asymptote} to an instantaneous version. Such idealization, find support in the invariance of the conditional PDF, which is shown in appendix \ref{sec:conditional_pdf}. 
	An other important point  to mention, is that although we assume that the local acceleration depends both on $K$ and $\varepsilon$ it is not assumed that those two variables are independent.

	The dissipation presents large fluctuations leading to very important accelerations and, even if the acceleration orientation is changing rapidly, it can cause a local increase of the velocity.
	When the kinetic energy becomes significantly larger than its averaged value, then the modulation of the acceleration by the exponential term becomes preponderant, thus offering a feedback mechanism allowing the obtention of the normal fluctuations of the velocity.
	This dynamic scenario appears consistent with the recent DNS analysis of \cite{Picardo:2020} showing that the fluid-particles can undergo energy gains in intense dissipative regions and is developed in the next section.

	\FloatBarrier


	\section{Stochastic modeling of the fluid-particle dynamics \label{sec:stoch_model}}
	
	\subsection{Model formulation}

    The foregoing multiplicative model suggests that the acceleration norm can be determined from the local kinetic energy and dissipation rate.
    The relation \eqref{eq:a2_cond_eps_k_asymptote} is pointwise and so it applies equally well to both Lagrangian and Eulerian  descriptions.
    However, in the Lagrangian framework, the kinematic relation between velocity and acceleration allows proposing a model for the acceleration depending only on the local dissipation rate.
    The evolution of the later along the particle path is to be obtained from a stochastic process.
    For the derivation of this model, we will rely on the relation \eqref{eq:a2_cond_eps_k_asymptote}, in which the contribution from low dissipative events are neglected:

	\begin{equation}
		a^2 = f(K,\varepsilon) = a_\eta^2 C \left(\dfrac{\varepsilon}{\langle \varepsilon \rangle}\right)^{\gamma} \exp\left(\alpha \dfrac{K}{\langle K \rangle}\right) \, .
		\label{eq:a2_start}
	\end{equation}
	We express the increments of $ a^2 $ as a second order Taylor expansion in $K$ and $\varepsilon$, 
	\begin{equation}
		da^2 = a^2 \left( \alpha \dfrac{dK}{\langle K \rangle} +\gamma \dfrac{d \varepsilon}{\varepsilon} + \dfrac{\alpha^2}{2} \dfrac{dK^2}{\langle K \rangle^2} + \dfrac{ \gamma(\gamma-1)}{2} \dfrac{d \varepsilon^2}{\varepsilon^2} + \gamma \alpha \dfrac{dK}{\langle K \rangle } \dfrac{d \varepsilon}{\varepsilon} \right) \, .
		 \label{eq:an2_1}
	\end{equation}
	We consider  $\varepsilon$ as stochastic variable reflecting the very large number of degrees of freedom that control them.
	In a fairly general way, we consider  that the dissipation $ \varepsilon $ follows a multiplicative stochastic process: 	
	\begin{equation}
		d \varepsilon = \varepsilon\Pi dt + \varepsilon \Sigma dW \, ,
		\label{eq:eps_mult}
	\end{equation}
	where $dW$ are the increments of the Wiener process ($\langle dW \rangle =0$ ; $\langle dW^2 \rangle = dt$).
	We specify the terms $ \Pi $ and $ \Sigma $ below. 
	Substituting \eqref{eq:eps_mult} into \eqref{eq:an2_1} one obtains, following the Ito calculus, at first order in $dt$ :
	\begin{equation}
		da^2 = a^2 \left[ \dfrac{\alpha}{\langle K \rangle} P + \gamma \Pi + \dfrac{\gamma(\gamma-1)}{2} \Sigma^2 \right] dt 
		 + \gamma a^2 \Sigma dW \, .
		 \label{eq:an2_2}
	\end{equation}
	We used the identity $dK = u_i du_i = u_i a_i dt = P dt $, where $P$ is the mechanical power per unit of mass exchanged by the fluid particle. 
	Even if $ \Pi $ and $ \Sigma $ are given, eq. \eqref{eq:an2_2} is not closed, as it remains to estimate $ P = a_i u_i $, 
	which requires the knowledge of $ a_i $ and $ u_i $.
	
	As mentioned in the introduction, we introduce a vectorial stochastic model for the dynamics of a fluid particle.
	We are looking for a stochastic process of the form: 	
	\begin{eqnarray}	
		d u_i &= & a_i dt \, ,
		\label{eq:u_stoch}		 \\
		d a_i &=& M_i dt + D_{ij} dW_j \, ,
		\label{eq:a_stoch}		
	\end{eqnarray}
	where $dW_j$ are the increments of the $j$th component of a tridimensional Wiener process ($\langle dW_j \rangle =0$~; $\langle dW_i dW_j \rangle = dt \delta_{ij}$).	
	A priori, the vector $ M $ and the tensor $ D $ depend on the vectors $ a $ and $ u = \int a dt $.
	Indeed, $ M $ must depend on $ u $ to allow the particle velocity to reach a statistically steady state. 
	
	Now, we propose expressions for $ M_i $ and $ D_ {ij} $.
	For this, we want to impose, on the one hand, that the model is isotropic ($ \langle a_i a_j \rangle = 0 $ for $ i \neq j $) and, on the other hand,  that its norm $ a^2 = a_i a_i $ is compatible with the expression \eqref{eq:an2_2}.
	We therefore write the stochastic equation for $ a_{ij}^2 = a_ia_j $ derived from \eqref{eq:a_stoch}, thanks to the Ito formula: 
	\begin{eqnarray}
		da^2_{ij} &=& a_j da_i + a_i da_j + da_i da_j \nonumber 
		\\
		 &=& \left( M_i a_j + M_j a_i + D_{ik} D_{jk} \right) dt + \left( a_j D_{ik} +a_i D_{jk} \right) dW_k \, .
		\label{eq:a2ij}
	\end{eqnarray}
	For the square of the norm $ a^2 = a_ia_i $, we have:	
	\begin{equation}
		da^2 = \left( 2 a_i M_i + D_{i j } D_{i j } \right) dt + 2 a_i D_{ij} dW_j \, .
		\label{eq:an2}
	\end{equation}
	We then proceed by identification between \eqref{eq:an2} and \eqref{eq:an2_2}, in a similar way as \cite{Girimaji:1990} and \cite{Pereira:2018}, by identifying first the square of the diffusion term and then the drift term.

	\subsubsection{Identification of the diffusion term and the maximum winding hypothesis}
	
	Quite generally, we can decompose the diffusion tensor into:
	\begin{equation}
		D_{ij} = c_1 \delta_{ij} + S_{ij} + \Omega_{ij} \, ,
		\label{eq:Dij_1}
	\end{equation}
	where $ S_{ij} $ is a zero-trace symmetric tensor and $ \Omega_{ij} $ is an antisymmetric tensor.
	The latter can be written as $ \Omega_{ij} = \epsilon_{ijk} \omega_k $ with $ \epsilon_{ijk} $ the Levi-Civita permutation symbol and $ \omega_k $ a pseudo-vector.
	$ S_{ij} $ must be zero in order to guarantee the statistical isotropy of the acceleration.
	But $ \Omega_{ij} $ can be different from 0.
	Indeed, the experimental results of \cite{Mordant:2004} and numerical results of \cite{Pope:1990b} have shown that the acceleration presents a scale separation between the evolution of the components and its norm, and that this separation can be modeled using processes for the acceleration norm and its orientation vector \cite{Zamansky:2013b,Gorokhovski:2018,Sabelnikov:2019}. 
	A stochastic model for orientation can be formulated as a diffusion process with a rotational part in the diffusion tensor \cite{Gorokhovski:2018,Wilkinson:2011}. 
	Since the model for the dynamics \eqref{eq:u_stoch} - \eqref{eq:a_stoch} involves only two vectors, $ a $ and $u $, we propose to form the pseudo-vector $\omega$ from these two vectors in order to get a closed model: $ \omega_k = c_2 \epsilon_{klm } a_l u_m $.
	The model remains statistically isotropic and the chirality of the flow is not broken either since the odd moments of $ dW_j $ are zero (Gaussian with zero mean). 
	In other words, the sign of $ c_2 $ does not matters.
	We then have:
	\begin{equation}
		D_{ij} = c_1 \delta_{ij} + c_2 (a_i u_j-a_j u_i) \, .
		\label{eq:Dij_2}
	\end{equation}
	It is to be noted that $ c_1 $ and $ c_2 $ are not constant. 
		
	By identifying the square of the diffusion term between \eqref{eq:an2_2} and \eqref{eq:an2} we find: 	
	\begin{equation}
		 \gamma^2(a^2)^2\Sigma^2 = 4 a_i a_j D_{ik}D_{jk} \, .
	\end{equation}
	Expanding it by using expression \eqref{eq:Dij_2}, we find: 	
	\begin{equation}
		\gamma^2(a^2)^2\Sigma^2 = 4 a^2(c_1^2 + c_2^2 (2 a^2 K -P^2)) \, ,
	\end{equation}
	which gives for $c_1$:
	\begin{equation}
		c_1^2 = \dfrac{\gamma^2}{4} a^2 \Sigma^2 - c_2^2 (2 a^2 K - P^2) = a^2 \left(\dfrac{\gamma^2}{4} \Sigma^2 - 2c_2^2 K \left( 1 - \dfrac{a_T^2}{a^2} \right)\right) \, ,
		\label{eq:c_1}
	\end{equation}
	where we have introduced the tangential acceleration $ a_T $, as the projection of the acceleration vector in the direction of the velocity vector: $ a_T = a_i u_i / \sqrt{u^2} = P/\sqrt{2K}$. 	
	Equation \eqref{eq:c_1} imposes a constraint on $ c_2 $ in order to guarantee the positivity of $ c_1^2$: 
	\begin{equation}
		c_2^2 2K \leq \dfrac{\gamma^2}{4} \Sigma^2\, ,
	\end{equation} 
	since $ 0\geq 1 - \dfrac{a_T^2} {a^2} \geq 1$.
	So, in order to guarantee the positivity of $ c_1 ^ 2 $ whatever $ K $, $ c_2^2$ must be proportional to $ 1 / K $.
	Introducing a parameter $ c_R $ as $ c_2^2 = \dfrac{\gamma^2}{4} \Sigma^2 \dfrac{c_R^2}{2K} $, with the constraint $ c_R^2 \leq 1 $, we obtain:
	\begin{equation}
		c_1^2 = \dfrac{\gamma^2}{4}\Sigma^2 \left(a^2 (1-c_R^2) + c_R^2 a_T^2\right) \, .
		\label{eq:c_1_b}
	\end{equation}
	Subsequently, we only consider the limit $ c_R = 1 $ that corresponds to the maximum rotational part of the diffusion tensor. 
	We will discuss this choice in more detail below in section \ref{sec:results}, when presenting the results.

	Finally, from \eqref{eq:Dij_2} and the expressions of $ c_1 $ and $ c_2 $, we write the components of the diffusion tensor as 
	\begin{equation}
		D_{ij} = \sqrt{\dfrac{\gamma^2}{4}\Sigma^2} \left[ \sqrt{ a_T^2 } \delta_{ij} + \sqrt{ a_N^2} \epsilon_{ijk} b_k \right] \, ,
		\label{eq:Dij_5}
	\end{equation}
	where we introduced the normal component $ a_N $ of the acceleration $ a_N^2 = a^2 - a_T^2 $, and the bi-normal unit vector\footnote{\label{note:rot}To obtain this relation we notice that $ a_i u_j -a_j u_i = \epsilon_{ijk} \epsilon_{klm} a_l u_m $ and that the vector $ b_k $ is the unit vector collinear to $ \epsilon_{klm} a_l u_m $: $ b_k = \epsilon_{klm} a_l u_m / | \epsilon_{hij} a_i u_j | $. By expanding the norm, we have: $ (\epsilon_{hij} a_i u_j)^2 = 2a^2 K -P ^2 $. 
	We therefore write:
$ \epsilon_{klm} a_l u_m = b_k \sqrt {2a^2 K-P^2} = b_k \sqrt{2K} \sqrt{a^2-a_T^2} = b_k \sqrt{2K} \sqrt{a_N^2} $. }
	$ b_k = \epsilon_{klm} u_l a_m / | \epsilon_{hij} u_i a_j | $.

 Note that $ b_k $, $ a_T $ and $ a_N $ are not well defined when $ K = 0 $. 
 However, $c_2$ must vanish when $u=0$ and we can therefore consider that $ c_R = 0 $ in that case.

	\subsubsection{Determination of the drift term}
	
	Identifying the drift term between \eqref{eq:an2} and \eqref{eq:an2_2}, we get: 	
	\begin{equation}
	2 a_i M_i + D_{ij}D_{ij} = a^2 \left( \dfrac{\alpha}{\langle K \rangle}P + \gamma \Pi + \dfrac{ \gamma(\gamma-1)}{2}\Sigma^2\right)	 \, .
	\end{equation}
	From \eqref{eq:Dij_5} the term $D_{ij}D_{ij}$ is computed as \footnote{ $D_{ij} D_{ij} = \dfrac{\gamma^2}{4} \left( a^2_T \underbrace{\delta_{ij}\delta_{ij}}_{3} + a^2_N \underbrace{ \epsilon_{ijk} b_k \epsilon_{ijl} b_l}_{2\delta_{kl} b_k b_l} \right)$,  $\delta_{kl} b_k b_l = 1 $ since $b$ is a unit vector and with $a^2=a^2_T + a^2_N$, we obtain the result.} 
	\begin{equation}
		D_{ij}D_{ij} = \dfrac{\gamma^2}{4}\Sigma^2 \left(2a^2+ a_T^2 \right) \, .
	\end{equation}
	We then have
	\begin{equation}
		a_i M_i = a^2 \left( \dfrac{\alpha}{2\langle K \rangle}P +\dfrac{\gamma}{2}\Pi -\dfrac{\gamma}{4}\Sigma^2\right) - a_T^2 \dfrac{\gamma^2}{8}\Sigma^2
		\label{eqaimi}
	\end{equation}
	
	To go further we must now specify the terms $ \Pi $ and $ \Sigma $ used for the stochastic process for $ \varepsilon $.
	Various models for the dissipation have been proposed.
	Pope and Chen \cite{Pope:1990} proposed a simple model based on the exponential of an Orstein-Uhlenbeck process (see appendix \ref{sec:stoch_dissip}).
	Here, we rely on the model proposed in \cite{Chevillard:2017b, Pereira:2018}.
	This non-Markovian log-normal model presents a logarithmic decrease in the correlation of $ \varepsilon $  which is consistent with the idea of a turbulent cascade and a multiplicative process (see appendix \ref{sec:cascade_dissip}), unlike the Pope and Chen  model  which gives an exponential decrease, see also the discussion in \cite{Letournel:2021}.
	As presented in appendix \ref{sec:stoch_dissip}, the drift and diffusion terms are written respectively as:
	\begin{equation}
		\Pi = \dfrac{1}{\tau_\varepsilon} \Big( - \ln \dfrac{\varepsilon}{\langle \varepsilon\rangle} + \dfrac{\sigma^2}{2 \Lambda^{2}}\big( \dfrac{\tau_\varepsilon}{\tau_c} - \Lambda^2 \big) + \dfrac{\sigma}{\Lambda}\hat{\Gamma} \tau_{\varepsilon} \Big) \, ,
		\label{eq:Pi}
	\end{equation}
	and
	\begin{equation}
		\Sigma = \sqrt{\dfrac{\sigma^2}{\Lambda^2\tau_c}} \, ,
		\label{eq:Sigma}
	\end{equation}
	with $\sigma^2$ the variance of the logarithm of $\varepsilon$, 
	$\tau_\varepsilon$ the correlation time of $\varepsilon$,
	$\tau_c$ the regularization time scale of the process (taken equal to the Kolmogorov dissipative time $\tau_\eta$),
	$\Lambda^2$ a normalization factor,
	and $\hat{\Gamma}$ the convolution of the Wiener increments with a temporal kernel, ensuring the non-markovian property of the process.
	In the process for $\varepsilon$ proposed by \cite{Chevillard:2017b, Pereira:2018}, the latter corresponds to a fractional Gaussian noise with 0 Hurst exponent \cite{Mandelbrot:1968} regularized at scale $\tau_c$. 
	The expression of the convolution kernel proposed by \cite{Chevillard:2017b,Pereira:2018}, and recalled in the appendix \ref{sec:stoch_dissip}, applies to a scalar noise since the dissipation rate is a scalar, whereas the acceleration model involves a vectorial noise.
	Therefore, the kernel in $\hat{\Gamma}$ includes a projection in order to apply to the vectorial Wiener increments:
	\begin{equation}
		\hat{\Gamma} = - \dfrac{1}{2}\int_{-\infty}^{t} \dfrac{1}{(t-s+\tau_c)^{3/2}}\mathcal{P}_j dW_j(s)
		\label{eq:Gamma_hat_b}
	\end{equation}
	By proceeding in a similar way as \cite{Pereira:2018}, the projection operator is obtained by identification between the diffusion terms of \eqref{eq:an2_2} and \eqref{eq:an2}: 
	\begin{equation}
		\mathcal{P}_j = \dfrac{2a_i D_{ij}}{\gamma a^2 \Sigma}= \dfrac{ \sqrt{a_T^2}-a_T}{a^2}{a_j} + e_j
		\label{eq:proj_convol}
	\end{equation}
	where we have used the relation recall in footnote \ref{note:rot} and where $e_j$ is the unit vector tangent to the trajectory, $e_i=u_i/\sqrt{2K}$.
	It is interesting to remark that the rotational part of the diffusion tensor induces an asymmetry of the projector between positive and negative power exchange (recall that $P = \sqrt{2K} a_T$). 
	Indeed for $P\geq 0$, $\mathcal{P}_j =e_j$ while for $P<0$ one has $\mathcal{P}_j =(1-2p^2)e_j-2p\sqrt{1-p^2}b_j$ with $p=P/\sqrt{2K a^2}$. 
	In both cases, as it can be readily checked, $\mathcal{P}$ is a unit vector.
	 
	Substituting the expression \eqref{eq:Pi} and \eqref{eq:Sigma} for $ \Pi $ and $ \Sigma $ in \eqref{eqaimi} we have 	
	\begin{equation}
		a_i M_i = a^2 \left( 
		\dfrac{\alpha}{2\langle K \rangle}P - \dfrac{\gamma}{2\tau_\varepsilon}\big( \ln\dfrac{\varepsilon}{\langle \varepsilon \rangle} + \dfrac{1}{2}\sigma^2 - \dfrac{\sigma}{\Lambda} \hat{\Gamma} \tau_\varepsilon \big) 
		\right)
		 - a_T^2 \dfrac{\gamma^2}{8}\dfrac{\sigma^2}{\Lambda^2 \tau_c}  \, .
 		\label{eq:aM_0}
	\end{equation}
	
	According to \eqref{eq:a2_start}, we can write: 
	\begin{equation}
		\ln\left(\dfrac{\varepsilon}{\langle \varepsilon \rangle}\right) = \dfrac{1}{\gamma} \left( \ln \dfrac{a^2}{a_\eta^2} - \ln C - \alpha \dfrac{K}{\langle K \rangle} \right) \, ,
	\end{equation} 
	which gives, once substituted into \eqref{eq:aM_0},
	\begin{equation}
		a_i M_i =
			 	 a^2 \left[ \dfrac{\alpha}{2\langle K \rangle} \left(P+\dfrac{K}{\tau_\varepsilon}\right)
			 	- \dfrac{1}{2\tau_\varepsilon} \left( \ln \big( \dfrac{a^2}{a_\eta^2} \big) \underbrace{-\ln C + \dfrac{\gamma}{2} \sigma^2 - \gamma\dfrac{\sigma}{\Lambda}\hat{\Gamma}\tau_\varepsilon }_{-\hat{\Gamma}_*} \right) \right]
			 - \dfrac{a_T^2}{\tau_c} \underbrace{\dfrac{\gamma^2}{8}\dfrac{\sigma^2}{\Lambda^2 }}_{\sigma_*^2} \, .
		\label{eq:aM}
	\end{equation}
	In order to simplify the notations, we have introduced $ \hat{\Gamma}_* = \gamma\dfrac{\sigma}{\Lambda}\hat{\Gamma}\tau_\varepsilon+\ln C - \dfrac{\gamma}{2} \sigma^2 $ and $\sigma_*^2 = \dfrac{\gamma^2}{8}\dfrac{\sigma^2}{\Lambda^2}$.	
	It is interesting to notice that in \eqref{eq:aM} the terms $ P + \dfrac {K} {\tau_\varepsilon} = \dfrac{dK}{dt} + \dfrac{K} {\tau_\varepsilon} $ acts as a penalization leading the correlation of the kinetic energy to decay exponentially.
	
	We then propose for $ M_i $ an expression compatible with \eqref{eq:aM}.
	Proceeding by identification, we have the following relation:
	\begin{eqnarray}
	 M_i &=& 
	 		 \dfrac{\alpha}{2\langle K \rangle} \left( \lambda a_i P + (1-\lambda) a^2 u_i + a_i\dfrac{K}{\tau_\varepsilon} \right) \nonumber \\
	 		& &- a_i \left(\ln\big( \dfrac{a^2}{a_\eta^2} \big) - \hat{\Gamma}_* \right) \dfrac{1}{2 \tau_\varepsilon} \nonumber \\
	 		& & - \dfrac{\sigma_*^2}{\tau_c} \dfrac{a_T^2}{a^2}a_i + B_i \, ,
			\label{eq:M}
	\end{eqnarray}
	where we have introduced the vector $ B_i $, such that $ B_i a_i = 0 $ as well as the factor $ \lambda $ that both account for the indeterminacy inherent to the inverse projection.
	By assuming again that there are only two vectors at our disposal, we can take $ B_i = \dfrac {\alpha} {2 \langle K \rangle} \lambda' \left (P a_i - a^2 u_i \right) $ by introducing the factor $ \lambda'$.
	Note that from the point of view of the projection, the factors $ \lambda $ and $ \lambda'$ are arbitrary in the sense that the scalar product of $ a_i $ and \eqref{eq:M} gives \eqref{eq:aM} whatever their values.
	We can nevertheless notice that the terms involving $ \lambda $ and $ \lambda'$ can be combined, and, by noting $ c_u = \lambda + \lambda'$, we get:
	\begin{eqnarray}
	 M_i &=& 
	 		 \dfrac{\alpha}{2\langle K \rangle} \left(a_i \big( c_u P + \dfrac{K}{\tau_\varepsilon} \big) - (c_u-1) a^2 u_i \right) \nonumber \\
	 	& &	- a_i \left( \ln \big( \dfrac{a^2}{a_\eta^2}\big) - \hat{\Gamma}_* \right) \dfrac{1}{2 \tau_\varepsilon} \nonumber \\
	 	& & 	- \dfrac{\sigma_*^2}{ \tau_c} \dfrac{a_T^2}{a^2}a_i \, .
			\label{eq:M2}
	\end{eqnarray}
	We can notice that the terms of the first line correspond to the coupling with the velocity, those of the second take into account the log-normal character of the dissipation and the last term is due to the rotational part of the diffusion tensor.
	The diffusion term \eqref{eq:Dij_5} becomes, by using expression \eqref{eq:Sigma}:
	\begin{equation}
		D_{ij} = \sqrt{\dfrac{2 \sigma_*^2}{\tau_c}} \left[ \sqrt{ a_T^2 } \delta_{ij} + \sqrt{ a_N^2} \epsilon_{ijk} b_k \right] \, .
		\label{eq:Dij_6}
	\end{equation}
	We have thus specified our stochastic model for the dynamics of a fluid particle.
	It is given by \eqref{eq:u_stoch}, \eqref{eq:a_stoch}, \eqref{eq:M2} and \eqref{eq:Dij_6}.

	\subsection{Parameters and numerical approach}

	From a dimensional point of view, to determine the physical parameters of the stochastic model, one must specify time and velocity scales as well as a Reynolds number.
	This amounts for example to imposing the average kinetic energy $ \langle K \rangle $, the average dissipation rate $ \langle \varepsilon \rangle $ and the viscosity $ \nu $.
	From these physical parameters, we calculate $ a_\eta^2 = \langle \varepsilon \rangle^{3/2} \nu^{- 1/2} $, $ \tau_\eta = \langle \varepsilon \rangle^{- 1/2} \nu^{1/2} $.
	We can also get the Reynolds number based on the Taylor scale $ Re_\lambda = u' \lambda / \nu = 2 \sqrt{15}/3 \, \langle K \rangle / \sqrt{\langle \varepsilon \rangle \nu } $ with $ u' = \sqrt{2 \langle K \rangle / 3} $ and $ \lambda^2 = 15 \nu u'^2 / \langle \varepsilon \rangle $.
	We then deduce the Lagrangian integral times scale $ \tau_L $ as $ \tau_L = 0.08 Re_\lambda \tau_\eta $ from the DNS results reported by \cite{Hackl:2011, Sawford:2011}.
	
	The parameter $ \sigma^2 $ is estimated using the relation given by \cite{Yeung:2006}: $ \sigma^2 \approx 3/8 \ln Re_\lambda/R_c $ with $R_c\approx10$ compatible with the prediction of Kolmogorov and Obhoukov \cite{Kolmogorov:1962, Oboukhov:1962}.
	As mentioned in \cite{Kolmogorov:1962} and \cite{Monin:1981b}, the specific value of $R_c$ is depending on the large-scales.
	Since the influence of the large-scales is neglected in our modeling (see section \ref{sec:acc_var}), we choose in the following simply $ \sigma^2 \approx 3/8 \ln Re_\lambda$.
	We set as well $ \alpha = 1/3 $ and $\gamma=3/2+ \beta$ with 
	$\beta=-1/ \ln Re_\lambda$ 
	in accordance with the results of the DNS presented above.
	The prefactor $C$ is computed as 
	$C=c_0\, A \, B $
	 where $A=\left(1-\dfrac{2 }{3} \alpha \right)^{3/2}\approx 0.686$, 
	$B= 17.1 - 54.7 / \ln Re_\lambda$
	as determined by DNS.	
	The term $c_0$ is introduced so that the predicted acceleration variance follows \eqref{eq:a_var_RZ}, as one would expect from the construction of the stochastic model, despite the fact that we take $\sigma^2=3/8 \ln Re_\lambda$ instead of $\sigma^2=3/8 \ln Re_\lambda/R_c$.
	Consequently, we have $c_0= (1/Re_c)^{9/64+3\beta/8(1+\beta/2)}$.

	For simplicity we have used $\tau_c=\tau_\eta$ and $\tau_\varepsilon=\tau_L$.
	From $ \tau_\varepsilon $ and $ \tau_c $ we calculate the value of the normalizing constant $ \Lambda $ as explained in \ref{sec:stoch_dissip}.
	Finally, for the parameter $ c_u $, which is the only free parameter of the model, we have determined numerically that with $ c_u = 5.22 $ the ratio $ K / \langle K \rangle $ is 1 on average for all values of the Reynolds number.

	The sample paths of this model are obtained by numerical integration of the stochastic differential equation.
	Numerical integration is made with an explicit Euler scheme by taking a time-step $ dt = \tau_{\eta, min} / 100 $ with $\tau_{\eta, min} = \sqrt{\nu/\varepsilon_{max}}$, an estimation of the minimum dissipative time scale likely to happen during the simulation. This is estimated from the log-normal distribution of the dissipation: $ \tau_{\eta, min} = \tau_\eta \exp(- x \sigma / 2 +\sigma^2/4) $, with $x=6$ by considering that the probability that a random number following the normal distribution reaches a value of $6$ standard deviation is sufficiently low (see \eqref{eq:def_chi}). 

	For the calculation of the convolution term $ \hat{\Gamma} $ appearing in \eqref{eq:M2}, we propose in appendix \ref{sec:algo_dissip} an efficient algorithm.
	
	A simple Python script presenting the algorithm used to integrate the proposed stochastic model is available in supplemental material \cite{SupplementalMaterial}.

	\subsection{Results} \label{sec:results}
	
\begin{figure*}[h]
 	\centering
	\includegraphics[width=.49\textwidth,height=0.4\textheight,keepaspectratio, clip]{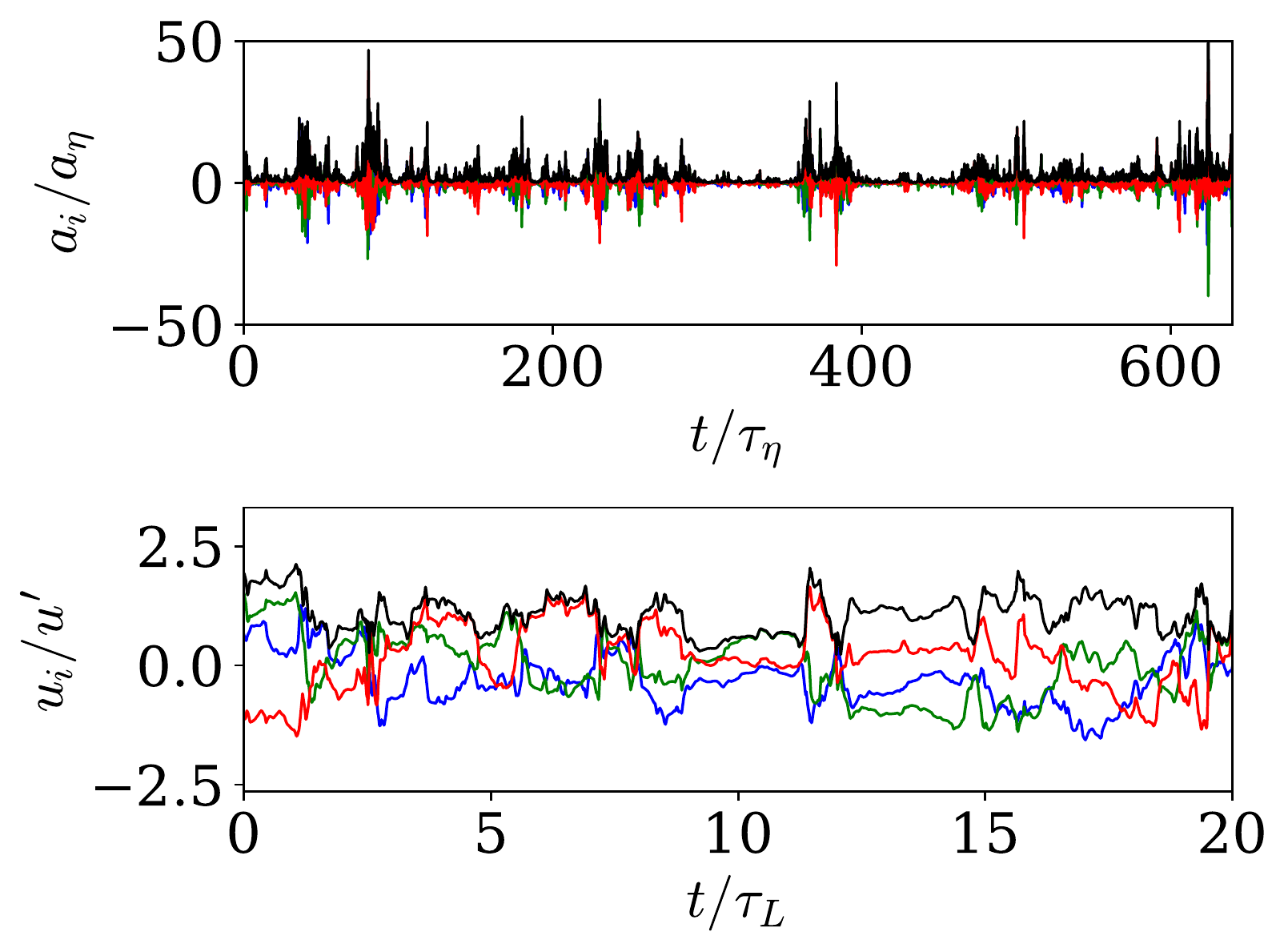}
	\includegraphics[width=.49\textwidth,height=0.4\textheight,keepaspectratio, clip]{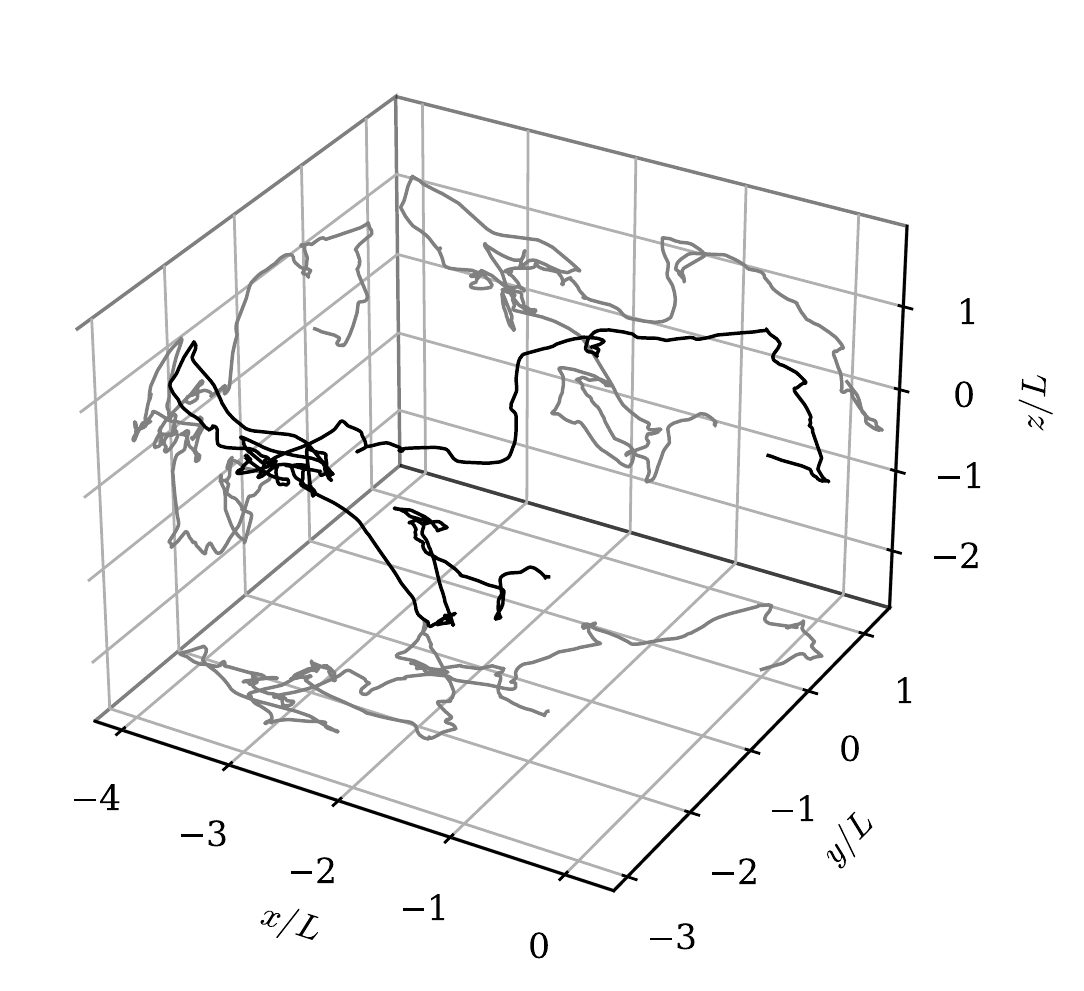}
	
 	\caption{A realization of the stochastic process for $Re_\lambda=1100$. 
	Top left: evolution of the acceleration with time, $a_x$: red, $a_y$: green, $a_z$: blue, $|a|$: black. 
	Bottom left: evolution of the velocity with time, $u_x$: red, $u_y$: green, $u_z$: blue, $|u|$: black.
	Right: 3D trajectory of a fluid particle for a duration of 100 $\tau_L$.
	}
 	\label{fig:model_rea} 
 	\end{figure*}

	We show in Fig. \ref{fig:model_rea} a realization of this process for $Re_\lambda=1100$.
	We see the temporal evolution of the components of acceleration and velocity.
	There is a very intermittent acceleration with an alternation of periods in which the acceleration of the fluid particle is almost zero with phases of very intense activity.
	This results in fluid-particle trajectories, obtained by integration of the velocity $ x_i = \int u_i (t) dt $, in long quasi-ballistic periods with typical length of the order of the integral scale ($ L \approx \langle K \rangle^{3/2} / \langle \varepsilon \rangle $) and short term disruptions during which the trajectory rolls up on itself.
	 
\begin{figure*}[h]
 	\centering
	\includegraphics[width=0.49\textwidth,height=0.49\textheight,keepaspectratio, clip]{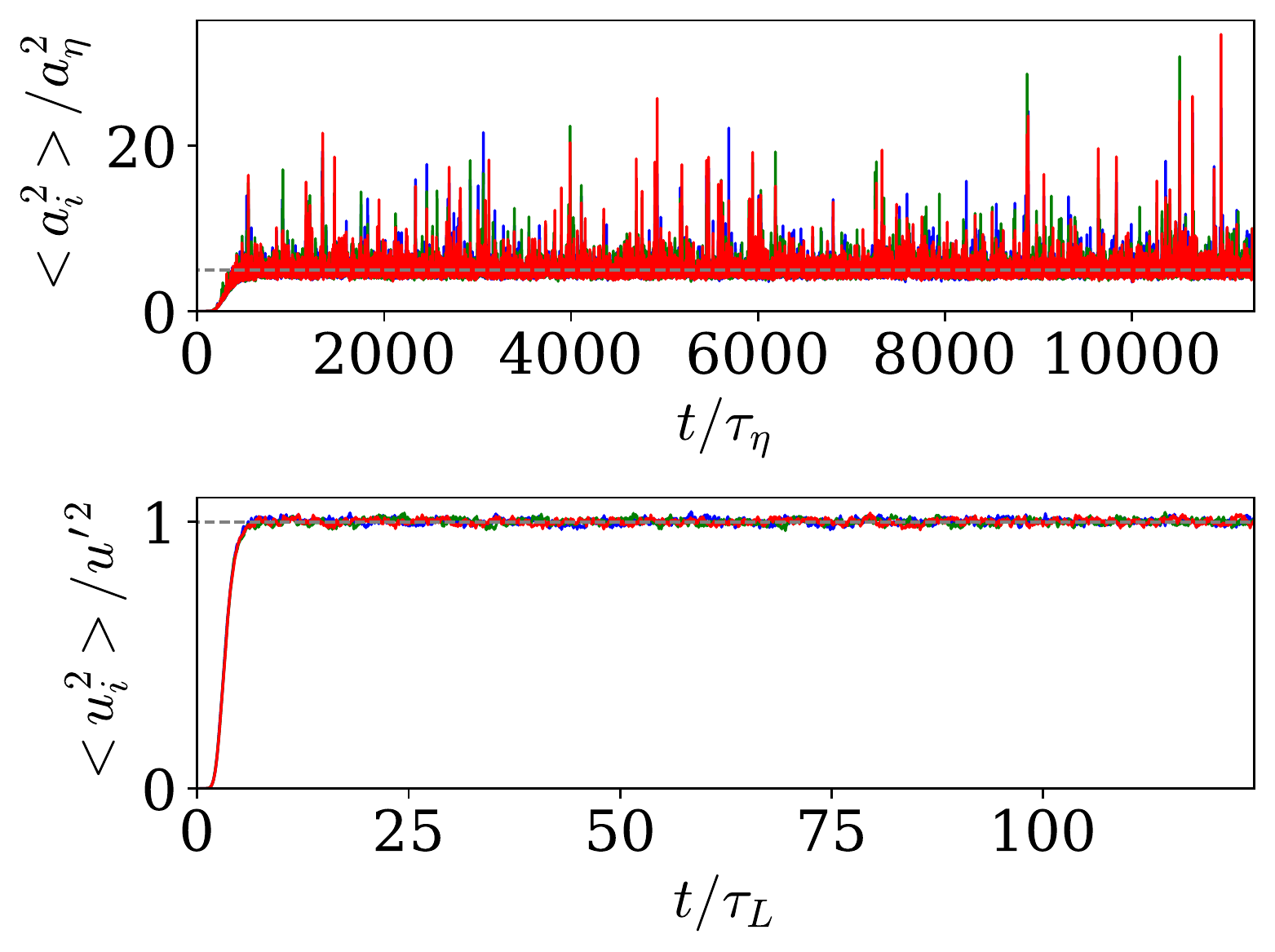}
 	\caption{Evolution of the acceleration and velocity variance with time for the stochastic model at $Re_\lambda=1100$ starting from an initial condition for the acceleration and velocity with very small magnitude. 
	Comparison in dashed gray line with the expected values: \eqref{eq:a_var_RZ} for the acceleration and the prescribe value of $u'=\sqrt{2\langle K \rangle /3}$ for the velocity. }
 	\label{fig:var_vs_time} 
 	\end{figure*}

	We have simulated the stochastic model for 15 different Reynolds numbers between $ Re_\lambda = 70 $ and $ 9000$.
	In each case, we have computed 26,000 realizations.
	The simulations are carried out over a period of $ 120 \tau_L $, over which we exclude an initial transitional regime of $ 20 \tau_L $ for the calculation of the statistics.
	In all cases, the initial value of the components of acceleration and velocity are sampled from the normal distribution having a standard deviation of $ 10^{-9} a_\eta $ for the acceleration and $ 10^{-9 } \sqrt {2 \langle K \rangle / 3} $ for the velocity.
	We can indeed notice from \eqref{eq:M2} and \eqref{eq:Dij_6} that, if the acceleration is exactly zero,  the stochastic model predicts that the acceleration would remain so.
	However, it should be noted that this event has a zero probability, and that for arbitrarily small, but non-zero, accelerations, the model presents an evolution towards a non-trivial stationary state.
	This is illustrated in Fig. \ref{fig:var_vs_time}, which presents the temporal evolution of the variance of the velocity and of the acceleration for $ Re_ \lambda = 1100 $, calculated from all the realizations. 
		
\begin{figure*}[h]
 	\centering
 	\includegraphics[width=0.49\textwidth,height=0.49\textheight,keepaspectratio, clip]{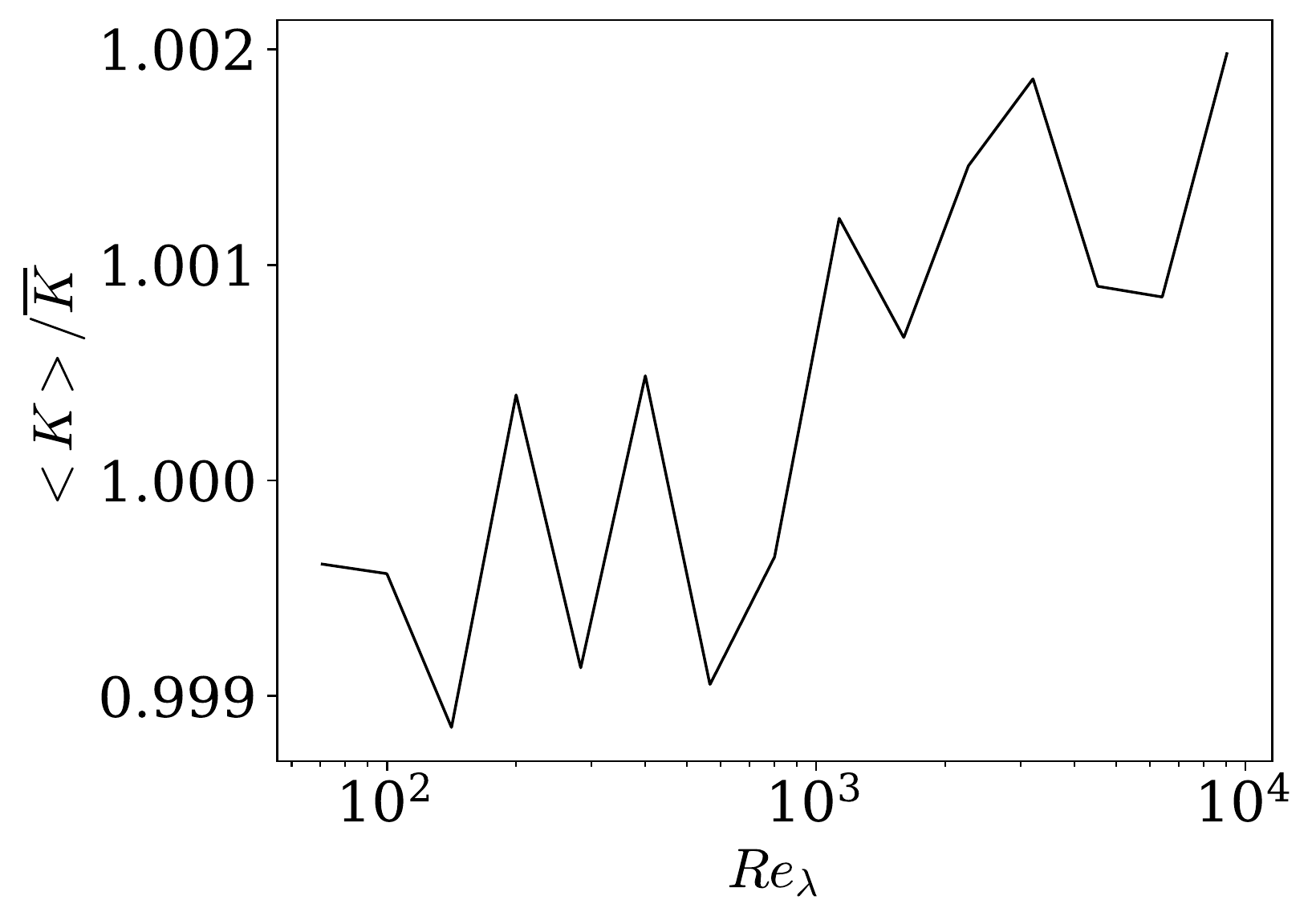}
 	\caption{Evolution with the Reynolds number, in the stationary regime, of the kinetic energy obtained by the stochastic model normalized by the prescribed kinetic energy $\langle K \rangle$.
	}
 	\label{fig:TKE_vs_Re} 
 	\end{figure*}

	Figure \ref{fig:TKE_vs_Re} shows the evolution with the Reynolds number of the mean kinetic energy in the stationary state.
	In this figure, we see that the average kinetic energy is equal, within the statistical convergence, to the value prescribed to the model.
	We note that the value of the average kinetic energy is directly related to the value of the parameter $c_u$ in \eqref{eq:M2} as mentioned above.
	
\begin{figure*}[h]
 	\centering
 	\includegraphics[width=0.49\textwidth,height=0.49\textheight,keepaspectratio, clip]{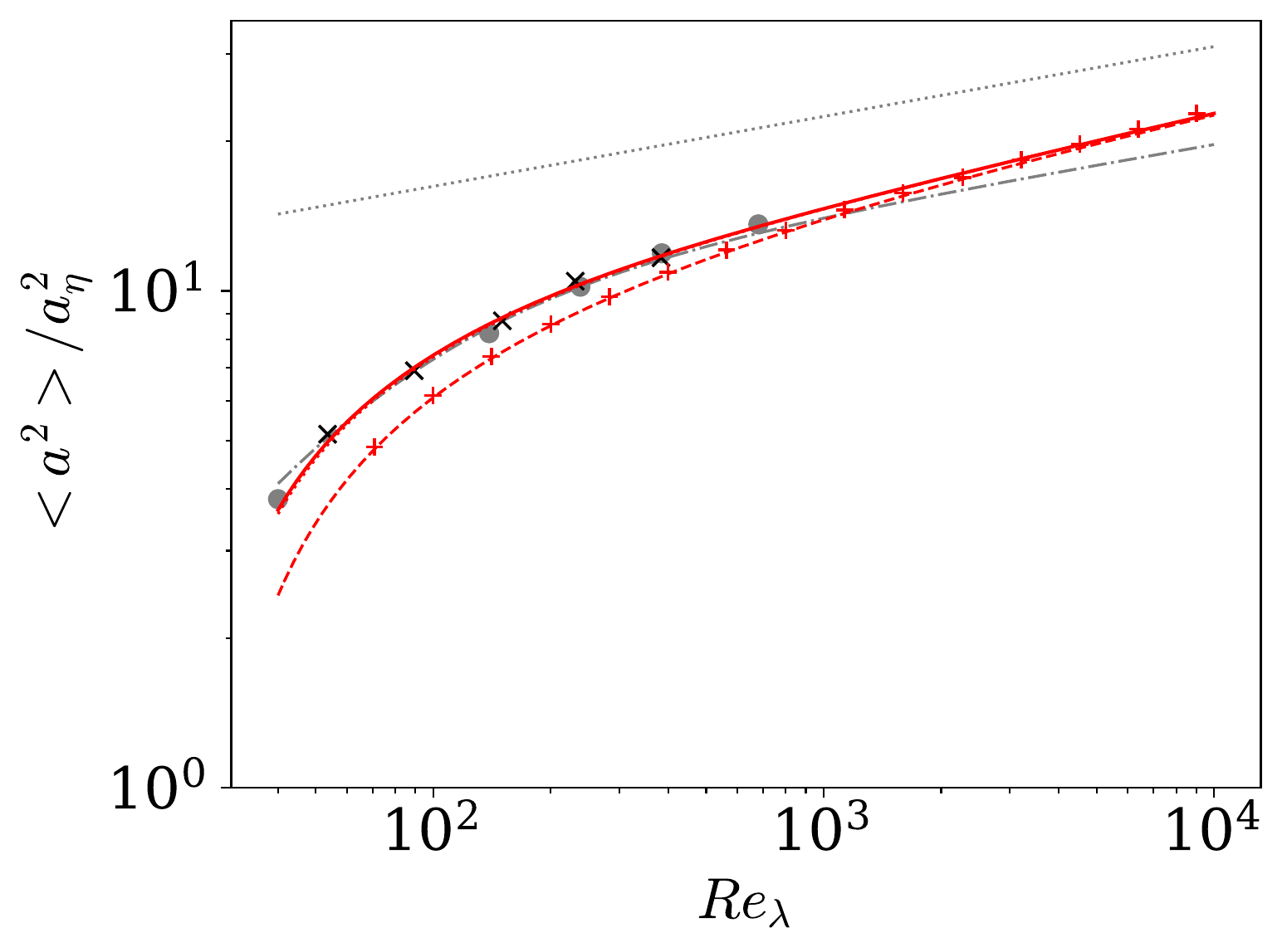}
 	\caption{Evolution with the Reynolds number, in the stationary regime, of the kinetic energy normalized by the prescribed kinetic energy $\langle K \rangle$ (left) and of the acceleration variance normalized by the Kolmogorov acceleration (right). Data from the stochastic  model (red plus) and  
	comparison with our DNS data (black crosses) and the DNS data from \cite{Yeung:2006} (gray dot-dash line), with the relation $3\times1.9 Re_\lambda^{0.135}(1+85Re_\lambda^{-1.35})$ from \cite{Sawford:2003} (gray dash dot line), with the numerical integration of $ \langle a^2 \rangle = \int \langle a^2| \varepsilon \rangle P(\varepsilon) d\varepsilon $ with $ \langle a^2| \varepsilon \rangle$ given by \eqref{eq:avar_cond_dissip_final} and $P(\varepsilon) $ log-normal (continuous red line), with its approximation of \eqref{eq:avar_1storder} (red dotted lines), with the large-Reynolds number limit relation \eqref{eq:a_var_RZ} (red dashed line) and with the asymptotic power law \eqref{eq:a_var_RZ_asymptotic} (gray dotted lines).
	}
 	\label{fig:avar_vs_Re} 
 	\end{figure*}

	Regarding the variance of the acceleration, we expect, by construction of the stochastic model, that the predicted value follows the log-normal relation \eqref{eq:a_var_RZ}.
	We observe in Fig. \ref{fig:avar_vs_Re}, that it is indeed the case, with only slight deviations for the largest Reynolds numbers which are attributed to numerical errors.
	We recall that the underestimation of the acceleration variance at small Reynolds numbers compared to the DNS or \eqref{eq:avar_1storder} stems from the fact that the model is based on the relation \eqref{eq:a2_cond_eps_k_asymptote} in which the effect of low dissipative and large-scale  structures are neglected (see the discussion in section \ref{sec:cond_dissip}). 
	This simplification enables to obtain the analytical formulation of the model proposed here.

\begin{figure*}[h]
 	\centering
 	\includegraphics[width=0.49\textwidth,keepaspectratio, clip]{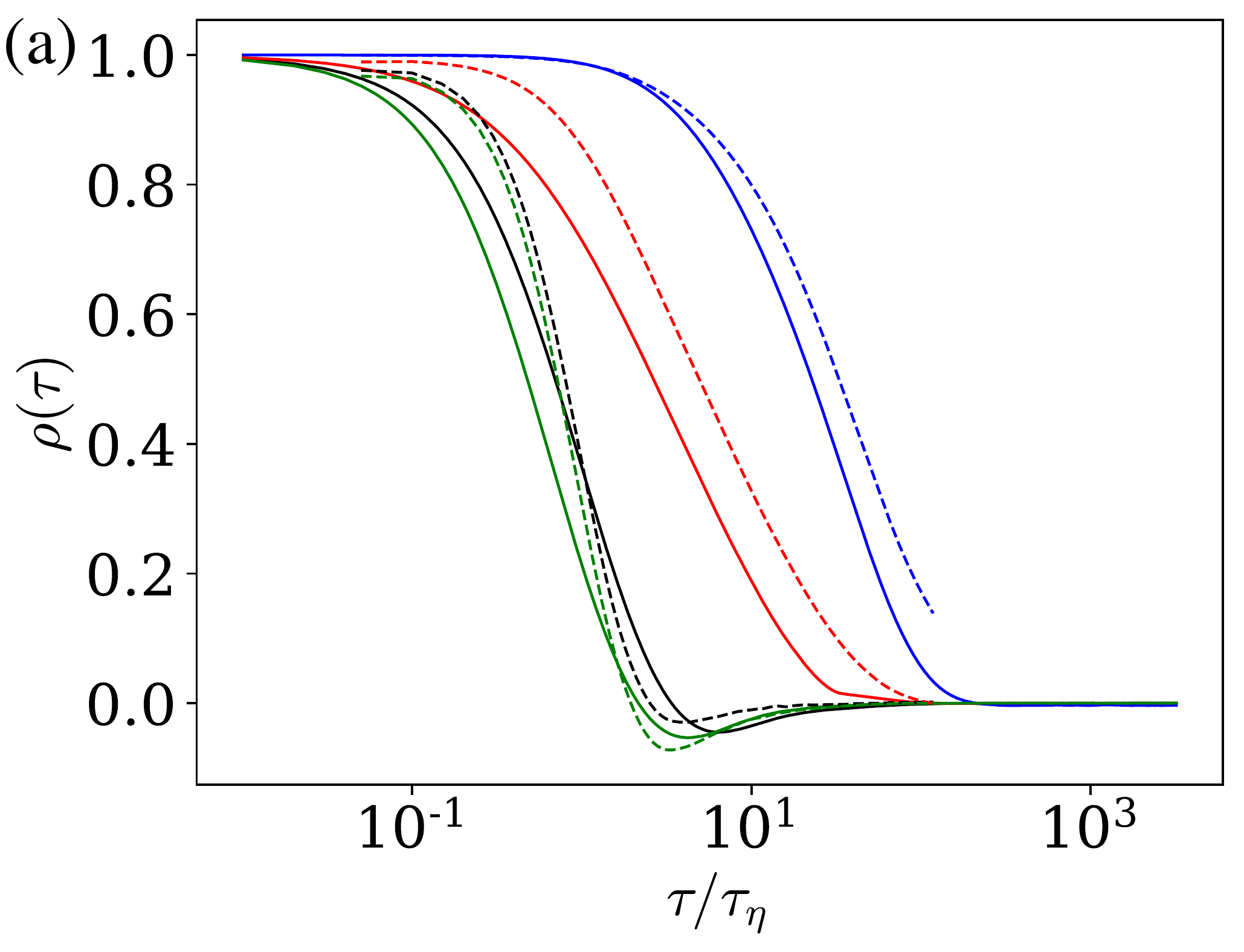}
 	\includegraphics[width=0.50\textwidth,keepaspectratio, clip]{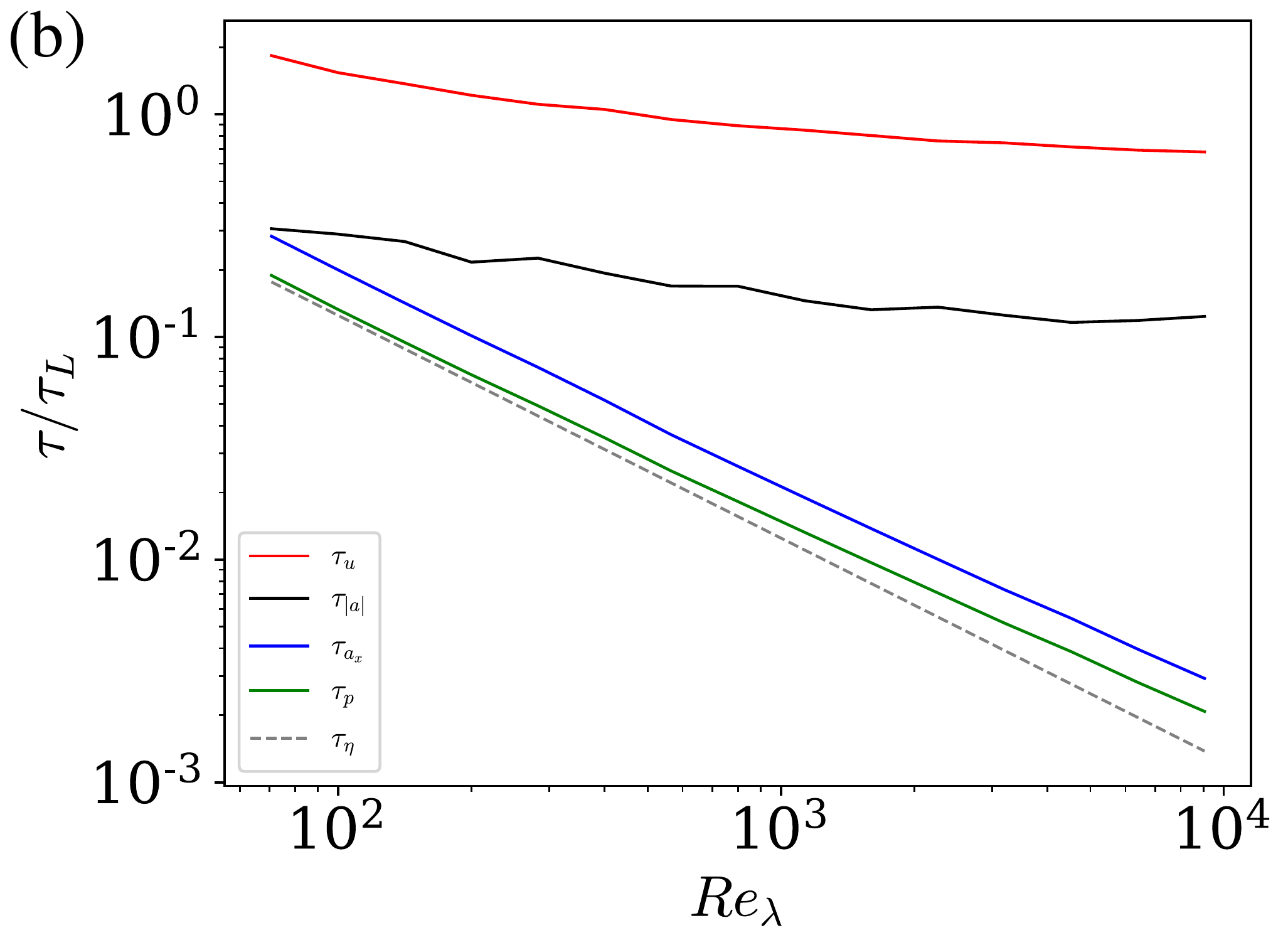}
	
	\includegraphics[width=0.49\textwidth,keepaspectratio, clip]{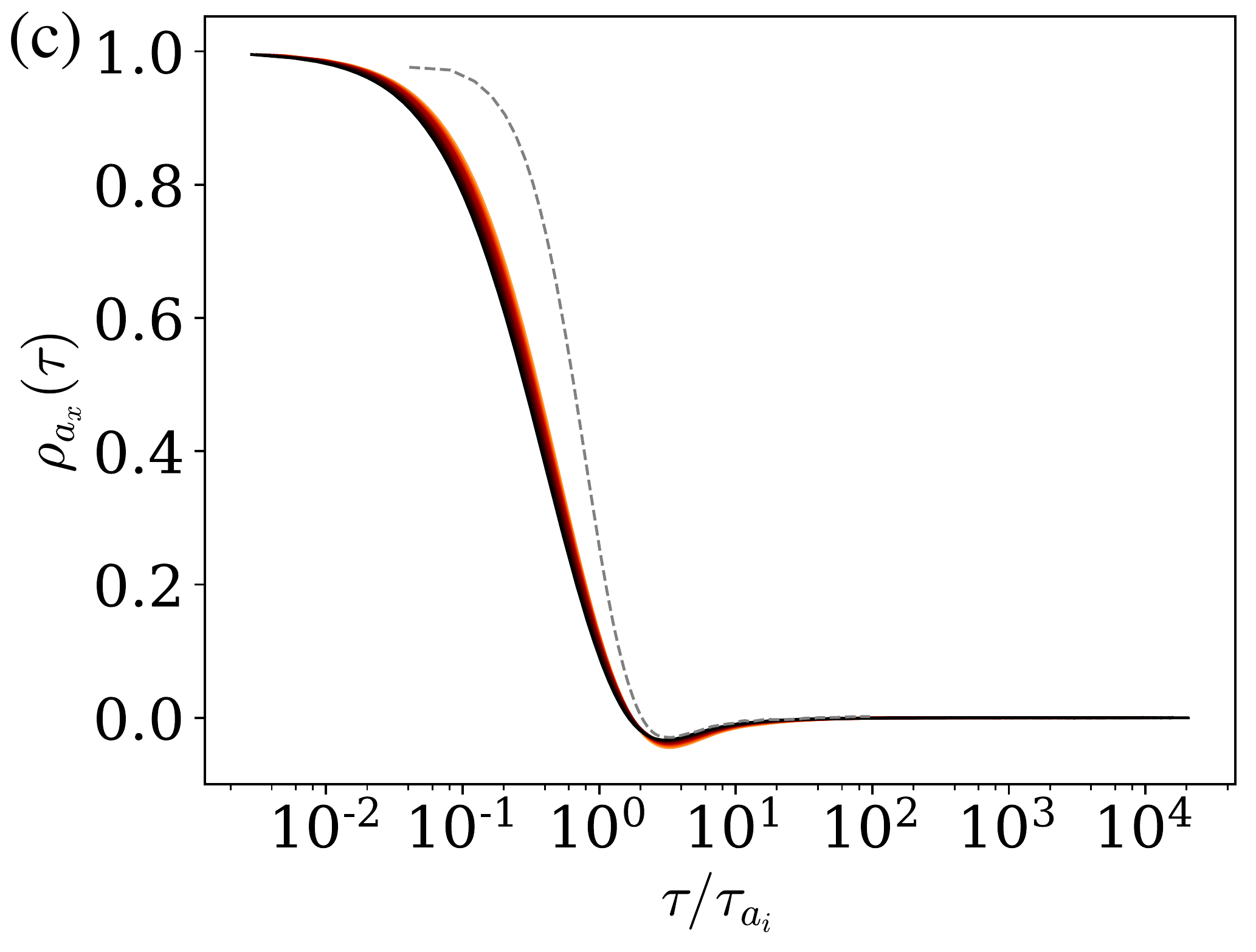}
	\includegraphics[width=0.49\textwidth,keepaspectratio, clip]{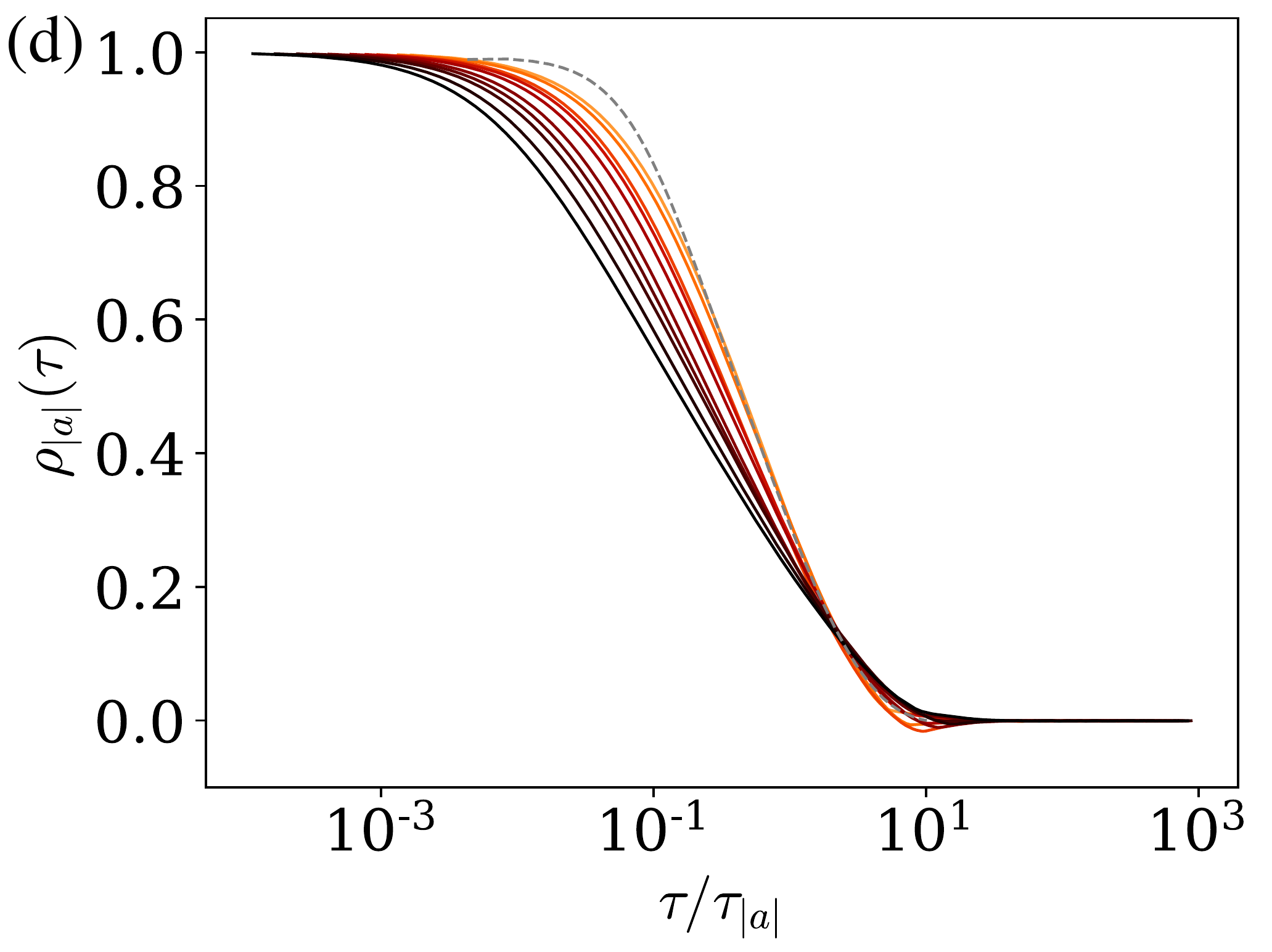}
		
	\includegraphics[width=0.49\textwidth,keepaspectratio, clip]{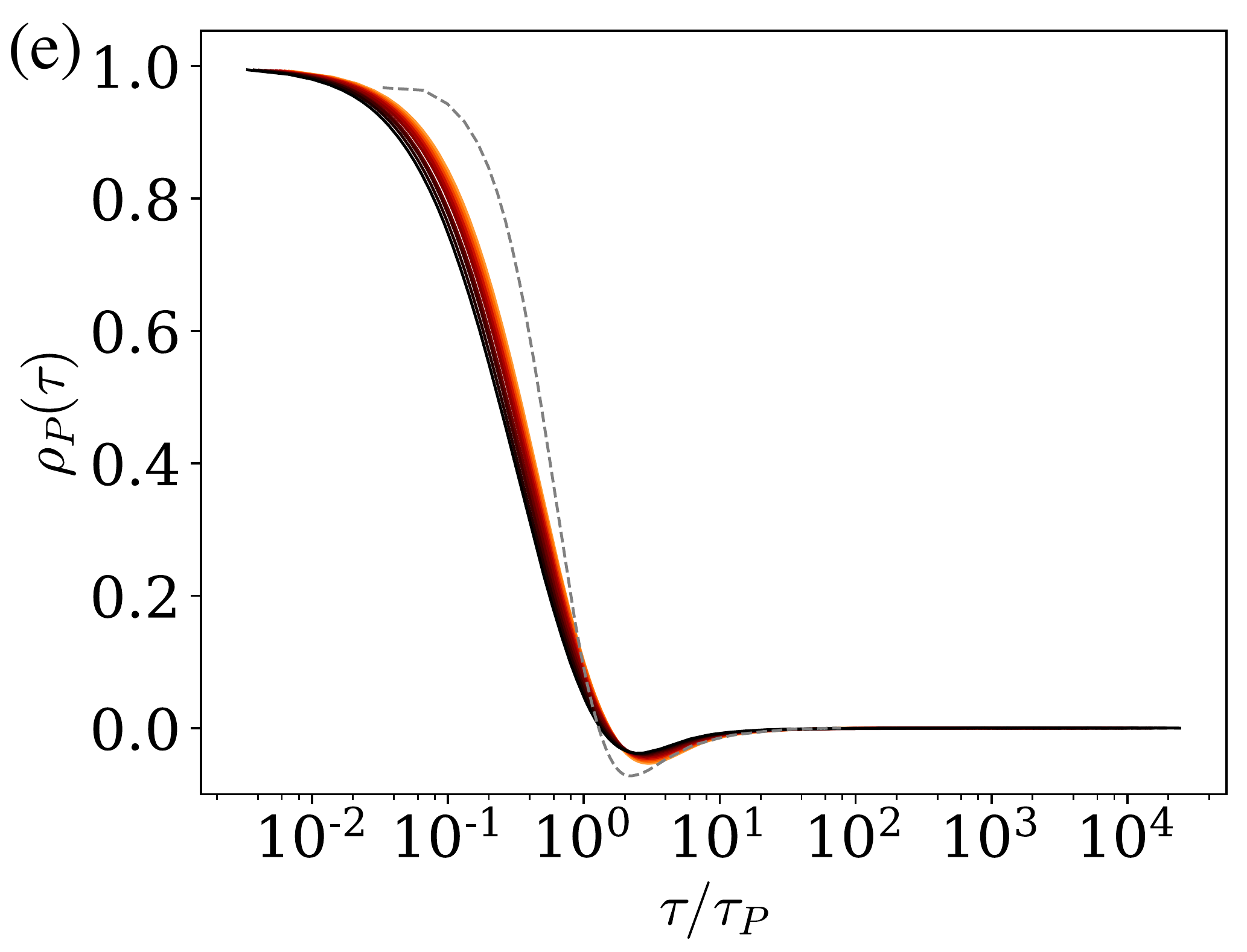}
	\includegraphics[width=0.50\textwidth,keepaspectratio, clip]{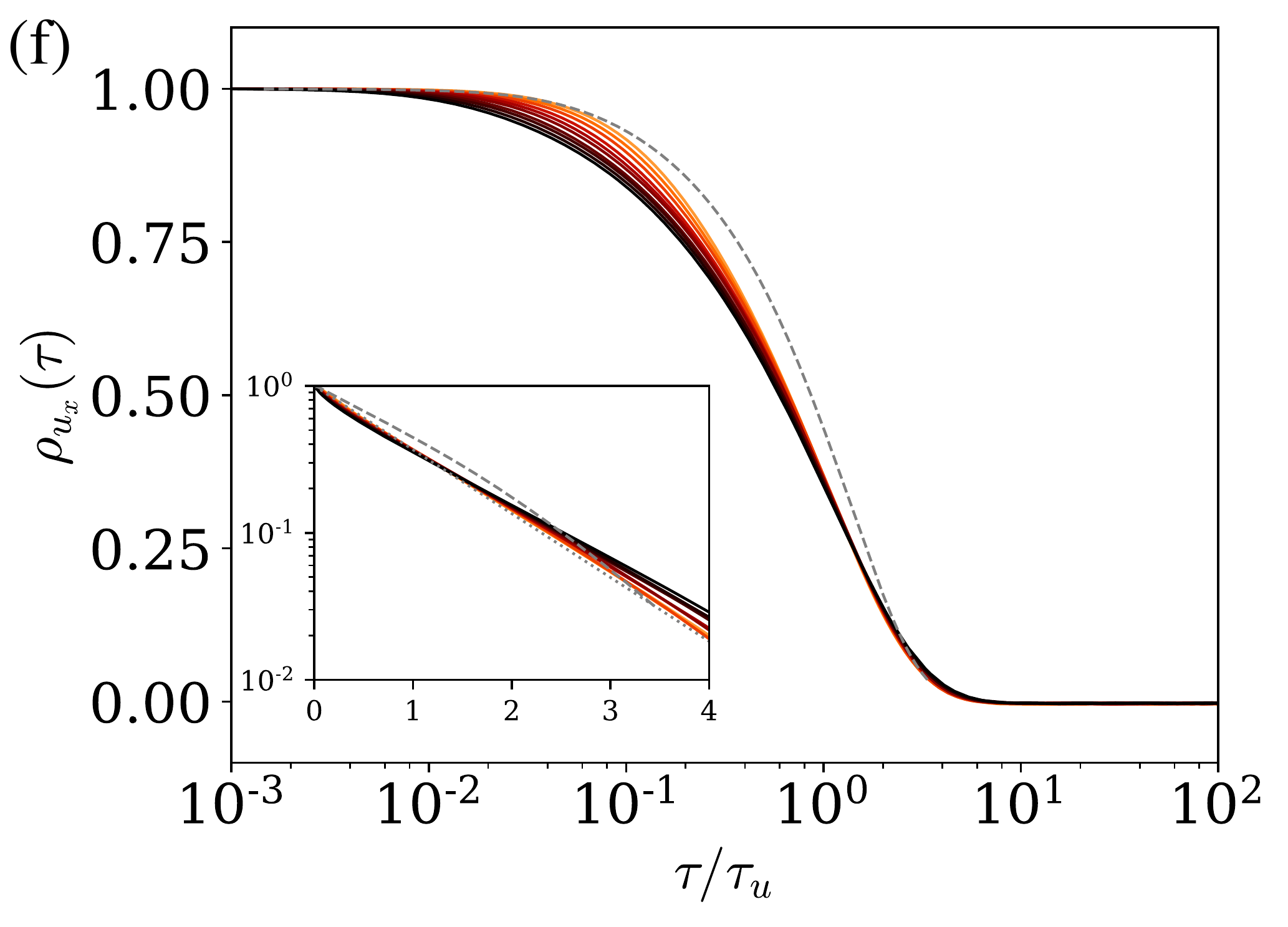}
	
 	\caption{(a) Evolution of the autocorrelation of $a_i$ (black), $(a^2)^{1/2}$ (red), $u_i$ (blue) and $P=a_iu_i$ (green) from the stochastic model for $Re_\lambda=400$ and comparison with the DNS data from \cite{Bec:2010} in dashed lines. 
	(b) Evolution of the integral time scale of $a_i$ (black), $(a^2)^{1/2}$ (red), $u_i$ (blue) and $P=a_iu_i$ (green) normalized by $\tau_L$ with the Reynolds number.
	(c,d,e,f) Evolution of the autocorrelation of $a_i$, $(a^2)^{1/2}$, $P=a_iu_i$ and $u_i$ respectively, for $Re_\lambda=400$, 567, 800, 1130, 1600, 2263, 3200, 4526, 6400 and 9051 from orange to black and comparison with the DNS data from \cite{Bec:2010} in dashed lines. In these plots the time lag is normalized by the corresponding integral time scale.
	For panel. f, inset: logarithmic scaling of the y-axis and comparison with $\exp(-\tau/\tau_u)$ in dotted line.
	}
 	\label{fig:model_autocorr} 
 	\end{figure*}

	Figure \ref{fig:model_autocorr} compares the autocorrelation of the components of the acceleration and of its norm calculated from the stochastic model for $ Re_\lambda = 400 $ with the calculations from the DNS of \cite{Lanotte:2011, Bec:2010}.
	It can be seen that the characteristic times of these two quantities are very different and that it is in good agreement with the DNS.
	It should be mentioned that the scale separation between the components and the norm results from the rotational part of the diffusion tensor.
	Indeed, no scale separation is found when $c_R$ is set to zero in  equation \eqref{eq:c_1_b} (corresponding then to a diagonal diffusion tensor).
	We see in \eqref{eq:Dij_5} that considering this rotational part, leads to the decomposition of the acceleration into its normal and tangential component. The former is associated with the intense rotation that rapidly changes the acceleration direction, whereas the second is associated with the variation of the kinetic energy of the particle.
	
	Figure \ref{fig:model_autocorr} also presents the evolution of the autocorrelation coefficient of the velocity components and of the power.
	It can be seen here also that the agreement with the DNS is relatively good.
	In Fig. \ref{fig:model_autocorr}, we also show the evolution of the characteristic correlation times for these four quantities with Reynolds numbers in the range $ Re_ \lambda = 70-9000 $ as predicted by the stochastic model.
	The characteristic correlation time for the velocity, the acceleration norm, the acceleration components and the power are 
	$ \tau_{u} = \int \rho_{u_i} (\tau) d \tau $, 
	$ \tau_{| a |} = \int \rho_{| a |} (\tau) d\tau $, 
	$ \tau_{a_i}= \int |\rho_{a_i}| (\tau) d\tau$ and
	 $ \tau_{P}= \int |\rho_{P}| (\tau) d\tau$.
	It can be seen in Fig. \ref{fig:model_autocorr} that the scales for the norm of the acceleration and for the velocity normalized by $ \tau_L $ remains quasi-constant with the Reynolds number and that the ratio between the correlation scale for the velocity and $ \tau_L $ is of order 1.
	Note that the characteristic time entering the model formulation is $ \tau_\varepsilon$ (the correlation time of the dissipation rate  following the path of a fluid particle). 
	For the calculation of the model, we simply set $\tau_\varepsilon=\tau_L$ arguing that the two quantities should be closed.
	It is therefore interesting to remark that the  integral time of the velocity is very close to the prescribe one $\tau_{u} \approx \tau_L$.
	Regarding the correlation scales for a component of acceleration and for the power normalized by $\tau_L$, they both present a variation close to $ 1 / Re_\lambda $, as expected.

	We also show in figure \ref{fig:model_autocorr} autocorrelation coefficient of $a_i$, $(a^2)^{1/2}$, $P=a_iu_i$ and $u_i$ for Reynolds numbers in the range  $ Re_ \lambda = 70-9000 $ obtained from the model.
	It is seen that, when the time shift is normalized by the corresponding integral time scale, the correlation coefficients of the power and of the acceleration component remains nearly unchanged with the Reynolds number.
	We observe as well that the shape of the autocorrelation obtained from DNS is well reproduced, although the decay predicted by the model is too fast at very short time lag. 
	This is attributed to the fact that the dissipative region is only taken into account in the model via the cutoff $\tau_c=\tau_\eta$ of the kernel $\hat{\Gamma}$.
	We observe that the correlation for the acceleration norm presents a logarithmic decrease, reflecting the absence of characteristic time for its evolution. 
	As expected, the correlation norm exhibits a lower slope as the Reynolds number increases.
	This is directly attributed to the use of the non-Markovian process of \cite{Chevillard:2017b} for the dissipation rate,  which proposes a logarithmic evolution of the autocorrelation in agreement with the underlying model of the turbulent energy cascade as discussed in appendix \ref{sec:model_dissip}.
	
	The shape of the velocity correlation from the model is overall close to the DNS.
	At small $\tau$, it presents some dependence on the Reynolds number, while at large time shift (i.e. $\tau$ of the order of $\tau_L$) the correlation decreases exponentially, as it can be seen in the inset of figure \ref{fig:model_autocorr}f, in agreement with DNS and experiments.
	The exponential relaxation results from the terms $ P + K / \tau_\varepsilon = dK / dt + K / \tau_\varepsilon $ appearing in the drift part of the stochastic model \eqref{eq:M}.
	It is interesting to remark that the presence of this term in the model is a direct consequence of the exponential dependence of the conditional acceleration variance on the kinetic energy \eqref{eq:a2_cond_eps_k_asymptote}.
	This term leads to the Reynolds number dependence on the velocity correlation observed at small $\tau$, which is connected to the logarithmic decorrelation of the acceleration norm, to vanish at large $\tau$ at which it relaxes exponentially.  
	This suggests therefore anomalous scaling at intermediate time lag.

 	\begin{figure*}[h]
 	\centering
	\includegraphics[width=0.49\textwidth,height=0.4\textheight,keepaspectratio, clip]{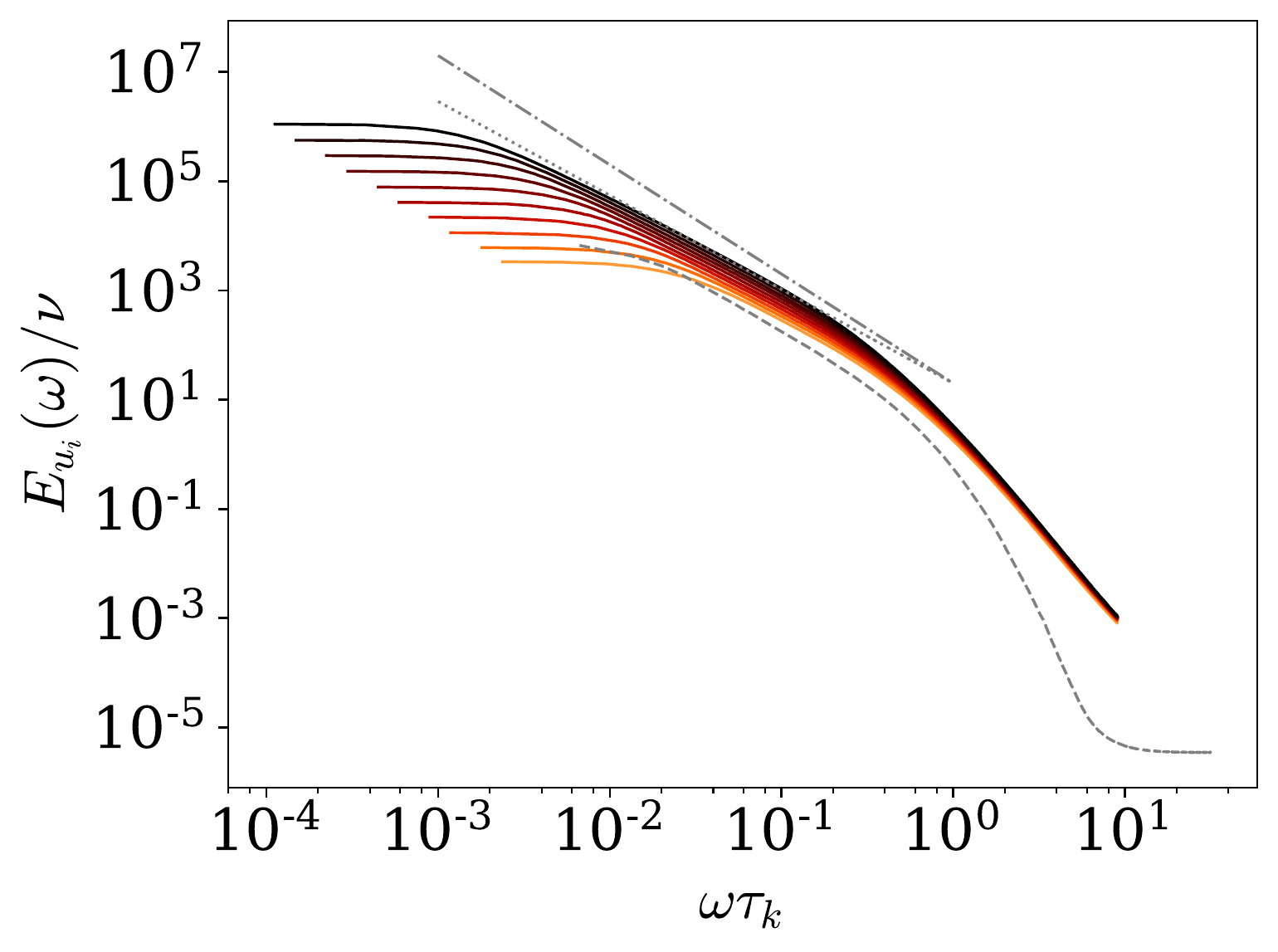}
 	\caption{Velocity spectra from the stochastic model for $Re_\lambda=400$ to $Re_\lambda=9000$ from orange to black and comparison with the DNS data from \cite{Bec:2010} at $Re_\lambda=400$ (gray dashed line), with the Hinze spectra $\omega^{-2}$ (gray dot-dashed line), and with the power law with anomalous exponent $\omega^{-2 + 9/64 }$  (gray dotted lines).}
 	\label{fig:model_spectra} 
 	\end{figure*}

	We show in Fig. \ref{fig:model_spectra} the velocity spectrum for $ Re_\lambda $  between  400  and 9000, which we compare with the DNS of \cite{Bec:2010} for $ Re_\lambda = 400 $.
	We see a good agreement between the DNS and the stochastic model.
	For higher Reynolds numbers, we clearly see that a power law behavior develops at intermediate scales.
	We see that the slope of the power law deviates from the Hinze spectra \cite{Tennekes:1972} predicted by dimensional arguments similar to those presented by Kolmogorov, with spectra less stiff than $ \omega^{-2} $.
	This shows that the proposed stochastic model leads to an anomalous scaling that reflects the persistent influence of the Reynolds number in the inertial-scales.
	We further notice that the slope that develops at intermediate scales are close to $ -2 + 0.14$,
	where $0.14$ is the exponent of the asymptotic power law of the acceleration variance with the Reynolds number determined in \eqref{eq:a_var_RZ_asymptotic} (see also Fig. \ref{fig:avar_vs_Re}). 
	We see here a confirmation of the relation between the acceleration scaling and the anomalous scaling of the velocity spectra proposed by \cite{Falkovich:2012}.

 	\begin{figure*}[h]
 	\centering	
	\includegraphics[width=0.49\textwidth,height=0.4\textheight,keepaspectratio, clip]{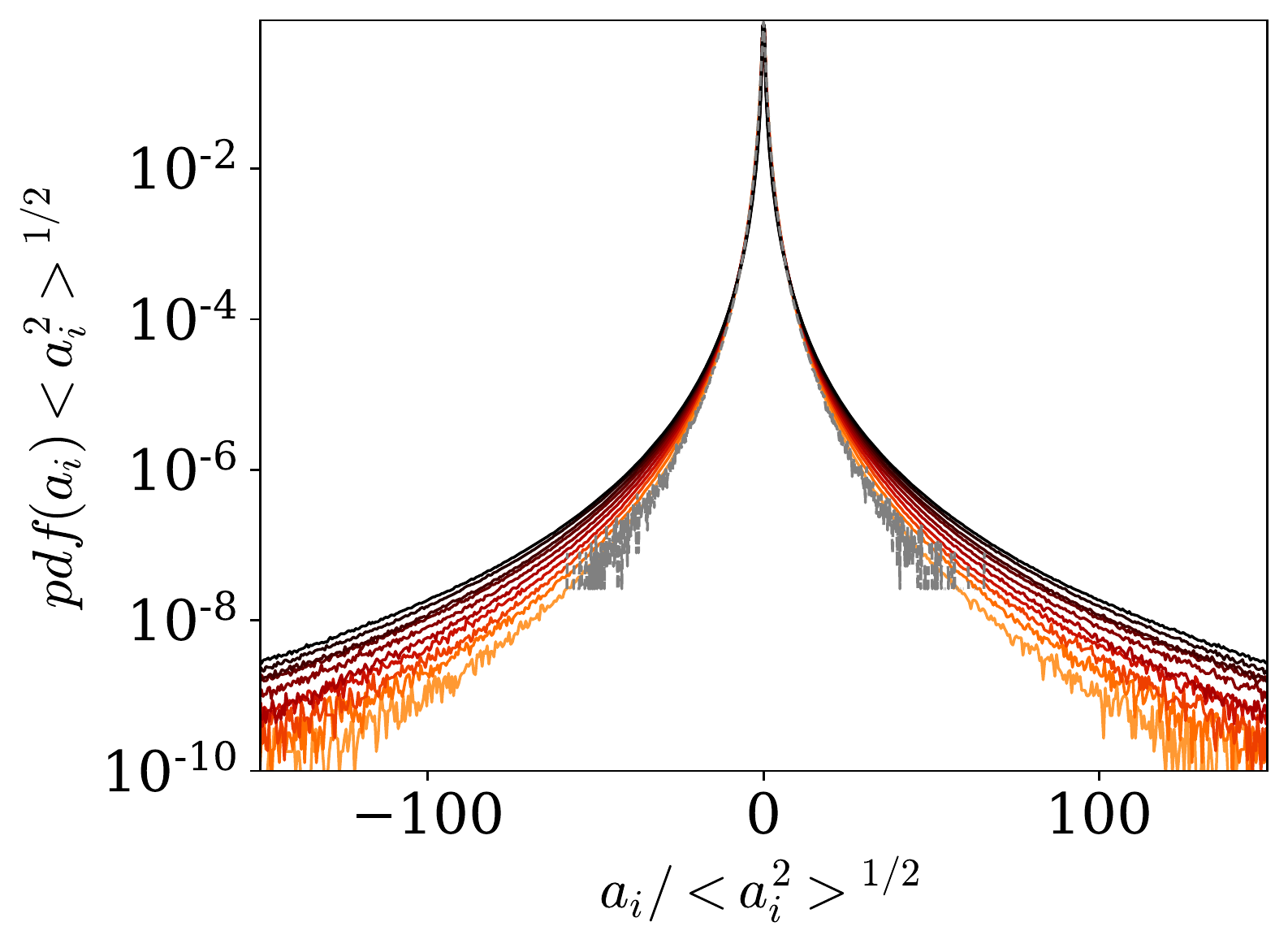}
	\includegraphics[width=0.49\textwidth,height=0.4\textheight,keepaspectratio, clip]{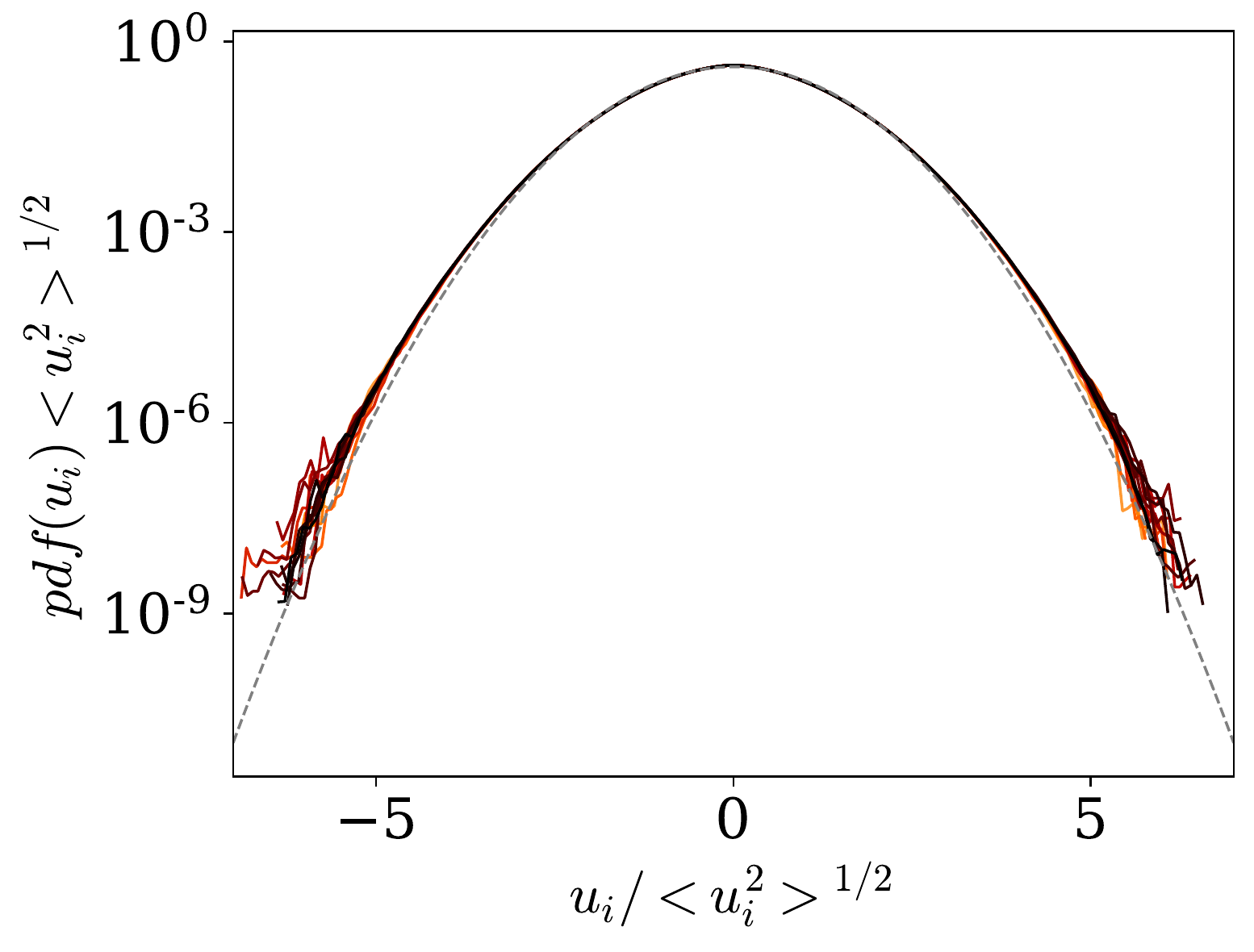}
	\includegraphics[width=0.49\textwidth,height=0.4\textheight,keepaspectratio, clip]{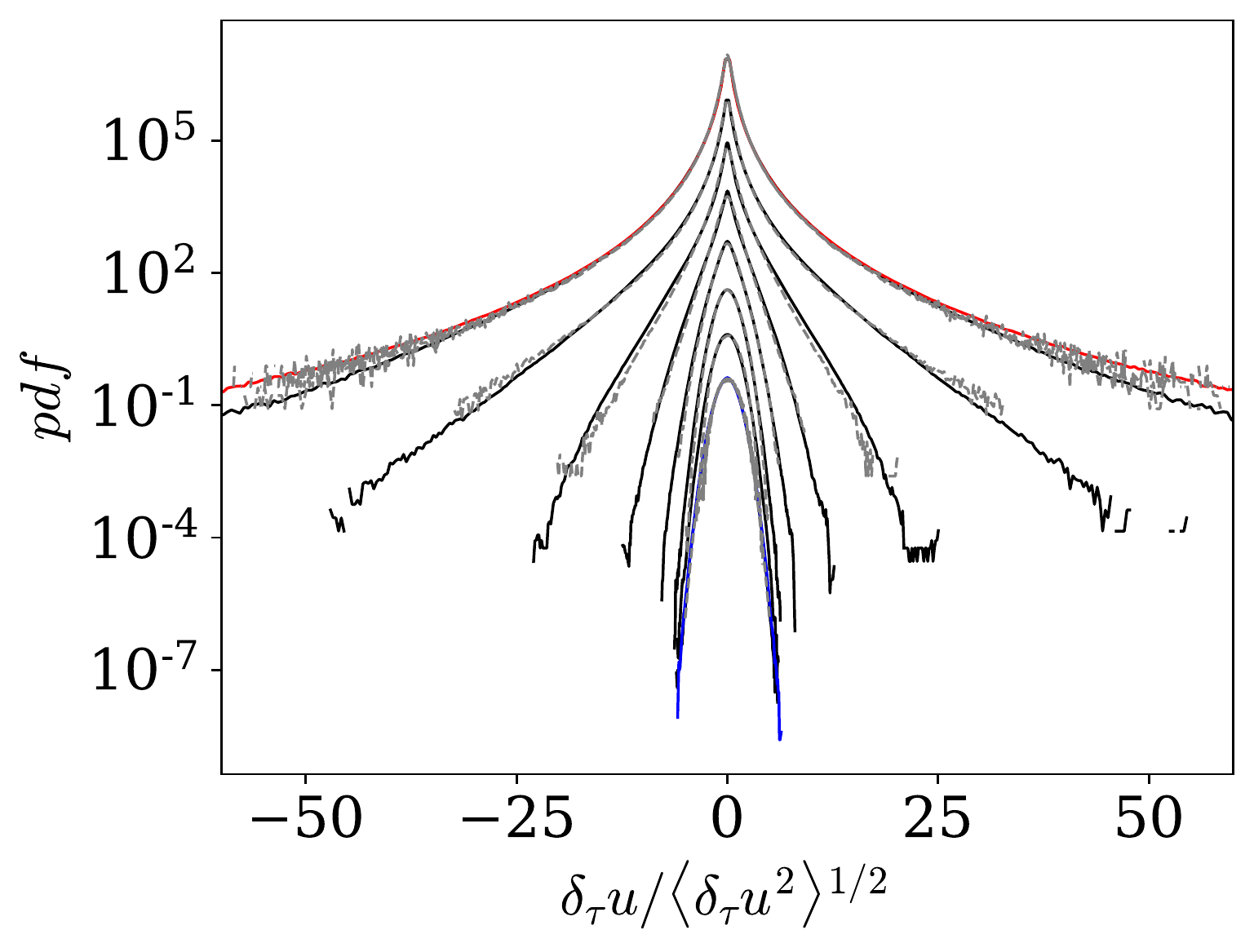}
	\includegraphics[width=0.49\textwidth,height=0.4\textheight,keepaspectratio, clip]{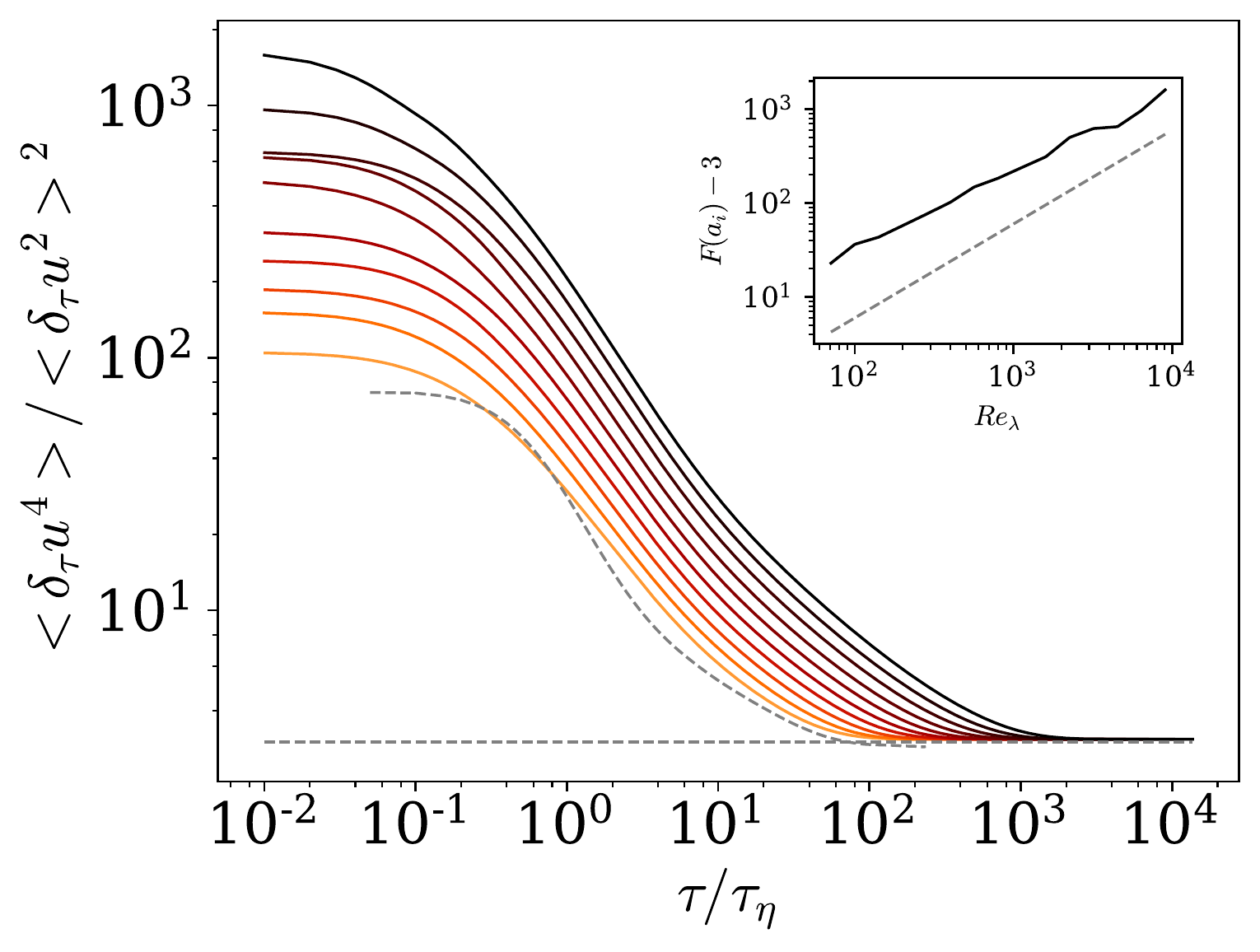}
	
 	\caption{PDF of $a_i $(top left) and comparison with the DNS data from \cite{Bec:2010}, and 
	 PDF of $u_i$ (top right) and comparison with the normal distribution, for $Re_\lambda=400$ to 9000 from orange to black.
	(Bottom left) PDF of the velocity increments for various times shift: $\tau/\tau_\eta=610$, 200, 70, 22, 7.3, 2.4, 0.8 and 0.25 each time lag is shifted upward by one decade for $Re_\lambda=400$ comparison with the PDF of the acceleration from the model (red) and from the DNS of \cite{Bec:2010} (gray) and the PDF velocity (blue). (Bottom right) Evolution of the Flatness of the velocity increments versus the time shift for $Re_\lambda=70$ to 9000 from orange to black and evolution of the flatness of the acceleration with the Reynolds number and comparison with the linear law in the inset. }
 	\label{fig:model_pdf} 
 	\end{figure*}

	We present in Fig. \ref{fig:model_pdf} the PDFs of the velocity and of the acceleration for $ Re_\lambda = 400 \sim 9000 $, as well as the comparison with the DNS of \cite{Bec:2010}.
	First, we find that the velocity distribution is very close to a Gaussian distribution for all Reynolds numbers, while the acceleration presents a much more stretched distribution. 
	For $ Re_\lambda = 400 $ the acceleration PDF is in very good agreement with the DNS, and, the model predicts an increase of the stretching of the tails with increasing the Reynolds number.
	We also show in this figure the PDF of the velocity increments for different time shifts $\delta_\tau u_i=u_i(t+\tau)-u_i(t)$ at $Re_\lambda=400$. 
	We observe that the distribution gradually returns to a Gaussian distribution as the time shift increases, and that at each time shift the agreement with the DNS of \cite{Bec:2010} is very good. 
	This is confirmed by the presentation of the flatness of the velocity increments for $ Re_\lambda = 400 \sim 9000 $, which reflects the strongly non-Gaussian behavior on small-scales which decreases to 3 for the larger-scales.
	Here also we notice a good agreement with the DNS of \cite{Bec:2010} for $Re_\lambda=400$.
	We also show in the inset, a quasi-linear increase of the flatness of the acceleration with the Reynolds number.

 	\begin{figure*}[h]
 	\centering
	\includegraphics[width=0.49\textwidth,height=0.4\textheight,keepaspectratio, clip]{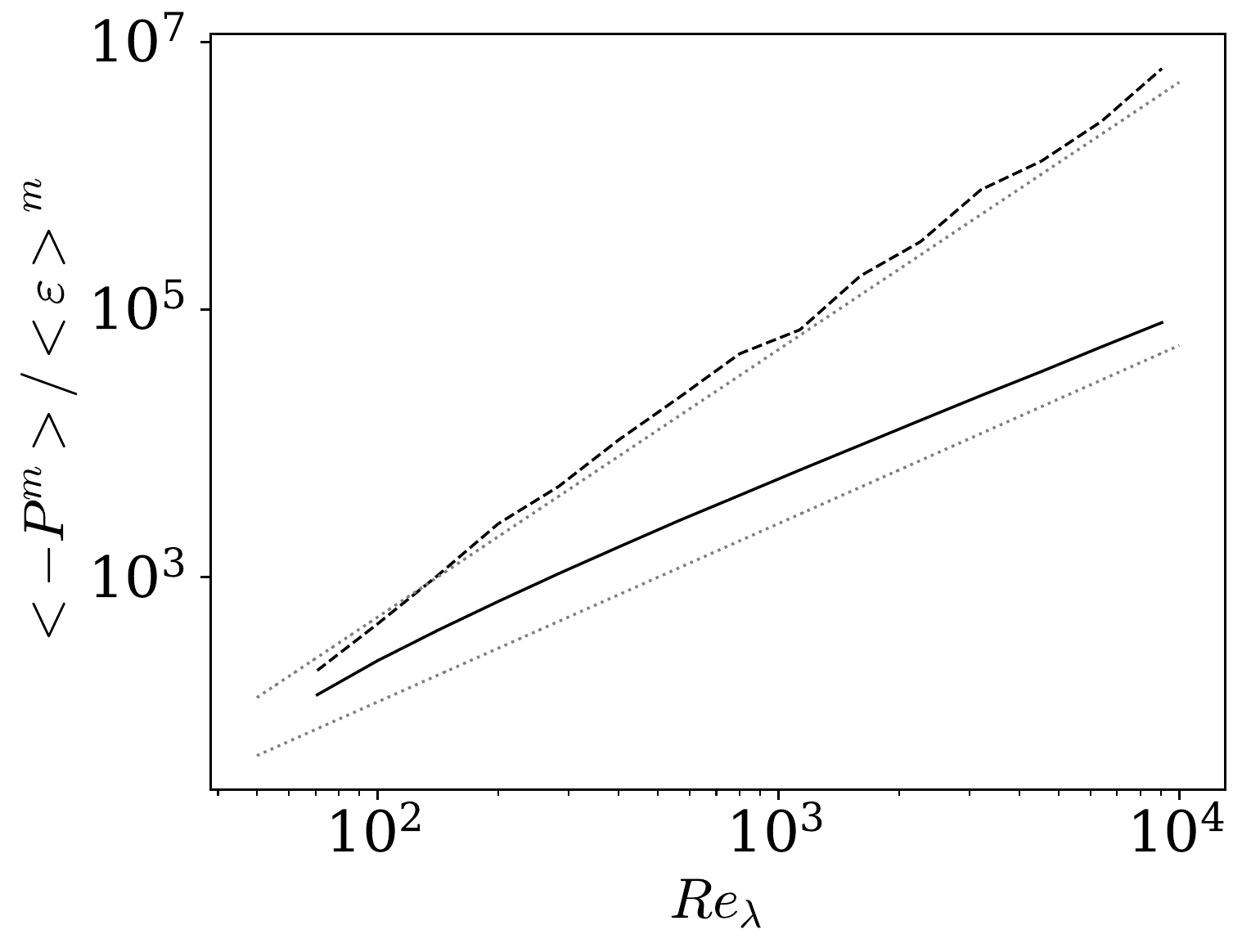}
 	\caption{Evolution of $\langle P^2\rangle / \langle \varepsilon \rangle^2 $ (continuous black line) and $-\langle P^3\rangle / \langle \varepsilon \rangle^3 $ (dashed black line) with the Reynolds number, and comparison with the power laws $Re_\lambda^{4/3}$ and $Re_\lambda^2$.
	}
 	\label{fig:model_power_mmt} 
 	\end{figure*}

	Finally, in Fig. \ref{fig:model_power_mmt} we show the second and third moments of the power $P=a_i u_i$. 
	It is observed that the increases of both moments with the Reynolds number are in close agreement with the power law supported by the DNS results of \cite{Xu:2014}, $\langle P^2\rangle / \langle \varepsilon \rangle^2 \sim Re_\lambda^{4/3}$ and $-\langle P^3\rangle / \langle \varepsilon \rangle^3 \sim Re_\lambda^{2}$ . 
	Clearly, the third order moment is negative, meaning that the time irreversibility of the dynamics of a fluid particle in a turbulent flow is correctly reproduced by the proposed stochastic model. 	
	The skewness of the power, $S=\langle P^3\rangle/\langle P^2\rangle^{3/2}$, seems to converge to -0.5 as the Reynolds number increases, as reported in \cite{Xu:2014}.

	\section{Discussion and final remarks}

	In this paper, we have analyzed the behavior of the acceleration statistics conditioned on both local dissipation rate and local kinetic energy, which to our knowledge have not been considered before.
	We have reported that the doubly-conditional variance is proportional to the acceleration variance conditional on the dissipation rate solely, with the proportionality factor depending exponentially on the kinetic energy: $\langle a^2 | \varepsilon, K\rangle = A \exp(\alpha K/ \langle K \rangle )\, \langle a^2 | \varepsilon \rangle$. 
	For large enough Reynolds number we show that $ A= \left(1- 2 \alpha/3 \right)^{3/2}$ and we proposed that the $\alpha$ coefficient is independent of the Reynolds number and its value $\alpha = 1/3$ was obtained from the DNS. 
	
This expression shows a direct effect of the kinetic energy, a large-scale quantity, on the Lagrangian acceleration. 
Furthermore since the argument of the exponential depends on $K/\langle K \rangle$, not on a local Reynolds number, it suggests a kinematic effect for the acceleration which may be due to the non-locality of the pressure. 
More specifically, these effects can be due to the interaction of vorticity and strain \cite{Douady:1991}.
In case of persistent large-scale strain,  intense vorticity tube would be generated and  align with the principal direction of the strain \cite{Moffat:1967}. It was shown that such vortical structure can produce significant acceleration in the direction of the vorticity \cite{Barge:2020,Lee:2004}.
Anyway, although the proper physical mechanism leading to the exponential dependence of the acceleration on the kinetic energy deserve further studies it is an additional effect to the influence of the large-scales on the acceleration through the intermittency of the dissipation rate.
	To study this later effect, we subsequently have proposed to account for the Reynolds number dependence of the acceleration variance conditional on the dissipation rate within the intermediate asymptotic framework \cite{Barenblatt:2004} leading to: $\langle a^2 | \varepsilon \rangle = a_\eta^2 B (\varepsilon/\langle \varepsilon \rangle)^{3/2+\beta} $ 
	for $\varepsilon \gg \langle \varepsilon\rangle$ with $B$ and $\beta$ depending logarithmically on the Reynolds number as the signature of the intermittency and the persistence of viscous effects.
	Further, we advance an expression for the conditional acceleration variance valid for the whole range of fluctuations of $\varepsilon$ by accounting for the dominant effect of the large-scale structures in low dissipative regions (see equation \eqref{eq:avar_cond_dissip_final}).
	From this finding we determine the evolution of the unconditional acceleration variance with the Reynolds number (equation \eqref{eq:avar_1storder}) and show that it is in good agreement with DNS, which gives another empirical validation of the incomplete similarities assumption used to obtain these results.
	
	Eventually, for large Reynolds numbers, we propose to express the doubly-conditional variance as $\langle a^2 | \varepsilon, K\rangle = C a^2_\eta \exp(\alpha K/ \langle K \rangle + \gamma \ln \varepsilon \langle \varepsilon \rangle)$, $\gamma=3/2+\beta$, which can be viewed as the results of a multiplicative process for the acceleration. 
	Such process can be interpreted as a momentum fluctuation cascade that includes kinematic effects by eddies all along the turbulence spectrum.

	Based on these results we propose a 3D stochastic model for the dynamics of a fluid particle that reproduce the essential features of the Lagrangian dynamics observed from DNS and experiments. 
	To obtain such model, (i) we have assumed, inline with the Kolmogorov universality hypothesis, that the dynamics can be described as a set of stochastic differential equation $d a_i = M_i dt + D_{ij} dW_j$ ; $d u_i=a_i dt$, with $M_i$ and $D_{ij}$ depending on the velocity and acceleration along with Reynolds number dependent parameters.
	(ii) We used the doubly-conditional acceleration variance obtained in this paper, to model the instantaneous relation of the dynamics between acceleration (or force), kinetic energy, and energy dissipation. This amounts to consider that the remaining degree of freedom can be discarded in procedure similar to an adiabatic elimination \cite{Gardiner:1985} as discussed by \cite{Castaing:1990}.
	(iii) We introduce a non-diagonal diffusion tensor along with a maximum winding hypothesis to ensure its physical realizability. 
	(iv) We consider that the dissipation rate along the trajectory is given by the non-Markovian log-normal process proposed recently by \cite{Chevillard:2017b}, giving logarithmic correlation consistently with the turbulent cascade picture.
	The model is closed by using the relation $DK/Dt=P=a_iu_i$. 
	For the model, it implies dependence of $K$ on $\varepsilon$ through  the dependence of $a^2$ on $\varepsilon$.
	This can be interpreted as feedback of the small scales on the large scales.
	On the other hand, the influence of the large-scales on the small-scales is accounted for in the model through the intermittent cascade model for the dissipation rate.
	With these 4 hypothesis, we obtain the model given by equations \eqref{eq:u_stoch}, \eqref{eq:a_stoch}, \eqref{eq:M2} and \eqref{eq:Dij_6} which reads:
	\begin{eqnarray}	
	d a_i &= & \Bigg[ \dfrac{\alpha}{2\langle K \rangle} \left(a_i \big( c_u P + \dfrac{K}{\tau_\varepsilon} \big) - (c_u-1) a^2 u_i \right)	- a_i \left( \ln \big( \dfrac{a^2}{a_\eta^2}\big) + \hat{\Gamma}_* \right) \dfrac{1}{2 \tau_\varepsilon} 	- \dfrac{\sigma_*^2}{\tau_c} \dfrac{a_T^2}{a^2}a_i \Bigg] dt \nonumber \\
	& & + \sqrt{\dfrac{\sigma_*^2}{\tau_c}} \left[ \sqrt{ a_T^2 } \delta_{ij} + \sqrt{ a_N^2} \epsilon_{ijk} b_k \right] dW_j 
		\label{eq:a_stoch_final}		
	\end{eqnarray}
	We show that the proposed model predicts Lagrangian dynamics presenting non-gausssianity, long-range correlations, anomalous scaling and time irreversibility.
	Moreover statistics obtained from the stochastic model are in good agreement with the DNS.

	In \eqref{eq:a_stoch_final} the term proportional to $\alpha$, which follows directly from the exponential dependence of the conditional acceleration on the kinetic energy, involves the coupling between velocity and acceleration and  
	leads to the exponential relaxation of the velocity correlation for large time lag along with one-time Gaussian distribution for the velocity.
	Introducing a rotational part in the diffusion tensor naturally leads to decomposition of the acceleration vector into its tangential part and its normal components. Since the normal part is associated with the curvature of the trajectory, the rotational part of the diffusion leads to the emergence of a time-scale separation between the correlation of the norm and the components of the acceleration.
	The term associated with the non-Markovianity of the dissipation along with the rotational part produce  irreversible dynamics, as seen by the skewness of the exchanged power
	and ensures a scale separation between velocity and acceleration. 
These three points can be easily checked, by taking $ \alpha = 0 $ or $ c_R = 0 $ in \eqref{eq:c_1_b} or by using for $ \Pi $ the Markovian log-normal dissipation model proposed by \cite{Pope:1990} rather than the non-Markovian one of \cite{Chevillard:2017b}.
			
	It is worth noting that from the conditional acceleration statistics obtained from DNS of the Navier-Stokes equation, it is possible to establish, in a fairly natural way, that is to say without using any other hypothesis than the cascade picture, a link between the refined Kolmogorov assumption and the dynamics of fluid particles. 
	It would be interesting to analyze further the stochastic equation to demonstrate the irreversibility of the dynamics, the emergence of anomalous scaling or to study the geometry of the particle trajectory e.g. its curvature and torsion, as well as to further test the conditional statistics between the acceleration and the velocity.
	Also interesting could be the improvement of the modeling of the high frequency part of the spectrum. Indeed the dissipative part of the spectrum is not well reproduced by the model of \cite{Chevillard:2017b} which intends to model the dissipation rate in the inertial range.
	
	In order to simplify the construction of the model, we have not taken into account the non-local effects of the largest structures of the flow, arguing that their effect vanish as the Reynolds number increases (term with $ a_0^2 $ in eq. \eqref{eq:avar_cond_dissip_final}).
	Based on the relation \eqref{eq:avar_cond_dissip_final} it is possible to account for the large-scale in the stochastic modeling.
	However, since this term is dependent on the Reynolds number, it is likely that it also depends on the flow configuration and boundary conditions.
	The proposed stochastic model could be further generalized to address shear flows \cite{Barge:2020} and improve Reynolds-averaged simulations \cite{Pope:1994,Innocenti:2020}.
	This model could be used among other things to improve the calculation of the dynamics of a dispersed phase with the large eddy simulation (LES) approach  \cite{Zhang:2019b,Gorokhovski:2018,Zhang:2021}.
	Finally, let us mention that an interesting extension could be the coupling of the proposed model with stochastic model for the evolution of the velocity gradients as proposed in \cite{Girimaji:1990,Meneveau:2011,Johnson:2016,Pereira:2018}.

	
	\begin{acknowledgments}
	We are very grateful to M. Gorokhovski for his comments on the manuscript.
	This work has benefited from stimulating discussions with M. Gorokhovski, R. Letournel, L. Gouden\`ege, A. Vi\'e, A. Richard and M. Massot.
	Also A. Lanottte, E. Calzavarini, F. Toschi, J. Bec, L. Biferale and M. Cencini are thanked for making the DNS dataset “Heavy particles in turbulent flows” (2011) publicly available from the International CFD Database \cite{Lanotte:2011}.
	Our DNS simulations were performed using high-performance computing resources from GENCI-CINES and GENCI-IDRIS (grant A0112B07400) and the CALMIP Centre of the University of Toulouse (grant P0910).
	\end{acknowledgments}

	\appendix
	
	\section{Estimation of $ c_\varepsilon $}
	\label{sec:c_eps}
 	To evaluate the $ c_\varepsilon $ factor appearing in \eqref{eq:c_eps}, we use the following relation between the conditional averages 
 	\footnote{	
 	This relation is simply obtained from the relation between the joint PDF and the conditional PDF: $P(a^2,\varepsilon, K) = P(a^2 | \varepsilon, K) P(\varepsilon, K) = P(a^2 | \varepsilon, K) P(K | \varepsilon) P(\varepsilon)$
 	and the relation between the joint probability density of $a^2,\varepsilon, K$ and $a^2,\varepsilon$: $P(a^2,\varepsilon) = \int dK P(a^2,\varepsilon, K)$.
 	}:
 	\begin{equation}
 		\langle a^2 | \varepsilon \rangle = \int dK \langle a^2 | \varepsilon, K \rangle P(K | \varepsilon)
 		\label{eq:cond}
 	\end{equation}
	
\begin{figure*}[h]
 	\centering
	
 	\includegraphics[width=0.49\textwidth,height=0.4\textheight,keepaspectratio, clip]{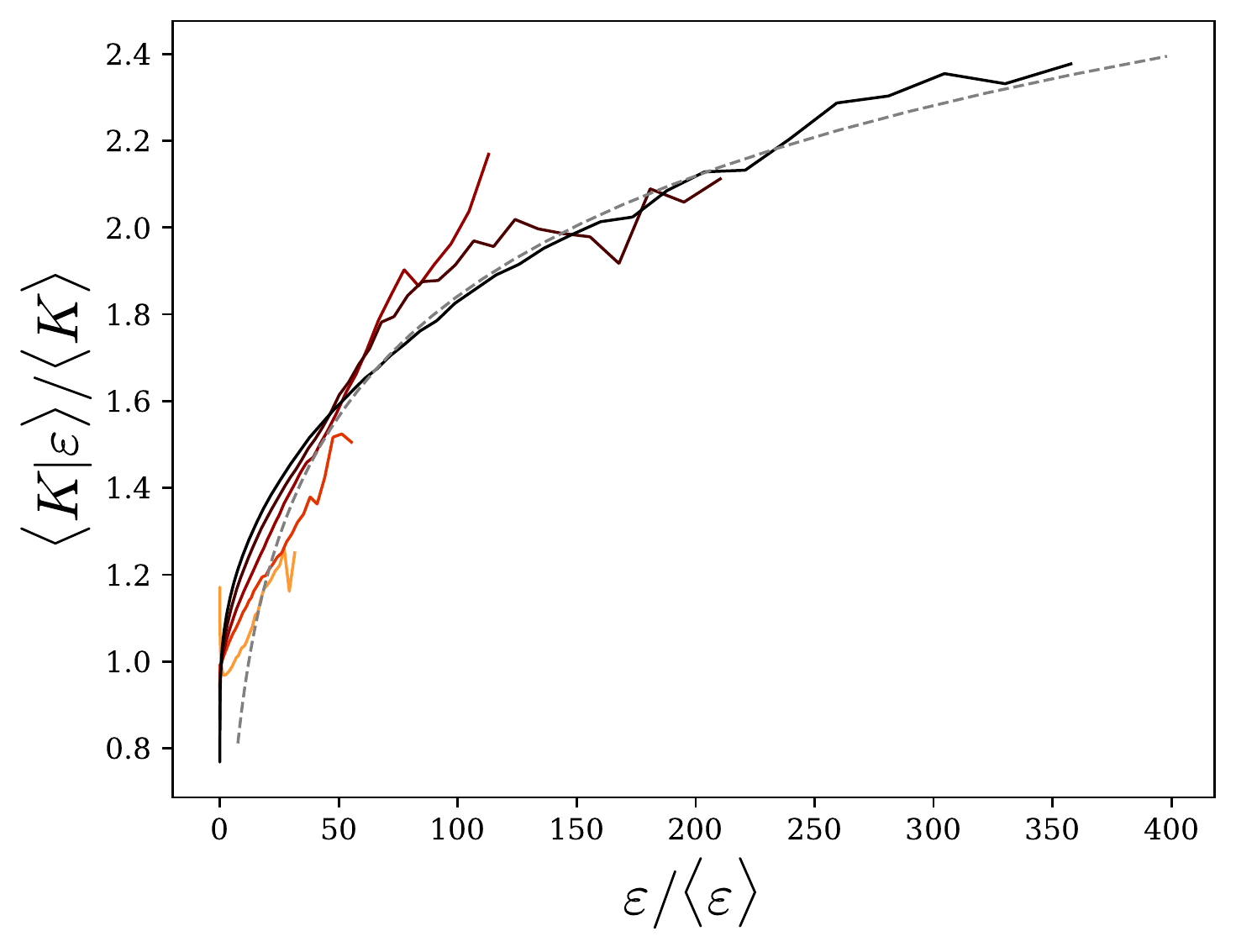}
 	\includegraphics[width=0.49\textwidth,height=0.4\textheight,keepaspectratio, clip]{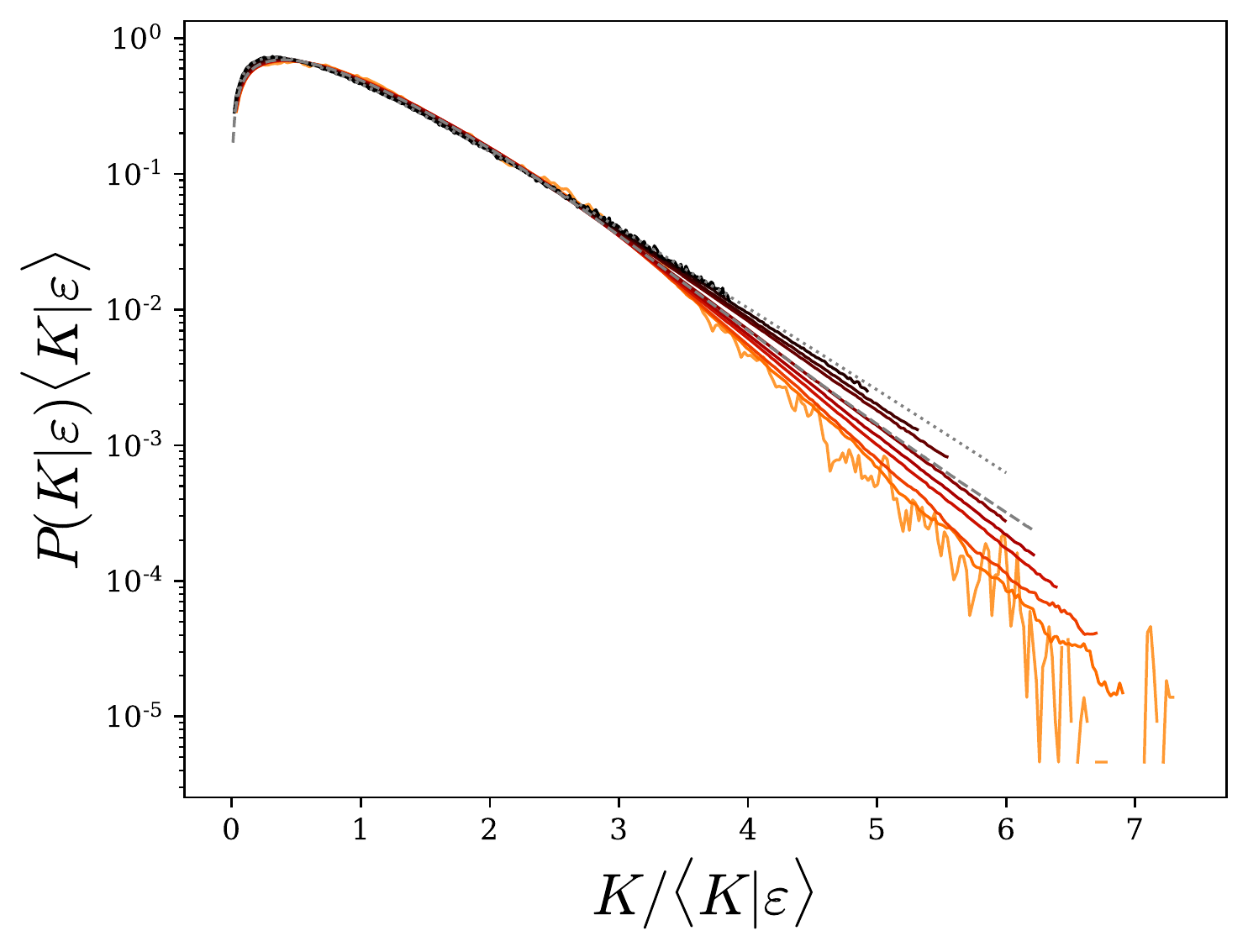}
	
 	\caption{ (Left) Average of the kinetic energy conditioned on the local dissipation rate from our DNS for  $Re_\lambda = 50$, 90, 150, 230 and 380 in continuous lines from orange to black respectively, 
	and comparison with the relation $\langle K | \varepsilon\rangle/ \langle K \rangle = 0.4 \ln(\varepsilon/ \langle \varepsilon\rangle)$.
 	(Right) Probability density function of the kinetic energy conditioned on the dissipation rate, normalized by its average $\langle K | \varepsilon\rangle$, from our DNS at $Re_\lambda=380$ 
	for $\varepsilon/\langle \varepsilon\rangle = 0.01, 0.05, 0.1, 0.3, 0.5, 1, 3, 5, 10, 50 \pm30\% $ from orange to black. 
	Comparison with the marginal PDF of the kinetic energy in dashed line and with the PDF \eqref{eq:PG} in dotted line.}
 	\label{stat_K_cond_eps} 
 	\end{figure*}

 	Substituting relation \eqref{eq:c_eps} in \eqref{eq:cond}, we find, assuming that $c_\varepsilon $ is independent of $ K $ 
 	\begin{equation}
 		\langle a^2|\varepsilon \rangle / a_\eta^2 = c_\varepsilon \int dK \exp(\alpha K/\langle K \rangle) P(K | \varepsilon)
 		\label{eq:cond_b}
 \end{equation}
	If the kinetic energy is statistically independent of the dissipation rate (\textit{i.e.} $ P(K | \varepsilon) = P (K) $) the integral in the previous relation takes a constant value and $ c_\varepsilon $ is proportional to $ \langle a^2 | \varepsilon \rangle $.
	However such an assumption is only approximate at moderate Reynolds numbers as shown from our DNS.
 	Indeed, it is seen in Fig. \ref{stat_K_cond_eps}, that the average of $ K $ conditioned on $ \varepsilon $ has a weak logarithmic dependence on $ \varepsilon $.
	Note that the average dissipation rate conditional on the kinetic energy can be found in \cite{Wilczek:2011}.
 	We also present in Fig. \ref{stat_K_cond_eps} the probability density of the kinetic energy conditioned on the dissipation rate. 
 	In this figure, the PDF is normalized by its mean value $ \langle K | \varepsilon \rangle $.
	It is to note that for large values of the dissipation rate, the conditional PDF takes a self-similar form:
 	\begin{equation}
 		 P(K / \langle K|\varepsilon\rangle | \varepsilon) = P_G( K / \langle K|\varepsilon\rangle)
 	\end{equation}

 	 Thus by combining the previous relations, one can write:
 \begin{equation}
 	\langle a^2|\varepsilon\rangle / a_\eta^2 = c_\varepsilon \int d K^* \exp(\alpha^* K^* ) P_G(K^*) 
 		\label{eq:c_eps1}
 \end{equation}
 where we have introduced $K^* = K / \langle K|\varepsilon\rangle$ and $\alpha^* (\varepsilon) = \alpha \, \langle K|\varepsilon\rangle/\langle K \rangle $.
 	 In addition, we see in Fig. \ref{stat_K_cond_eps} that the self-similar form of the distribution of $ K^* $ knowing $ \varepsilon $ is well approximated by the following distribution obtained from the Maxwell distribution (i.e. assuming that the three components of the velocity are Gaussian and independent): 
 
 \begin{equation}
 	 	P_G(x) = \dfrac{3}{ \sqrt{\pi } } \sqrt{\dfrac{3}{2} x } \, \exp\left( - \dfrac{3}{2} x\right)
 \label{eq:PG}
 \end{equation}
 	Note that the average of this distribution is indeed unity: $ \int x P_G(x)dx = 1$.
 With this expression for $P_G$ the integral of equation \eqref{eq:c_eps1} can be computed as: 
 \begin{equation}
 	A_\varepsilon^{-1} = \int d K^* \exp(\alpha^* K^* ) P_G(K^*) = \left(1-\dfrac{2 }{3}\alpha^*\right)^{-3/2}  \, .
 \end{equation}
 	Thus, according to \eqref{eq:cond_b}, we have for $c_\varepsilon$:
 	\begin{equation}
 		c_\varepsilon = A_\varepsilon \, \langle a^2|\varepsilon \rangle / a_\eta^2
 	\end{equation}
	The dependence of $A_\varepsilon$ with $\varepsilon$ explains the deviation of the power law behavior between $\langle a^2 | \varepsilon, K\rangle $ and $\langle a^2 | \varepsilon\rangle $ observed in Fig. \ref{fig:cond_acc_dissip_tke}b for $\varepsilon \gg \langle \varepsilon \rangle$.
		
 	It is to note that the integral $A_\varepsilon$ converges only if $\alpha^*<3/2$. 
 	This observation suggests that the dependence of $ \langle K | \varepsilon \rangle / \langle K \rangle $ on $\varepsilon$ should decrease as the Reynolds number increases to allow $\alpha^*$ to remains lower than $3/2$ even for the most intense fluctuations of $\varepsilon/ \langle \varepsilon\rangle$, and thus ensuring the convergence of the integral. 
 	Therefore, the larger the Reynolds number, the weaker the dependence of $\langle K | \varepsilon \rangle / \langle K \rangle $ on $\varepsilon$.
	This is consistent, with the statistical independence between the local  values of the kinetic energy and of the dissipation at large Reynolds numbers, in line with scale separation of the turbulent cascade.
	Accordingly, we simply propose to write: 
 	\begin{equation}
 		c_\varepsilon \approx A \, \langle a^2|\varepsilon \rangle / a_\eta^2
 	\end{equation}
 	 where $A= \left(1-\dfrac{2 }{3} \alpha \right)^{3/2}$, which is equal to $A=7\sqrt{7}/27\approx 0.686$, for $\alpha = 1/3$, neglecting the small logarithmic dependence in $ \varepsilon / \langle \varepsilon \rangle $.
	
	
	\section{Analytical estimation of the acceleration variance}
	\label{sec:acc_var_serie_expension}

	Using expression \eqref{eq:avar_cond_dissip_final} for the conditional acceleration variance, we write the acceleration variance as
	\begin{equation}
		\dfrac{\langle a^2 \rangle}{a_\eta^2} = B  \int_0^{\infty}  \left( Z + \varepsilon_* \right )^{\gamma}  P(\varepsilon_*) d\varepsilon_*\ ,
		\label{eq:a_var_int}
	\end{equation}
	 where we introduced $\gamma = 3/2+\beta$,  $Z= \left( \dfrac{1}{B}\dfrac{a_0^2}{a_\eta^2}\right )^{1/\gamma}$ and $\varepsilon_*=\varepsilon/\langle \varepsilon \rangle$ for simplicity.
	 We can expand the term within the integral using the generalized binomial series:
	 \begin{equation}
	 	\left( Z + \varepsilon_* \right )^{\gamma}  = 
	    \left\{ 
			\begin{array}{rcl}
	           \displaystyle  \sum_{k=0}^{\infty} C_{\gamma,k} Z^{\gamma-k} \varepsilon_*^k & \mbox{for} & \varepsilon_*< Z 
				\\[4ex] 
				\displaystyle  \sum_{k=0}^{\infty} C_{\gamma,k} Z^{k}         \varepsilon_*^{\gamma-k} & \mbox{for} & \varepsilon_*> Z 
	        \end{array}
		\right.
	 \end{equation}
	 with $\displaystyle C_{\gamma,k}=\begin{pmatrix} \gamma \\k \end{pmatrix} = \dfrac{1}{k!}\prod_{i=0}^{k-1} (\gamma-i)$ the generalized binomial coefficient.
	In particular, we have  $C_{\gamma,0}=1$ and  $C_{\gamma,1}=\gamma$.
	Splitting the integral in \eqref{eq:a_var_int} enables to write:
	\begin{equation}
		\dfrac{\langle a^2 \rangle}{a_\eta^2} = B \sum_{k=0}^{\infty} C_{\gamma,k}
		 \left[
		Z^{\gamma-k}  \int_0^{Z} \varepsilon_*^{k}  P(\varepsilon_*) d\varepsilon_*
		+
		Z^{k}  \int_{Z}^{\infty}  \varepsilon_*^{\gamma-k}  P(\varepsilon_*) d\varepsilon_*
		 \right] \, .
	\end{equation}
	The two integrals are partial moments of the normalized dissipation rate.
	Considering that $\varepsilon_*$ follows the lognormal distribution with parameters $\mu$ and $\sigma^2$, with the change of variable  $x=(\ln{\varepsilon_*}-\mu)/\sqrt{\sigma^2}$, we can express the partial moments of order $n$ as
	\begin{equation}
	\int_{Z}^{\infty} \varepsilon_*^{n}  P(\varepsilon_*) d\varepsilon_* = \exp(n \mu + n^2 \sigma^2/2)	\int_{(\ln Z - \mu)/\sqrt{\sigma^2} }^{\infty}  \exp( - (x -n\sqrt{\sigma^2})^2/2) dx = \langle \varepsilon_*^n \rangle \Phi(Z,n )
	\end{equation}
	with $\Phi(Z,n) = \dfrac{1}{2}-\dfrac{1}{2} \erf\left(\dfrac{\ln Z -\mu}{\sqrt{2\sigma^2}}-\dfrac{n\sqrt{\sigma^2} }{\sqrt{2}} \right)$.
	Note that since $\langle \varepsilon_* \rangle =1$,  we have $\mu=-\sigma^2/2$ which gives for the regular moments $\langle \varepsilon_*^n \rangle = \exp(\sigma^2 n (n-1)/2 ) $ and 
	$\Phi(Z,n) = \dfrac{1}{2}-\dfrac{1}{2} \erf\left(\dfrac{\ln Z +\sigma^2(n-1/2)}{\sqrt{2\sigma^2}} \right)$.
	As well, we have for the other partial moments  
	\begin{equation}
		\int_0^{Z} \varepsilon_*^{n}  P(\varepsilon_*) d\varepsilon_* = \langle \varepsilon_*^n \rangle (1- \Phi(Z,n )) \, .
	\end{equation} 
	
	For the acceleration variance, we obtain eventually the following analytical series expansion
	\begin{equation}
		\dfrac{\langle a^2 \rangle}{a_\eta^2} = B \sum_{k=0}^{\infty} C_{\gamma,k} 
		 \left[
		Z^{\gamma-k}  \langle \varepsilon_*^k \rangle (1- \Phi(Z,k ))
		+
		Z^{k}  \langle \varepsilon_*^{\gamma-k} \rangle \Phi(Z,\gamma-k )
		 \right] \, .
 		\label{eq:a_var_RZ_full_annalytic}
	\end{equation}
	This expression is observed to converge very rapidly to the numerical evaluation of the integral \eqref{eq:a_var_int}, as there is just minute differences when considering only the first three elements of the sum.
	
	For $Z$ and $k$ small, we have $\Phi(Z,k) = O(1)$, which allow us to simplify the relation for the acceleration variance:
	\begin{equation}
		\dfrac{\langle a^2 \rangle}{a_\eta^2} = B \sum_{k=0}^{\infty} C_{\gamma,k} Z^{k}  \langle \varepsilon_*^{\gamma-k} \rangle \, .
	\end{equation}
	The successive terms of the series correspond to corrections of low-Reynolds number effects of increasingly high order, since $Z \sim Re_\lambda^{-1/\gamma}$.
	At leading order we have
	\begin{equation}
		\dfrac{\langle a^2 \rangle}{a_\eta^2} = B \langle \varepsilon_*^{\gamma} \rangle \, , 
		\label{eq:a_var_RZ_eps}
	\end{equation} 
	while the first order correction gives:
	\begin{equation}
		\dfrac{\langle a^2 \rangle}{a_\eta^2} = 
		B \left(\langle \varepsilon_*^{\gamma} \rangle + \gamma Z \langle \varepsilon_*^{\gamma-1} \rangle \right)\, .
	\end{equation} 
	With $\sigma^2 = 3/8 \ln{Re_\lambda/R_c}$ as proposed by \cite{Yeung:2006}, the leading order expression \eqref{eq:a_var_RZ_eps}  gives  \eqref{eq:a_var_RZ}, and the first order expression reads:
	\begin{equation}
		\dfrac{\langle a^2 \rangle }{a_\eta^2} = B \left( \dfrac{Re_\lambda}{R_c} \right)^{3/16 \gamma (\gamma-1)} \left( 1+ \gamma \left( \dfrac{1}{B}\dfrac{a_0^2}{a_\eta^2}\right )^{1/\gamma} \left( \dfrac{Re_\lambda}{R_c} \right)^{-3/8 (\gamma-1)}  \right) \, .
		\label{eq:avar_1storder_appendix}
	\end{equation}
	This later expression is seen in fig. \ref{fig:avar_vs_Re} to give a very good approximation of \eqref{eq:avar_cond_dissip_final}.

	\section{Conditional PDF of the acceleration} \label{sec:conditional_pdf}
	
\begin{figure*}[h]
    	\centering
    	\includegraphics[width=0.49\textwidth,height=0.4\textheight,keepaspectratio, clip]{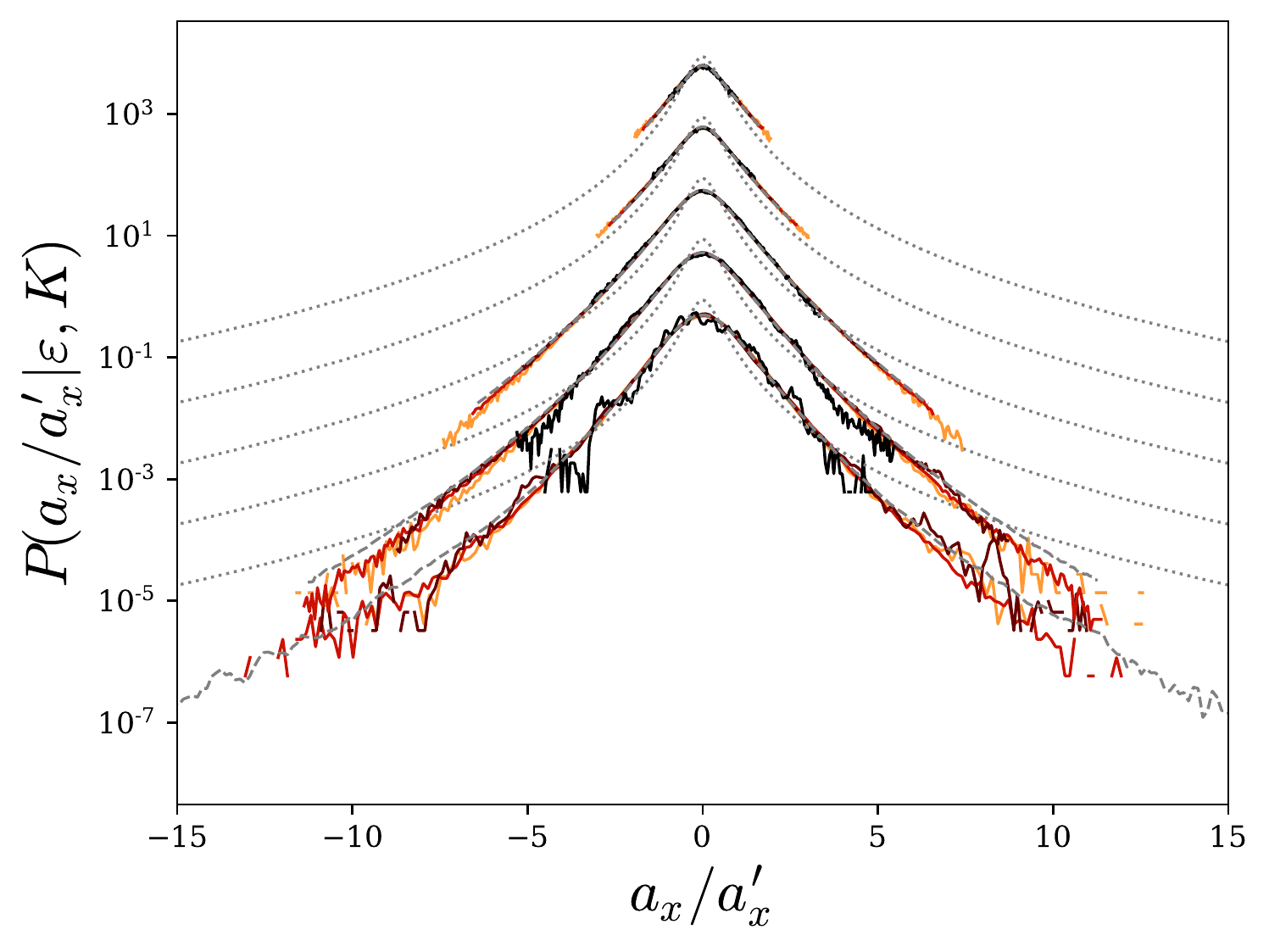}
    	\caption{ PDF of the acceleration conditional on the dissipation and the kinetic energy $P(a_i | \varepsilon, K)$ for various values of $\varepsilon$ and $K$: 
   	$K/\langle K \rangle=0.1$, 1, 2.5 and $5\pm30\%$ from orange to black respectively and 
    $\varepsilon/\langle \varepsilon \rangle=0.05$, 0.3, 1, 5 and $10 \pm 30\%$ shifted respectively by one decade upward. 
   	Comparison with the acceleration PDF conditioned only by the dissipation in gray dashed line, and with the unconditional PDF in dotted gray line. Each PDF is normalized by its standard deviation. 
   	Data from our DNS at $Re_\lambda=380$. 
   	}
    	\label{fig:pdf_cond_acc_2}
    	\end{figure*}

   	We complete the statistical description of the conditional acceleration by showing, in Fig. \ref{fig:pdf_cond_acc_2}, its probability density function (PDF).
   	In this figure, we compare the PDF of the acceleration conditional on the dissipation and the kinetic energy, with the PDF conditioned only by the dissipation and with the unconditional PDF obtained from our DNS at $Re_\lambda=380$. 
   	All the PDFs are normalized by their respective standard deviation.
   	It is observed that the conditional PDFs present much less developed tails than the unconditional PDF.
   	Moreover the doubly-conditional PDFs overlap with the simply-conditional PDF, showing that conditioning by the velocity does not alter the shape of the PDF.
   	As well the shape is observed to be invariant for all values of $\varepsilon$, 
   	supporting the idea of a canonical distribution presented in \cite{Castaing:1996}.
	
	\section{Modeling of the dissipation rate} \label{sec:model_dissip}
	
	\subsection{Dissipation as multiplicative cascade process} \label{sec:cascade_dissip}
	
	The image of the energy cascade is naturally associated with multiplicative processes \cite{Yaglom:1966,Kolmogorov:1941,Monin:1981b,Mandelbrot:1974,Benzi:1984,Frisch:1978}.
	Such model proposes to express the locally space-averaged dissipation over a volume of size $\ell= L \lambda^n$, with $L$ the large-scale of the flow and $\lambda<1$, as:
	\begin{equation}
		\varepsilon_n = \varepsilon_0 \dfrac{\varepsilon_1}{\varepsilon_0} \ldots \dfrac{\varepsilon_n}{\varepsilon_{n-1}} = \varepsilon_0 \prod^n_{i=1} \xi_i
	\end{equation}
	Assuming that $\xi_i = \varepsilon_i/\varepsilon_{i-1} $ are independent positive random numbers with identical distribution across scales we write:
	\begin{equation}
		\ln \dfrac{\varepsilon_n}{\varepsilon_0} = \sum_{i=1}^n \ln \xi_i
	\end{equation}
	Therefore according to the central limit theorem the term on the right must present a normal distribution with parameters $ \mu = n \mu_\xi $ and $\sigma^2 = n \sigma_\xi^2$. 
	The parameters $\mu_\xi$ et $\sigma_\xi^2$ appear as fundamental unknowns, but can nevertheless be related with the relation $ \mu_\xi = - \sigma_\xi^2/2 $ obtained from the moments of a log-normal variable in order to guarantee that the average energy flux is conserved throughout the cascade. 
	Setting $\ell = \eta$ (i.e. $n= \ln(\eta/L)/\ln\lambda \sim \ln Re_\lambda$) we obtain a model for the local dissipation rate.
	The log-normal distribution for $\varepsilon$ has been confirmed for example by DNS of \cite{Yeung:2006}.
	Moreover for the variance of the logarithm of the local dissipation rate is then $\sigma^2 = \dfrac{ \sigma_\xi }{ \ln \lambda} \ln \eta / L = A+B\ln Re_\lambda$ as predicted by Kolmogorov and Oboukhov \cite{Kolmogorov:1962, Oboukhov:1962}. 
	Such evolution for $\sigma^2$ has been also confirmed by the DNS of \cite{Yeung:2006} showing that $\sigma^2 \approx 3/8 \ln Re_\lambda/10$.
	
	Such multiplicative process also implies a logarithmic evolution of the spatial correlation of the dissipation rate as explained by Mandelbrot \cite{Mandelbrot:1972}.
	We consider the dissipation rate at two points $A$ and $B$, $\varepsilon^{A}_{n}$ and $\varepsilon^{B}_{n}$, both defined on the same scale $n$.
	The points $A$ and $B$ are separated by a distance $L>r>\eta $ from each other and we note $k = \ln (r/L) / \ln \lambda$, then $0<k<n$. 
	Clearly, the greater the distance between the two points, the larger the scale of their common root in the cascade:
	\begin{equation}
		\varepsilon^A_{n} = \varepsilon^{AB}_{ 0} \dfrac{\varepsilon^{AB}_{1}}{\varepsilon^{AB}_{0}} \ldots \dfrac{\varepsilon^{AB}_{k}}{\varepsilon^{AB}_{k-1} } \dfrac{\varepsilon^{A}_{k+1}}{\varepsilon^A_{k}} \ldots \dfrac{\varepsilon^A_{n}}{\varepsilon^A_{n-1} }
	\end{equation}
	\begin{equation}
		\varepsilon^{B}_{n} = \varepsilon^{AB}_{0} \dfrac{\varepsilon^{AB}_{1}}{\varepsilon^{AB}_{0}} \ldots \dfrac{\varepsilon^{AB}_{k}}{\varepsilon^{AB}_{k-1} } \dfrac{\varepsilon^{B}_{k+1}}{\varepsilon^{B}_{k}} \ldots \dfrac{\varepsilon^{B}_{n}}{\varepsilon^{B}_{n-1} }
	\end{equation}	
	In the two previous equations, we have distinguished by the exponents $ A $ and $ B $ the variables which are specific to points $A$ and $B$ and by $ AB $ those which are common. 
	This can be expressed as:
	\begin{equation}
		\ln \dfrac{\varepsilon^{A}_{n}}{\varepsilon_{0}} = \sum_{i=1}^{k} \ln \xi^{AB}_{i} + \sum_{i=k+1}^{n} \ln \xi^{A}_{i}
	\end{equation}
	\begin{equation}
		\ln \dfrac{\varepsilon^{B}_{n}}{\varepsilon_{0}} = \sum_{i=1}^{k} \ln \xi^{AB}_{i} + \sum_{i=k+1}^{n} \ln \xi^{B}_{i}
	\end{equation} 
	
	The correlation between $\ln\varepsilon^{A}_{n}$ and $\ln\varepsilon^{B}_{n}$ is defined as 
	\begin{equation}
		R_{\ln\varepsilon}(r)= \langle (\ln \dfrac{\varepsilon^{A}_{n}}{\varepsilon_{0}}-\mu) (\ln \dfrac{\varepsilon^{B}_{n}}{\varepsilon_{0}} -\mu)\rangle = \langle \ln \dfrac{\varepsilon^{A}_{n}}{\varepsilon_{*}} \ln \dfrac{\varepsilon^{B}_{n}}{\varepsilon_{*}} \rangle
	\end{equation}	
	 where we noted $\varepsilon_{*}=\varepsilon_{0}e^{\mu}$. 
	Introducing similarly $\xi_*=e^{\mu_\chi}$ and $\xi^{\prime}=\xi/\xi_*$ we express the correlation as:
	\begin{eqnarray}
		R_{\ln\varepsilon}
		 &=& \langle \sum_{i=1}^n (\ln \xi^A_i -\mu_\xi) \sum_{j=1}^n (\ln \xi^B_i -\mu_\xi) \rangle = \langle \sum_{i=1}^n \ln \xi^{\prime\, A}_i \sum_{j=1}^n \ln \xi^{\prime\, B}_j \rangle \nonumber \\
		 &=& \langle \left( \sum_{i=1}^{k} \ln \xi^{\prime\, AB}_i + \sum_{i=k+1}^n \ln \xi^{\prime\, A}_i \right) \left( \sum_{j=1}^{k} \ln \xi^{\prime\, AB}_j + \sum_{j=k+1}^n \ln \xi^{\prime\, B}_j \right) \rangle \nonumber \\
		&=& \sum_{i=1}^{k} \sum_{j=1}^{k} \langle \ln \xi^{\prime\, AB}_i \ln \xi^{\prime\, AB}_j \rangle 
		 			+ \sum_{i=1}^{k} \sum_{j=k+1}^n \langle \ln \xi^{\prime\, AB}_i \ln \xi^{\prime\, AB}_j \rangle\nonumber \\	
		 		& &	+ \sum_{i=k+1}^n \sum_{j=1}^k \langle \ln \xi^{\prime\, A}_i \ln \xi^{\prime\, AB}_j \rangle
					+ \sum_{i=k+1}^n \sum_{j=k+1}^n \langle \ln \xi^{\prime\, AB}_i \ln \xi^{\prime\, B}_j \rangle\nonumber \\		
		&=& 	\sum_{i=1}^{k} \sum_{j=1}^{k} \delta_{ij} \sigma^2_\xi = k \sigma^2_\xi 				 
	\end{eqnarray}
	To obtain this results we used the hypothesis that within the same branch, the events at a given scale are independent of those at another scale, $ \langle \ln \xi^{\prime\, AB}_i \ln \xi^{\prime\, AB}_j \rangle = \delta_{ij} \sigma^ 2_\xi $, as well as vanishing correlation between branches $ A $ and $ B $: $ \langle \ln \xi^{\prime\, A}_i \ln \xi^{\prime\, B}_j \rangle = 0 $.
	This gives a logarithmic evolution of the correlation coefficient $ \rho_{\ln \varepsilon} = R_{\ln\varepsilon}/\sigma^2 $, in the range $\eta<r<L $:
	\begin{equation}
		\rho_{\ln \varepsilon} = \dfrac{\langle \ln \dfrac{\varepsilon^{A}_{n}}{\varepsilon_{*}} \ln \dfrac{\varepsilon^{B}_{n}}{\varepsilon_{*}} \rangle}{\langle \ln^2 \dfrac{\varepsilon_{n}}{\varepsilon_{*}} \rangle } = \dfrac{k}{n} = \dfrac{\ln L/r}{\ln L/\eta} = 1- \dfrac{\ln r/\eta}{\ln L/\eta}
	\end{equation}
	
	Although not trivial, this result can be transposed for the temporal correlation along particle path \cite{Sawford:2015,Huang:2014,Meneveau:1994,Borgas:1993b}.
	The logarithmic behavior of the correlation is confirmed by DNS, as it can be seen in \cite{Letournel:2021} where the evolution of the Lagrangian correlation of the logarithm of the dissipation is presented.

	
	\subsection{Stochastic Modeling of the dissipation} \label{sec:stoch_dissip}
	
	It has been proposed to model the dissipation rate as stochastic multiplicative process. 
	Such process can be generically expressed as :
	\begin{equation}
		d \varepsilon = \varepsilon \Pi dt + \varepsilon \Sigma dW
	\end{equation}
	with $dW$ the increment of the Wiener process ($\langle dW=0$ and $\langle dW^2 \rangle = dt$) and where $\Pi$ and $\Sigma$ are to be determined.

	Considering that $ \varepsilon $ follows a log-normal distribution with parameter $\sigma^2$ and $\mu=-\sigma^2/2$, we define the standard normal variable $\chi$ (Gaussian random variable with zero mean and unit variance)	as:
	\begin{equation}
		\dfrac{\varepsilon}{\langle \varepsilon \rangle} = \exp\left( \sigma \chi - \sigma^2/2 \right) 
		\label{eq:def_chi}
	\end{equation}
	
	A stochastic process for $\chi$ has to be given in order to obtain the stochastic process for $\varepsilon$, via the Ito transformation.

	Pope and Chen \cite{Pope:1990} proposed to obtain $\chi$ from an Orstein-Uhnlebbeck process with a characteristic time $\tau_{\varepsilon}$:
	\begin{equation}
		d\chi = -\dfrac{\chi}{\tau_{\varepsilon}} dt + \sqrt{\dfrac{2}{\tau_{\varepsilon}}} dW
		\label{eq:OU_pope}
	\end{equation}
	According to the Ito formula, this gives for $\Pi$ and $\Sigma$ :
	\begin{equation}
		\Pi= -\left( \ln \varepsilon / \langle \varepsilon \rangle - \sigma^2/2 \right) / \tau_{\varepsilon}
		\label{eq:Pi_pope}
	\end{equation}
	
	\begin{equation}
		\Sigma=\sqrt{2\sigma^2 / \tau_{\varepsilon}}
		\label{eq:Sigma_pope}
	\end{equation}
	
	This process gives as expected log-normal distribution for $\varepsilon$ (normal distribution for $\chi$) as well as an exponential decrease of the correlation of $ \ln \varepsilon $ with a characteristic time $ \tau_\varepsilon $.
	This exponential behavior is not consistent with the multiplicative cascade model as discuss above. 
	It rather corresponds to a direct energy transfer from large to small-scales.

	To ensure a logarithmic decorrelation of the dissipation, Chevillard \cite{Chevillard:2017b} proposed to adapt the Gaussian multiplicative chaos introduced by Mandelbrot \cite{Mandelbrot:1972}.
	This leads to a multifractal model in which the increment of the Wiener process in \eqref{eq:OU_pope} is replaced by a fractional Gaussian noise:
	\begin{equation}
		d \chi = -\dfrac{\chi}{\tau_\varepsilon}dt + \dfrac{1}{\Lambda} dW^0_{\tau_c}
		\label{eq:x_chevi}
	\end{equation}
	here, $dW^0_{\tau_c}$ is formally a fractional Gaussian noise with a 0 Hurst exponent, regularized at a time scale $\tau_c$, and $\Lambda$ is a normalization factor ensuring unit variance for $\chi$. The value of $\Lambda$ is dependent on the specific regularization of $dW^0_{\tau_c}$.
	As explained in \cite{Chevillard:2017b} this process can be reexpressed as 
	\begin{equation}
		d\chi(t) = \left(-\dfrac{\chi}{\tau_\varepsilon}+ \dfrac{\Gamma}{\Lambda} \right) dt + \dfrac{1}{\sqrt{\Lambda^{2}\tau_c}} dW
		\label{eq:x_chevi2}
	\end{equation}
	with $dW$ the increments of a standard Wiener process and $\Gamma$ corresponds to a convolution of the Wiener increments:
	\begin{equation}
		\Gamma = -\dfrac{1}{2}\int_{-\infty}^{t} (t-s+\tau_c)^{-3/2} dW(s)
		\label{eq:Gamma_b}
	\end{equation}
	where $dW(s)$ is the increments of the same realization of the Wiener process as in \eqref{eq:x_chevi2}.
	In \eqref{eq:Gamma_b}, the regularization time $\tau_c$ prevent the divergence of the kernel when $s\rightarrow t$.
	The normalization factor $\Lambda$ is estimated as $\Lambda=\langle X^2 \rangle$ where $X$ obey the stochastic equation \eqref{eq:x_chevi2} in which $\Lambda$ has been set to 1.

\begin{figure*}[h!]
	\centering
	\includegraphics[width=0.49\textwidth,height=0.3\textheight,keepaspectratio, clip]{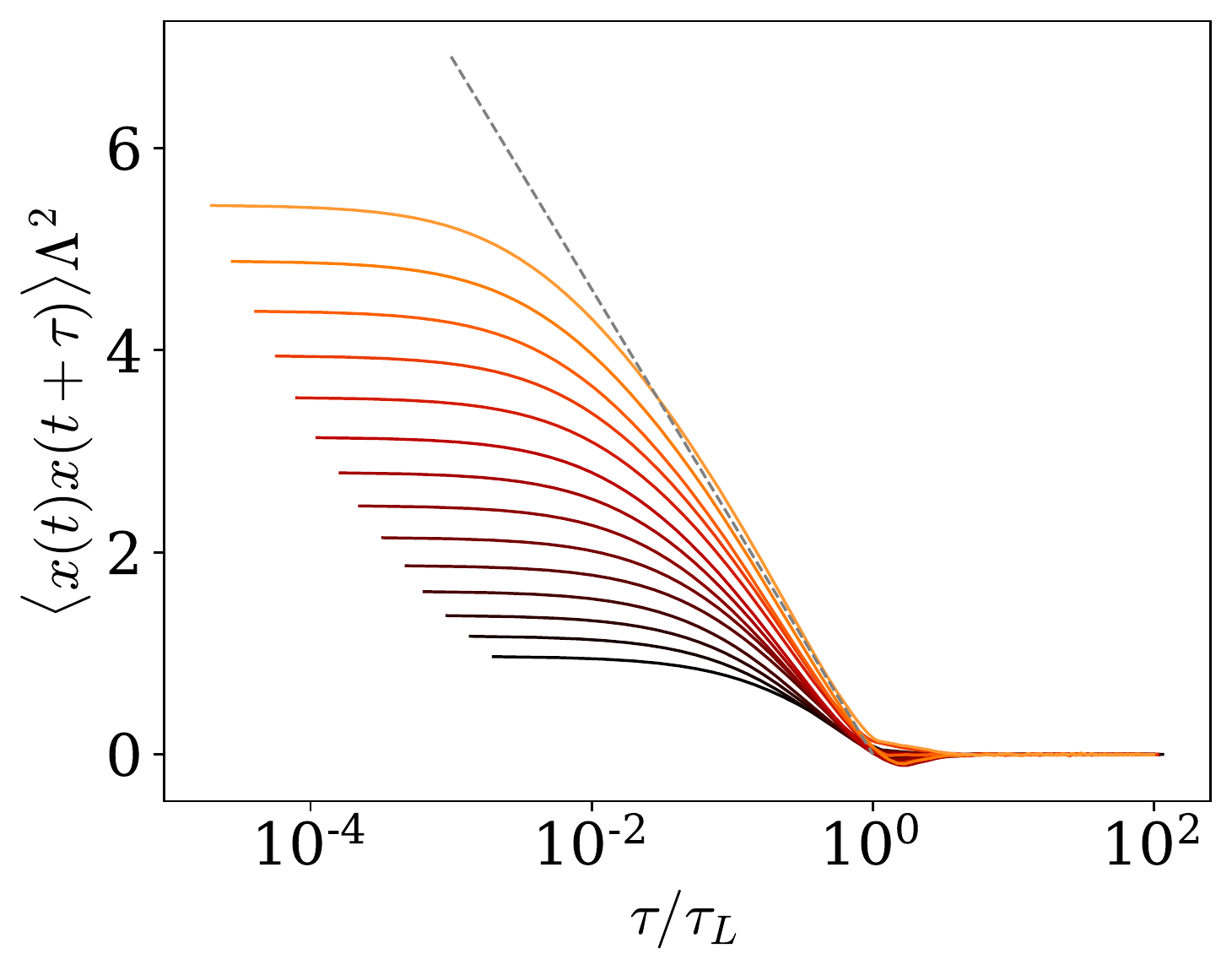}
	\caption{ Correlation of $\chi$ for various values of $\tau_\varepsilon/\tau_c$. }
	\label{fig:corr_x}
	\end{figure*}

	The stochastic process \eqref{eq:x_chevi2} gives a logarithmic correlation for $\chi$: $\langle \chi(t) \chi(t-s) \rangle \sim \ln \dfrac{\tau_\varepsilon}{s} $ for $\tau_c \ll s \ll \tau_\varepsilon$, as illustrated in the Fig. \ref{fig:corr_x}. 
	
	With the Ito transformation, we obtain the process for $ \varepsilon $ from \eqref{eq:x_chevi2}.
	This gives for $\Pi$ and $\Sigma$ :
	\begin{equation}
		\Pi= \left( - \ln \dfrac{\varepsilon}{\langle \varepsilon\rangle} + \dfrac{\sigma^2}{2\Lambda^{2}}\left( \dfrac{\tau_\varepsilon}{\tau_c} - \Lambda^{2} \right) + \dfrac{\sigma}{\Lambda}\Gamma \tau_{\varepsilon} \right) / \tau_\varepsilon
		\label{eq:Pi_chevi}
	\end{equation}
	
	\begin{equation}
		\Sigma= \sqrt{\dfrac{\sigma^2}{\Lambda^2\tau_c}}
		\label{eq:Sigma_pope}
	\end{equation}

		\subsection{Efficient calculation of the stochastic convolution $\Gamma$} \label{sec:algo_dissip}
	
\begin{figure*}[htb!]
			\centering
			\includegraphics[width=0.49\textwidth, keepaspectratio]{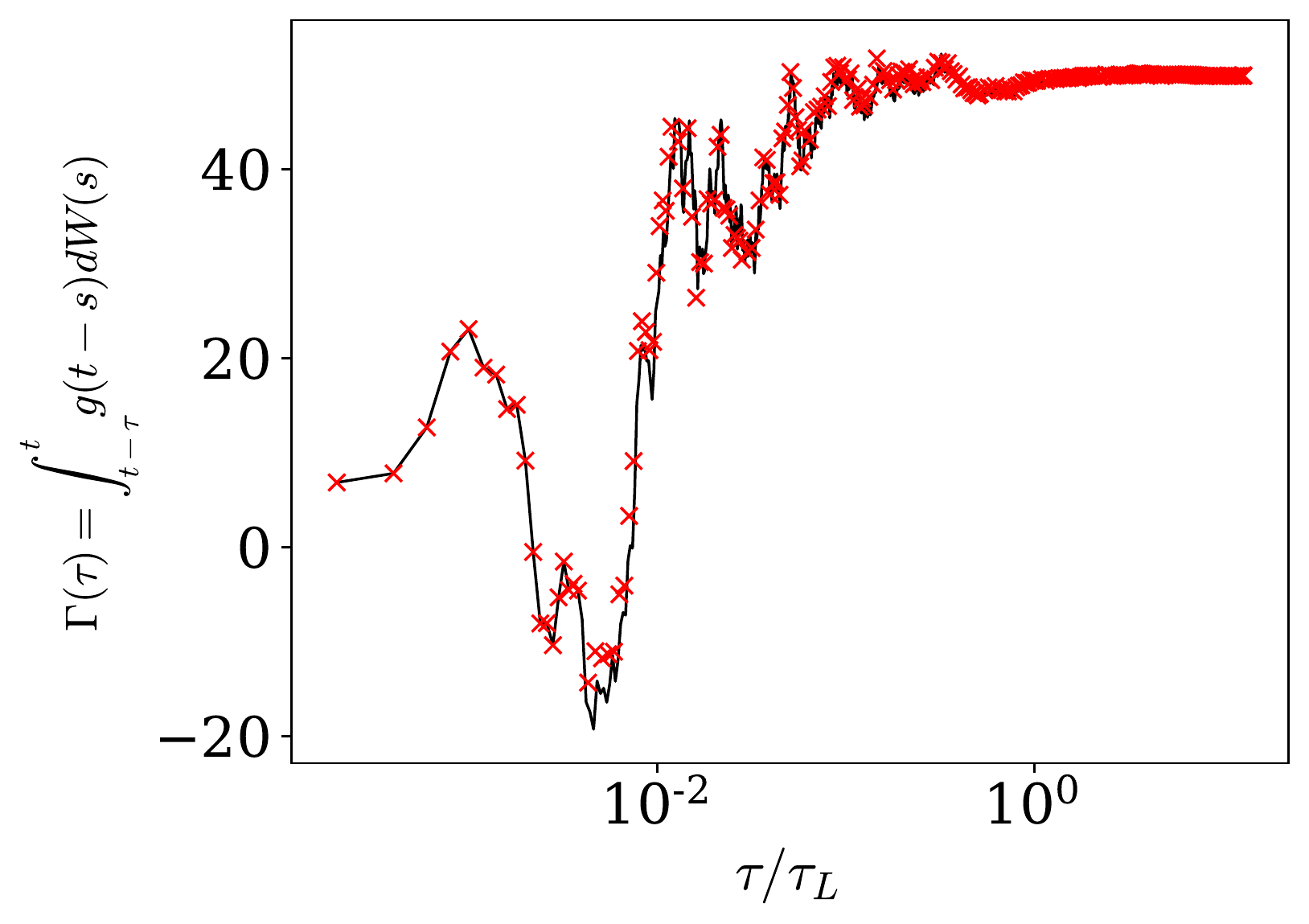}
			\caption{
			One realization of the integral  \eqref{eq:Gamma_b} as a function of the lower bound of the integral for  $dt = \tau_c/100$.
			Comparison between the direct calculation of the history integral \eqref{eq:Gamma_b} (black line) and the optimized calculation with $ N_s = 12$ (red crosses).}
			\label{fig:Gamma_optim}
	\end{figure*}

		In order to obtain a stationary process the lower bound of the integration is set to $-\infty$.		
		For the numerical computation of this integral, the lower limit has to be truncated.
		We present in figure \ref{fig:Gamma_optim} one realization of the evolution of the integral $ \Gamma(t, \tau) = - \dfrac{1}{2}\int^{t}_{t - \tau} (t-s + \tau_c)^{-3/2} dW(s) $ when the lower bound varies.
		We see that for values larger than $ \tau_\varepsilon $ the integral converges to a value (which remains random).
		In addition, the convergence threshold does not seem to depend on the time step used.
		So in practice $ \Gamma $ will be calculated with a lower-bound set to $t-5\tau_\varepsilon$.

		To obtain these calculations, the integral \eqref{eq:Gamma_b} giving the value of $ \Gamma $ at time $ t_n = n \, dt $ has been discretized as proposed by \cite{Chevillard:2017b} 		
	\begin{equation}
			\Gamma_n = -\dfrac{1}{2} \sum_{m=0}^{N_{hist}} (s_m+\tau_c)^{-3/2} \, dW_{n-m}
\label{eq:Gamma_num}
	\end{equation}
	with $s_m = t_n-t_{n-m} = m\, dt$, $N_{hist}=5\tau_\varepsilon/dt$ and $dW_{n-m}$ the increment of the Wiener process at time $(n-m)dt$.
	
	This direct calculation requires a lot of memory in order to keep the last $ N_{hist} $ instants and requires a very large number of operations, of the order of $ N_t \times N_{hist} $ where $ N_t $ is the number of time steps of the simulation. 
	Thus this direct method is difficult to use in practice when $ \tau_\varepsilon / \tau_\eta \sim Re_\lambda$ becomes large. 	
		
	For this reason \cite{Chevillard:2017b} proposed to speed up significantly the calculation using the fast Fourier transforms (FFT). 
	The integral at time $n\, dt$ is then computed as $\Gamma_n = -1/2 z_n$ where $z_n = FFT^{-1}(Z_k)$ is given by the inverse Fourier transform of $Z_k$.
	$Z_k=X_k \, Y_k$ is the convolution in spectral space between $x_n$ and $y_n$ ($X_k=FFT(x_n)$ and $Y_k=FFT(y_n)$) where $x_n$  and $y_n$ are the sequences $dW_n$ and $(s_n+\tau_c)^{-3/2}$ padded with zeros such that they have $N>N_{hist}+N_t$ points.
	This algorithm is indeed much faster. 
	Nevertheless, the memory occupation becomes more important since all the values of the sequence $ dW_n $ must be known simultaneously in order to calculate the Fourier transform, which limits the possibility of using this algorithm for large Reynolds numbers. 
	
	Such limitation can be overcome by using the approach proposed in \cite{Letournel:2021} based on the inverse Laplace transform of the convolution kernel.
	In this approach $\Gamma$ is estimated as a weighted sum of correlated Orstein-Uhlenbeck processes with characteristic time ranging from $\tau_c$ to $\tau_\varepsilon$.
	
	Despite its efficiency, this technic, nor the one based on FFT, cannot be used to determine the $\hat{\Gamma}$ that appears in the vectorial stochastic model for the acceleration, or as noted by \cite{Pereira:2018} for velocity gradients.
	Indeed in such cases it is not the increments of the components of the Wiener process which are convoluted, but a projection of them as shown in \eqref{eq:Gamma_hat_b}.
The issue is that the projection cannot by computed a priori, because it requires knowing $ a_i $ and $ u_i $, as seen in \eqref{eq:proj_convol}.

\begin{figure*}[htb!]
			\centering
			\includegraphics[width=0.8\textwidth, keepaspectratio]{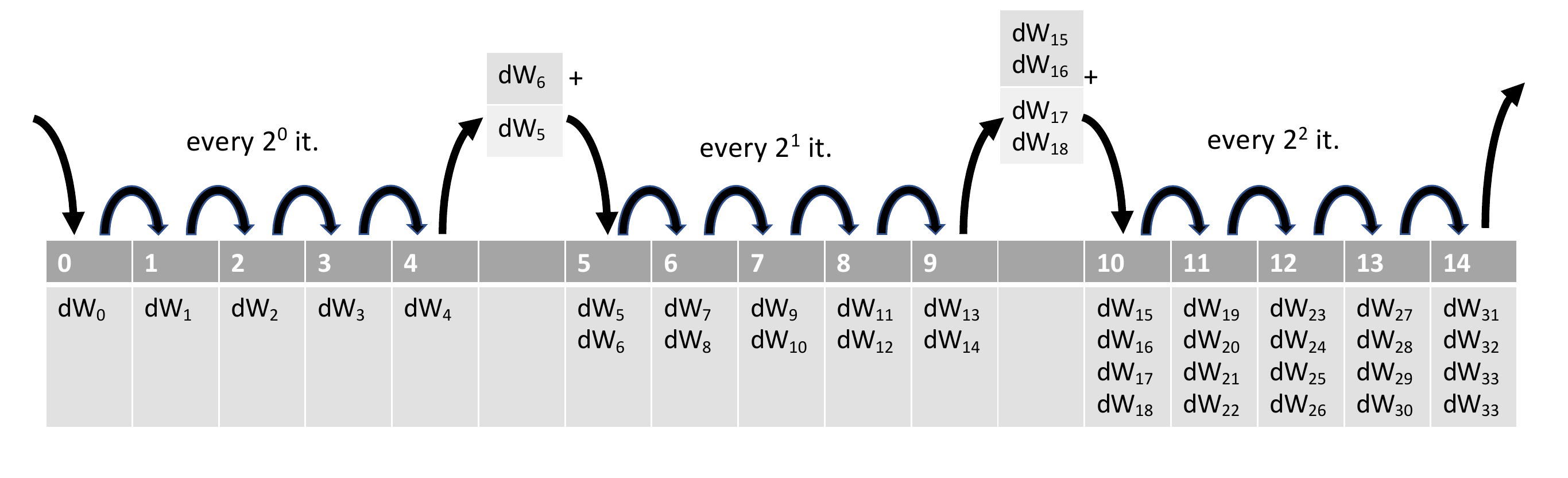}
			\caption{Diagram illustrating the coarse-graining of the integral \eqref{eq:Gamma_b} and the non-uniform update of the list.}
			\label{fig:Gamma_list}
	\end{figure*}

	For these reasons we propose a new algorithm which is fast, using a limited amount of memory and which only requires knowing the $ dW_n $ sequentially. 
	This algorithm is derived from the one introduced for non-stochastic integrals in \cite{Le-Roy-De-Bonneville:2021}. 	
	As we go back in the past, we can afford to remember with less precision the noise entering this integral, since the kernel decreases with the lag.
	We will thus proceed by progressive "coarse graining" and group together the oldest $ dW_n $, by introducing an increasingly extended local average.
	We then decompose the sum of \eqref{eq:Gamma_num} into sub-sums comprising an increasing number of terms:
	\begin{eqnarray}
					- 2 \Gamma_n &=& \sum_{m=0}^{N_{hist}} (s_m+\tau_c)^{-3/2} \, dW_{n-m} \nonumber \\
					&=& \sum_{m=m_{s1}}^{m_{e1}} (s_m+\tau_c)^{-3/2} \, dW_{n-m} + \ldots + \sum_{m=m_{sN}}^{m_{eN}} (s_m+\tau_c)^{-3/2} \, dW_{n-m} \nonumber \\
					&=& \sum_{j=1}^{N} \sum_{m=m_{sj}}^{m_{ej}} (s_m+\tau_c)^{-3/2} \, dW_{n-m}\nonumber \\
					&\approx & \sum_{j=1}^{N} (\overline{s}_j+\tau_c)^{-3/2} \, \overline{dW}_{j}
	\label{eq:Gamma_num_2}
	\end{eqnarray} 
	Where we introduced $\overline{s}_j = (m_{ej}+m_{sj})dt/2$ and
	$\overline{dW}_{j} = \sum_{m=m_{sj}}^{m_{ej}} dW_{n-m} $ such that 
	 $(\overline{s}_j+\tau_c)^{-3/2} \, \overline{dW}_{j} \approx \sum_{m=m_{sj}}^{m_{ej}} (s_m+\tau_c)^{-3/2} \, dW_{n-m}$. 
	 The bounds $ m_{ej} $ and $ m_{sj} $ are progressively spaced as $ j $ increases leading to an increasingly coarse splitting of the integral. 
	 This approximation of the integral can be carried out very efficiently by using a non-homogeneous list updating for $\overline{dW}$. 
	 The first elements of the list are updated every time steps and the older ones less and less regularly, as described in the diagram of Fig. \ref{fig:Gamma_list}).
	In detail, we update the first $ N_s $ elements of the list at each time step, the following $ N_s $ every two time steps, and the elements between $ i N_s $ and $ (i + 1) N_s $ are only updated every $ 2 ^ i $ iterations. 
	Thus, with $ n \times N_s $ elements in the list we can estimate the integral going up to $ \sum_{i=0}^{n} N_s \, 2^i dt = 2 N_s (2^n-1) dt $ in the past.
	This gives a considerable saving in computation time and memory with good accuracy as it is illustrated in Fig. \ref{fig:Gamma_optim}.
	For all the calculation presented in this paper we have used $N_s=12$.
	
\bibliographystyle{plain}
\bibliography{biblio}

\end{document}